\documentclass{article}
\usepackage{jheppub,esint,shuffle,psfrag}
 \usepackage[utf8]{inputenc}
\usepackage{xcolor}
\usepackage{varioref}
\usepackage{amsmath,amsfonts,amsthm,mathrsfs}
\usepackage{enumerate}
\usepackage{float}
\usepackage{fancyvrb}
\usepackage{verbatim}
\usepackage{wrapfig}
\usepackage{appendix}
\usepackage{amstext}
\usepackage{amssymb}
\usepackage{graphicx}
\usepackage{color}
\usepackage{varioref}
\usepackage{multirow,graphics}
\usepackage{epstopdf}

\usepackage{bbm}
\usepackage{slashed}
\usepackage{relsize}

\numberwithin{equation}{section}

\usepackage{tikz}
\usetikzlibrary{plotmarks,calc,decorations, decorations.pathmorphing, patterns}
\usetikzlibrary{arrows}
\tikzstyle dynkin node=[very thick,shape=circle,draw,inner sep=0pt,minimum size=5mm]
\tikzstyle dynkin line=[very thick]
\tikzstyle inverse line=[gray,line width=1.46pt,line cap=round, dash pattern=on 0pt off 2\pgflinewidth]
\tikzstyle red phase=[red,decoration={snake,amplitude=0.1mm,segment length=1.6mm},decorate]
\tikzstyle blue phase=[blue,decoration={snake,amplitude=0.1mm,segment length=0.9mm},decorate]
\tikzstyle green phase=[green,decoration={snake,amplitude=0.1mm,segment length=0.9mm},decorate]
\tikzstyle brown phase=[brown,decoration={snake,amplitude=0.1mm,segment length=0.9mm},decorate]
\newcommand{\boundellipse}[3]
{(#1) ellipse (#2 and #3)
}
\usetikzlibrary{decorations.pathmorphing}
\tikzstyle arrow=[thick,rounded corners=18pt,-latex]
\tikzstyle box=[draw,rounded corners,outer sep=4pt]
\tikzstyle B node=[outer sep=0pt]
\tikzstyle Q node=[inner sep=1pt,outer sep=0pt]
\definecolor{purple_nice}{rgb}{0.4,0.2,0.7}
\definecolor{fuel_blue}{RGB}{42,162,185}

\def\<{\langle}
\def\>{\rangle}

\makeatletter
\@ifundefined{usebibtex}{} {}
\makeatother

\def\Tr{\text{Tr}~}

\title{
\Large Hexagonalization of Fishnet integrals I: mirror excitations}
\author{Enrico Olivucci}
\affiliation{Perimeter Institute for Theoretical Physics, Waterloo, Ontario N2L2Y5, Canada.}

\usepackage{esint}
\usepackage{breqn}
\def \Tr {\mathop{\rm Tr}\nolimits}

\newcommand{\rmat}{\mathcal{R}}

\def\numberbysection{\@addtoreset{equation}{section}
                     \def\theequation{\thesection.\arabic{equation}}}

\begin{document}
\abstract{
In this paper we consider a conformal invariant chain of $L$ sites in the unitary irreducible representations of the group $SO(1,5)$. The $k$-th site of the chain is defined by a scaling dimension $\Delta_k$ and spin numbers $\frac{\ell_k}{2}$, $\frac{\dot{\ell}_k}{2}$. The model with open and fixed boundaries is shown to be integrable at the quantum level and its spectrum and eigenfunctions are obtained by separation of variables. The transfer matrices of the chain are graph-builder operators for the spinning and inhomogeneous generalization of squared-lattice ``fishnet" integrals on the disk. As such, their eigenfunctions are used to diagonalize the {mirror channel} of the the Feynman diagrams of Fishnet conformal field theories. The separated variables are interpreted as momentum and bound-state index of the \emph{mirror excitations} of the lattice: particles with $SO(4)$ internal symmetry that scatter according to an integrable factorized $\mathcal{S}$-matrix in $(1+1)$ dimensions.}
\maketitle
\section{Introduction}
The ultimate motivation of this work is to achieve a better understanding of integrable conformal field theories in $4d$, without appealing to the holographic correspondence with $2d$ world-sheet integrable models \cite{Beisert:2010jr}. The most suitable toy model that we know in order to pursue the investigation is the family of planar (multicolour) theories dubbed as Fishnet conformal field theories (CFTs) \cite{Gurdogan:2015csr,Caetano:2016ydc}. The Fishnet CFTs are related to the integrable supersymmetric $\mathcal{N}=4$ Yang-Mills theory by a limit of weak coupling $g$ and strong twists $\gamma_j$ of its $R$-symmetry deformation \cite{Lunin:2005jy,Frolov:2005dj,Beisert:2005if,Fokken:2014soa,Sieg:2016vap}
\begin{equation}
\label{DS}
g \to 0\,,\,\,\,\,\,\gamma_j \to i\infty\,,\,\,\,\xi_j^2 = g^2 \exp(i \gamma_j)\,;\,\,\,\, j=1,2,3\,.
\end{equation}
Moreover, they are conformal symmetric theories for every value of their couplings $\xi_j$, well defined in the double-scaling limit \eqref{DS}. Their main feature is the extremely simple content of planar Feynman diagrams entering the weak coupling expansion of correlators, which allows to handle all the contribution at once, and to compute quantities at finite coupling. In this respect, further simplification happens by setting one or two couplings $\xi_j$ to zero. 
The topology of such diagrams in the bulk is constrained to be that of a regular lattice, namely a squared-lattice for the bi-scalar Fishnet ($\xi_1=\xi_2=0$)
\begin{equation}
\label{biscalar_intro}
\begin{aligned}
    \mathcal{L}_{bi-scalar} ={}N
\,\Tr\Bigl[\xi_3^2\,\phi_1^\dagger \phi_2^\dagger
\phi_1\phi_2\!\,\Bigr]\,,\,\,\,\,\,\, \phi_i \in (N,\bar N)\,,
\end{aligned}
\end{equation}
that of an honeycomb for the $\mathcal{N}=2$ analogue of the theory \cite{Pittelli2019}, or a mix of them for a more general doubly-scaled deformation ($\xi_3=0$)
\begin{equation}
\label{chiCFT4_intro}
\begin{aligned}
    \mathcal{L}_{\chi_0} ={}N
\,\Tr\Bigl[\xi_1^2\,\phi_2^\dagger \phi_3^\dagger &
\phi_2\phi_3\!+\!\xi_2^2\,\phi_3^\dagger \phi_1^\dagger
\phi_3\phi_1\!
 +i\sqrt{\xi_1\xi_2}(\psi_2 \phi_3 \psi_{ 1}+ \bar\psi_{ 2} \phi^\dagger_3 \bar\psi_1 )\,\Bigr]\,.
\end{aligned}
\end{equation}
These properties allow for a non-perturbative re-summation of the Feynman integrals of two-point functions of single trace operators, and establish a map between the spectral problem of the dilation operator and that of the integral ``Bethe-Salpeter" kernel of the Dyson equation. The latter happens to coincide with the transfer matrix of an integrable model - the periodic non-compact conformal spin chain $SO(1,5)$ \cite{Gromov:2017cja,GrabnerGromovKazakovKorchemsky,Kazakov_2019}. 
\begin{figure}[H]
\begin{center}
\includegraphics[scale=0.62]{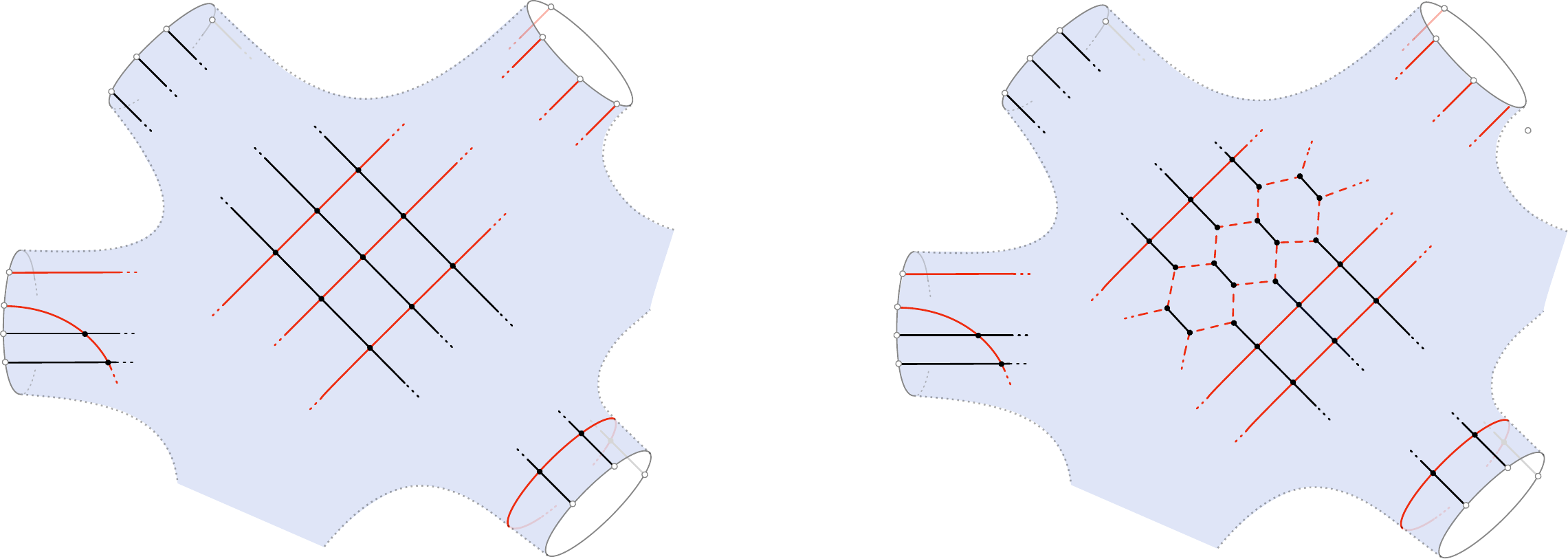}
\end{center}
\caption{\textbf{Left:} scheme of a planar Feynman integral of an $n$-point function of single-trace operators in the bi-scalar fishnet theory. The Feynman integral lies on an $n$-punctured sphere and its bulk - far from the punctures, is a square-lattice of scalar propagators. \textbf{Right:} analogue in the fishnet theory with two coupling $\xi_1,\xi_2$. The bulk of the Feynman integral mixes the topology of hexagonal lattice of Yukawa vertices and the square-lattice of quartic scalar interactions.}
\end{figure}
This statement of integrability, extensible to fishnet theories in any space-time dimension \cite{Kazakov:2018qez}, matches with the expectations and data from $\mathcal{N}=4$ SYM (see the review \cite{Kazakov_QSC}), but it is a first-principle observation and does not rely on whatsoever holography. The present work is meant to be the first paper of a series devoted to the extension of the Feynman integrals/spin-chain approach to the integrability of $n$-point functions of single-trace operators. In practise, we aim to a rigorous derivation of the hexagonalization techniques for correlation functions developed in \cite{Basso:2015zoa,Eden:2016xvg,Fleury:2016ykk, Fleury:2017eph} and deformed to the bi-scalar fishnet limit in \cite{Basso_2019}. Namely, we are going to show that in the theories at hand the entire picture of cutting-and-gluing planar correlators into hexagonal patches can be realized  at the level of Feynman graphs, and the picture of mirror excitations as particles that scatter in $(1+1)$ dimensions emerges from the computation of the graphs. The latter, as already mentioned, is mapped to the spectral problem for specific $SO(1,5)$ spin chain transfer matrices. In this paper we describe the eigenfunctions and spectrum of the transfer matrices. In fact, a fishnet graph on the disk can be regarded as a product of integral graph-building operators that are transfer matrices for the conformal spin-chain with non-periodic (\emph{``open"}) fixed boundaries.
\begin{figure}[H]
\begin{center}
\includegraphics[scale=0.8]{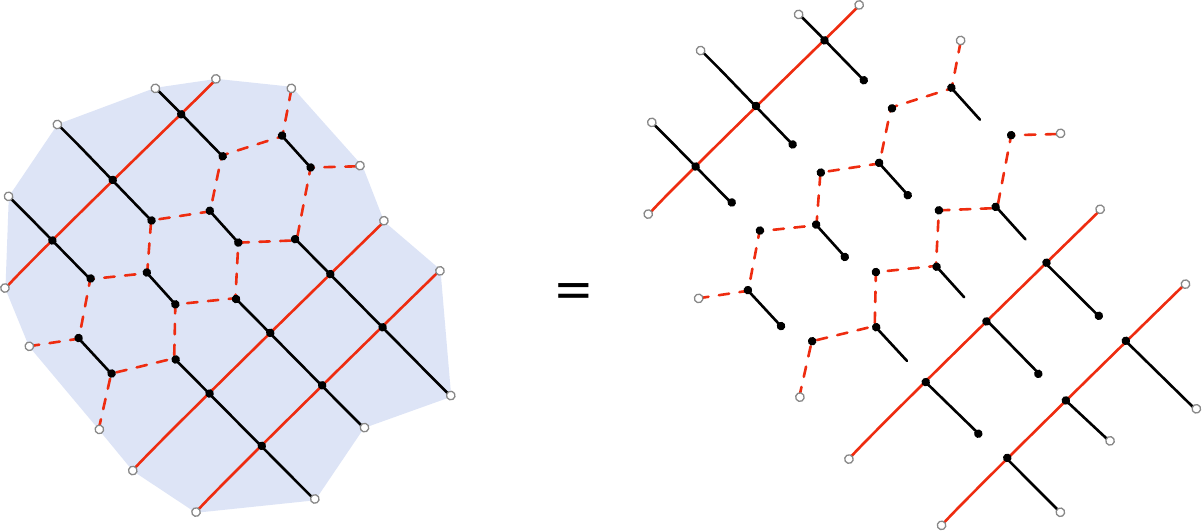}
\end{center}
\caption{Cutting a disk out of the bulk of a Feynman integral, and setting the cut legs to external - in general different - points, we have a generalized fishnet on the disk. This Feynman integral is the unique contribution to $n$-point single-trace correlators of type $\langle\text{Tr}\left[\Phi_1(x_1) \Phi_2(x_2) \cdots\Phi_n(x_n) \right]\rangle$, where $\Phi_j(x_j)$ is an elementary field of the theory. The correlators with point-split fields inside a $SU(N)$ trace are allowed by the Fishnet theories since the limit \eqref{DS} breaks the gauge invariance of $\mathcal{N}=4$ SYM. The diagrams can be decomposed into integral kernels with external fixed points (white blobs), and such \emph{graph-building} operators are in fact spin chain transfer matrices.}
\label{disk_chi}
\end{figure}
Each transfer matrix can be decomposed over a basis of \emph{mirror} excitations, whose quantum numbers are the separated variables of the lattice. The main result of the paper are the wave-functions of mirror excitations, studied in section \ref{sect:eigenf} in the general spinning and inhomogeneous chain. This model is introduces in section \ref{sect:fixed_boundary} by an integrable twist to fixed boundaries (mirror model) of the periodic chain of \cite{Derkachov:2021rrf}. The reason of such generalization is twofold: first, it tells about the complete integrability of the spin-chain model as we diagonalize transfer matrices with higher-spin auxiliary space (section \ref{sect:spinning}). Secondly, we need a more general lattice in order to describe the more general fishnet theories, i.e. of an inhomogeneous chain whose sites are in the representations of a scalar and fermion fields. This fact is established in section \ref{sect:graph_builder} where the graph-builder transfer matrices for fishnet graphs are listed, relating Feynman integrals to conformal chains. The picture that emerges from the Feynman graph/spin-chain computations is that of a $(1+1)d$ integrable (factorized) scattering between mirror excitations with internal symmetry $SO(4)\sim SU(2)\times SU(2)$. Its discussion, in the last section \ref{sect:scatter}, links our language to that proper of hexagonalization \cite{Basso:2015zoa}. The computation of hexagon form factors, the explicit tessellation of planar $n$-point correlators and the gluing-back of hexagonal patches, rely entirely on the knowledge of a basis of mirror excitations and their overlap, and is the topic of the part II of the series \cite{Olivucci_hex_II}. To start with, in the next section \ref{sect:spin_chain} we introduce the relevant spin chain model $SO(1,5)$ and the graphical computation technique based on the star-triangle duality of a conformal invariant vertex - largely recalling the results of \cite{Chicherin:2012yn, Derkachov:2021rrf}. We dedicate a few appendices to list useful integral identities and to the most cumbersome proof of equations of the main text. 
\section{Conformal quantum chain}
\label{sect:spin_chain}
Let us consider a quantum particle moving in the $4d$ Euclidean space with left/right spins $\frac{\ell}{2}$ and $\frac{\dot{\ell}}{2}$. Its state is described by a wave function of the position $x=(x_1,x_2,x_3,x_4)$ with $\ell$-fold and $\dot{\ell}$-fold symmetric spinor indices
\begin{align}
\begin{aligned}
\label{wave_f}
 \Phi_{\mathbf{a\dot{a}}}(x)=\Phi_{(a_1\dots a_{\ell})(\dot{a}_1\dots\dot{a}_{\dot{\ell}})}(x)\,,\,\,\,\,\, a_i,\dot{a}_i 
\in \{1,2\}\,.
\end{aligned}
\end{align}
The  wave-functions \eqref{wave_f}  belong to the Hilbert space
\begin{equation}
\label{Hilbert}
\mathbb{V}\simeq L^2(d^4x)\otimes \text{Sym}_{\ell}[\mathbb{C}^2] \otimes \text{Sym}_{\dot{\ell}}[\mathbb{C}^2]\,,
\end{equation} 
for $\text{Sym}_{
\ell}[\mathbb{C}^2]\subset (\mathbb{C}^{2})^{\otimes \ell} $ the space of complex symmetric spinors $u_{\mathbf{a}}=u_{(a_1,\dots,a_{\ell})}$.  The scalar product on \eqref{Hilbert} is inherited from the standard ones defined on its factors, that is
\begin{equation}
\label{scalar_product}
\langle F , G\rangle_{\mathbb{V}} \,=\,  \int d^4 x\, \langle F(x) |G(x) \rangle\, \equiv \,\int d^4 x\, (F^*)^{\mathbf{a}\mathbf{\dot{a}}}(x) \,G_{\mathbf{a}\mathbf{\dot{a}}}(x)\,,
\end{equation}
where repeated spinor indices are summed over.
An isometry of the Euclidean space $x^{
\mu} \mapsto \Lambda_{\nu}^{\mu}x^{\nu}$ rotates the wave-function according to the decomposition $SO(4)\sim SU(2)\times SU(2)$, by two matrices $U,V$ acting on left/right spinor indices (see appendix \ref{app:spinors})
\begin{equation}
 \Lambda_{\nu}^{\,\mu} (\boldsymbol{\sigma}_{\mu})_{a}^{\dot a} \,=\,U_{a}^{b} (\boldsymbol{\sigma}_{\mu})_{b}^{\dot b} (V^{\dagger})_{\dot b}^{\dot a}\,, \,\,\,\,\text{where}\,\,\,\boldsymbol{\sigma}_k = i\sigma_k\,,\,\,\, \overline{\boldsymbol{\sigma}}_0=\mathbbm{1}\,,\,\overline{\boldsymbol{\sigma}}_k={\boldsymbol{\sigma}}_k^{\dagger}\,,
\end{equation}
\begin{align}
\label{spinor_rotation}
\begin{aligned}
&u_{\mathbf{a}} \mapsto u_{\mathbf{a}} '=[U]^{\mathbf{b}}_{\mathbf{a}} u_{\mathbf{b}}\,,\,\,\,\,\,\,\,\, v_{\mathbf{\dot a}} \mapsto v_{\mathbf{\dot a}} '=[V]^{\mathbf{\dot b}}_{\mathbf{\dot a}} v_{\mathbf{\dot b}}\,,
\end{aligned}
\end{align}
and the following notation for symmetric representations will be frequently used throughout the text
\begin{align}
\begin{aligned}
\label{spin_notation}
([U]^{\ell})_{\mathbf{a}}^{\mathbf{b}} = [U]_{\mathbf{a}}^{\mathbf{b}} = U_{(a_1}^{(b_1}\cdots U_{a_{\ell})}^{b_{\ell})}\,.
\end{aligned}
\end{align}
We are interested in quantum systems with conformal symmetry,  that is into particles evolving under the action of an Hamiltonian operator $\mathcal{H}$ which is invariant under a conformal change of coordinates $x^{\mu} \mapsto y^{\mu}(x)\,|$
\begin{equation}
\frac{\partial y^{\mu}}{\partial x^{\kappa}}\frac{\partial y_{\nu}}{\partial x_{\kappa}}= \lambda(y)\,\delta^{\mu}_{\nu}\,,\,\,\,\,\,\, \lambda(y) =\left|\det\left(\frac{\partial{y^{\mu}}}{ \partial x^{\nu}}\right)\right|\,.
\end{equation}
The eigenstates of a conformal invariant system transform under an irreducible unitary representation  \cite{Tod:1977harm} of the conformal group
\begin{equation}
\label{princ_rep}
\Phi_{\mathbf{a\dot{a}}}(x)\,\mapsto\, {\Phi'}_{\mathbf{a\dot{a}}}(x)\,=\lambda(y)^{\Delta}\,[U]_{\mathbf{a}}^{\,\mathbf{b}} [V]^{\,\mathbf{\dot{b}}}_{\mathbf{\dot{a}}} \,\Phi_{\mathbf{b \dot{b}}}(y) \,,
\end{equation}
and without loss of generality we can restrict to the representations of the principal series,  that means $\Delta=2+i\nu$ with $\nu$ on the real line\footnote{The unitary irreducible representations of $SO(1,5)$ would include also the complementary series,  for which $0<\Delta<4$. Our results throughout the paper can extended to this latter case by analytic continuation of $\nu$ to the imaginary axis.}. The evolution of the quantum system respect to a certain parameter $u$ is given by the action of an evolution operator $t(u)$ - that in a chain model is dubbed \emph{transfer matrix operator} -
which in the case at hand is a conformal invariant function of operators position $x^{\mu}$ and momenta $\hat p^{\mu}$, hence the evolution is unaffected by a conformal transformation $x^{\mu} \mapsto y^{\mu}(x)$
\begin{equation}
\Phi'_{\mathbf{a\dot{a}}}(u,x^{\mu}) = (t(u)\Phi_{\mathbf{a\dot{a}}})' (0,x^{\mu}) = t(u)\Phi'_{\mathbf{a\dot{a}}}(0,x^{\mu})\,.
\end{equation}
In the next sections we consider the integrable conformal chain studied in \cite{Chicherin:2013rma,Derkachov:2021rrf}, that is a system of $L$ particles with a nearest-neighbour interaction defined starting from a solution of the Yang-Baxter equation for the conformal group $SO(1,5)$. 
\subsection{Conformal $\rmat$-operator and Star-Triangle duality}
Let us summarize here the properties of the $\rmat$-operator and the underlying star-triangle dualities following the analysis of \cite{Derkachov:2021rrf}. An explicit solution $\rmat_{ij}(u)$ of the Yang-Baxter equation
\begin{equation}
\label{YBE}
\rmat_{12}(u-v) \rmat_{13}(u) \rmat_{23}(v)= \rmat_{23}(v) \rmat_{13}(u) \rmat_{12}(u-v)\,,
\end{equation}
acting on the space $\mathbb{V}_i\otimes \mathbb{V}_j$ of two particles in the principal series representations $(\Delta_i,\ell_i,\dot\ell_i)$ and $(\Delta_j,\ell_j,\dot\ell_j)$ can be expressed following \cite{Chicherin:2013rma} in terms of the operators position $x^{\mu}$  and momentum $\hat p_{\mu}=i \partial_{\mu}$ of the two particles. For this and later scope, we introduce a notation for (operator valued) $SU(2)$ matrices \begin{equation}
\label{SU(2)_x_p}
\mathbf{x}=\boldsymbol{\sigma}_{\mu}\frac{x^{\mu}}{(x^2)^{\frac{1}{2}}}\,,\,\,\,\, \mathbf{\overline x}=\boldsymbol{\overline{\sigma}}_{\mu}\frac{x^{\mu}}{(x^2)^{\frac{1}{2}}}\,,\,\,\,\,\;\mathbf{p}=\boldsymbol{\sigma}_{\mu}\frac{\hat p^{\mu}}{(\hat p^2)^{\frac{1}{2}}}\,,\,\,\,\,\mathbf{\overline p}=\boldsymbol{\overline{\sigma}}_{\mu}\frac{\hat p^{\mu}}{(\hat p^2)^{\frac{1}{2}}}\,,
\end{equation}
and with the notation $[\cdot]$ of formula \eqref{spin_notation} the same group elements in the $n$-symmetric representation read $[\mathbf{x}]^n$ and $[\mathbf{\bar x}]^n$.
A non-integer power $u$ of the laplacian $-\hat{p}^2$, as the terms appearing in \eqref{SU(2)_x_p}, is defined via Fourier transformation by the integral operator
\begin{equation}
\label{p_def}
\hat{p}^{2u} f(x)=4^u\frac{\Gamma(
u+2)}{\pi^2\Gamma(-u)}\int d^4 x \frac{f(y)}{(x-y)^{2(u+2)}}\,.
\end{equation}
The solution $\rmat_{12}(u)$ then can be compactly written as
\begin{align}
\begin{aligned}
\label{R_mat}
\rmat_{12}(u)  = &\mathbb{P}_{12}
\frac{[\mathbf{\overline{x}}_{12}]^{\dot{\ell}_1}
\mathbf{R}_{\dot{\ell}_1\ell_2}(u-\Delta_{+})
[\mathbf{x}_{12}]^{\ell_2}}{x_{12}^{2\left(-u+\Delta_{+}\right)}}
\,\frac{[\mathbf{\overline{p}}_{2}]^{\ell_2}
\mathbf{R}_{\ell_2\ell_1}(u+\Delta_{-})
[\mathbf{p}_{2}]^{\ell_1}}{\hat{p}_{2}^{2(-u-\Delta_{-})}} \times \\ &\times\frac{[\mathbf{\overline{p}}_{1}]^{\dot{\ell}_2}
\mathbf{R}_{\dot{\ell}_2\dot{\ell}_1}(u-\Delta_{-})
[\mathbf{p}_{1}]^{\dot{\ell}_1}}{\hat{p}_{1}^{2(-u+\Delta_{-})}}
\frac{[\mathbf{\overline{x}}_{12}]^{\ell_1}\mathbf{R}_{\ell_1\dot{\ell}_2}(u+\Delta_{+})
[\mathbf{x}_{12}]^{\dot{\ell}_2}}{x_{12}^{2\left(-u-\Delta_{+}\right)}}
\,,
\end{aligned}
\end{align}
where $\Delta_{-}=\frac{\Delta_1-\Delta_2}{2}$,
$\Delta_{+}=\frac{\Delta_1+\Delta_2}{2}-2$.
The matrix $\mathbf{R}_{mn}(u)$ that multiplies $[\mathbf{x}]^n,\,[\mathbf{\bar x}]^m$ and $[\mathbf{p}]^n,\,[\mathbf{\bar p}]^m$ in the numerators of \eqref{R_mat} is the $SU(2)$-invariant solution of Yang-Baxter equation acting on the $n$-fold and $m$-fold symmetric representation, and defined via \emph{fusion} procedure \cite{Kulish:1981gi} (see also the appendix of \cite{Derkachov:2020zvv})
\begin{equation}
\label{RR_fused}
(\mathbf{R}_{mn})_{\mathbf{ac}}^{\mathbf{bd}}(u) =
\mathbf{R}_{(a_1\dots a_m)(c_1\dots c_n)}^{(b_1\dots b_m)(d_1\dots d_n)}(u)\,,
\end{equation}
\begin{align}
\begin{aligned}
\label{Rmat_YBE}
\mathbf{R}_{nm}(u-v)\mathbf{R}_{n\ell}(u)\mathbf{R}_{m\ell}(u)=
\mathbf{R}_{m\ell}(v)\mathbf{R}_{n\ell}(u)\mathbf{R}_{nm}(u-v)\,.
\end{aligned}
\end{align}
The matrix \eqref{RR_fused} satisfies also unitarity 
\begin{align}
\begin{aligned}
\label{R_unit}
&\mathbf{R}_{nm}(u)\mathbf{R}_{nm}(-u)= \mathbbm{1}_{n}\otimes \mathbbm{1}_{m}\,,\,\,\,\,\left(\mathbf{R}_{nm}(u)\right)^{\dagger} = \mathbf{R}_{nm}(u^*)\,,\\ 
\end{aligned}
\end{align}
and the crossing equation
\begin{align}
\begin{aligned}
\label{R_cross}
\left(\mathbf{R}_{mn}(u)\right)^{t_n}=\left(\mathbf{R}_{mn}(u)\right)^{t_m}=r_{mn}(u)  (\boldsymbol{\sigma}_2)^{\otimes n} \mathbf{R}_{mn}(-u-1)(\boldsymbol{\sigma}_2)^{\otimes n}\,,
\end{aligned}
\end{align}
where $t_m$, $t_n$ are the transposition in the space of $m-$ or $n-$ symmetric spinors, $\boldsymbol{\sigma}_2$ defines the charge conjugation and the crossing factor $r_{mn}(u)$ is
\begin{equation}
r_{mn}(u)=(-1)^{m n}\frac{\Gamma\left(u+\frac{m+n}{2}+1\right)\Gamma\left(u-\frac{m+n}{2}+1\right)}{\Gamma\left(u+\frac{m-n}{2}+1\right)\Gamma\left(u+\frac{n-m}{2}+1\right)}\,.
\end{equation}
From the definition of $\rmat$-operator \eqref{R_mat} and the properties \eqref{R_unit} it follows that
\begin{equation}
\label{R_hc}
\rmat_{12}(-u) = \rmat_{12}^{-1} (u)= \rmat_{12}(u) \,,\,\,\,\,\,\,\,\,\,\rmat_{12}(u)^{\dagger} \,\equiv\,\rmat_{12}(u)^{\dagger_1\dagger_2} \,=\,\rmat_{\dot1 \dot 2}(u^*)\,,
\end{equation} 
where the dotted subscripts denote the exchange of left/right spins $\ell_1\leftrightarrow \dot{\ell}_1$ and $\dot{\ell}_2 \leftrightarrow {\ell}_2$. 
Recalling the representation \eqref{princ_rep}, it is straightforward to check the conformal invariance of the operator $\rmat_{12}(u)$
\begin{equation}
\label{rmat_inv}
\rmat_{12}(u)\, \mapsto \left( \prod_{k=1,2} \lambda(x_k)^{\Delta_k}\, [U]^{\ell_k}  [V]^{\dot \ell_k}\right) \rmat_{12}(u) \left( \prod_{k=1,2} [U^{\dagger}]^{\ell_k}[V^{\dagger}]^{\dot \ell_k}\,\lambda(x_k)^{-\Delta_k}\right)\,,
\end{equation}
based on the $SU(2)$ invariance of the fused $\mathbf{R}$-matrix and on the transformation property of a conformal propagator
\begin{equation}
\frac{[\mathbf{x}]^{\ell}[\mathbf{\overline x}]^{\dot \ell} }{(x^2)^{\Delta}}\,\longrightarrow \,\lambda(x)^{-2\Delta} \,\frac{[U\mathbf{x}V^{\dagger}]^{\ell}[V\mathbf{\overline x}U^{\dagger}]^{\dot \ell} }{(x^2)^{\Delta}} \,.
\end{equation}

\begin{figure}[H]
\begin{center}
\includegraphics[scale=1.8]{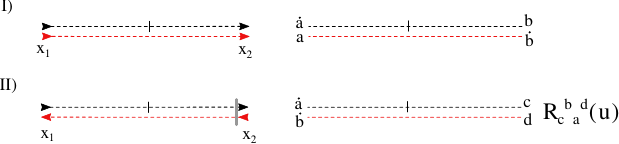}
\caption{\textbf{(I) Left}: Graphic representation of the conformal propagator $x^{2u}\,[\mathbf{\overline{x}}]^{\ell}
[\mathbf{x}]^{\dot{\ell}}$ with conformal dimension $-u$ and spins $\ell,\dot{\ell}$. The two dashed lines stand for the two spinorial matrices $\mathbf{x}$ and $\mathbf{\overline{x}}$, the one with the bar denoting ${\mathbf{\overline x}}$.  The arrows denote the flow of spinor indices. \textbf{(I) Right}: Explicit spinor structure corresponding to the notation of arrows and dashed lines.
\textbf{(II) Left}: Graphic representation of $\mathbf{R}$-matrix in $x^{2u}\,[\mathbf{\overline{x}}]^{\ell} 
\mathbf{R}_{\ell\dot \ell}(u)[\mathbf{x}]^{\dot{\ell}}$. The two dashed lines stand for the two spinorial matrices $\mathbf{x}$ and $\mathbf{\overline{x}}$. The arrows denote the flow of spinor indices and the grey thick segment defines where $\mathbf{R}(u)$ is inserted along the spinorial matrix structure. \textbf{(II) Right}: Explicit spinor structure.}
\label{fig_propag1}
\end{center}
\end{figure}

\begin{figure}[H]
\begin{center}
\includegraphics[scale=1.2]{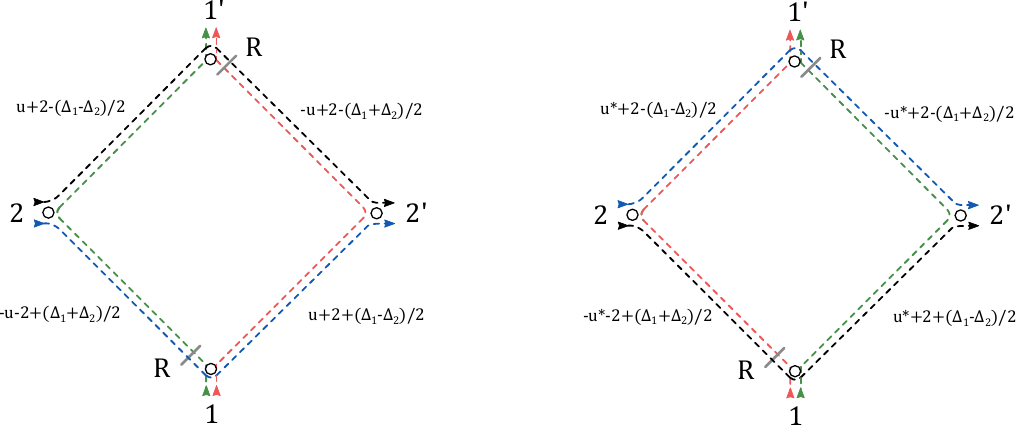}
\end{center}
\caption{\textbf{Left:} diagram representation of the kernel $\rmat_{12}(x_1,x_2|x_1',x_2')$ of the operator $\rmat_{12}(u)$. \textbf{Right:} diagram representation of the kernel of the operator $\rmat_{12}(u)^{\dagger}=\rmat_{21}(u^*)$. The position of different spinor spaces - identified by differenet colours of dashed lines - gets exchanged by the hermitian conjugation.
}
\label{R_mat_YBE}
\end{figure}

\begin{figure}[H]
\begin{center}
\includegraphics[scale=1.3]{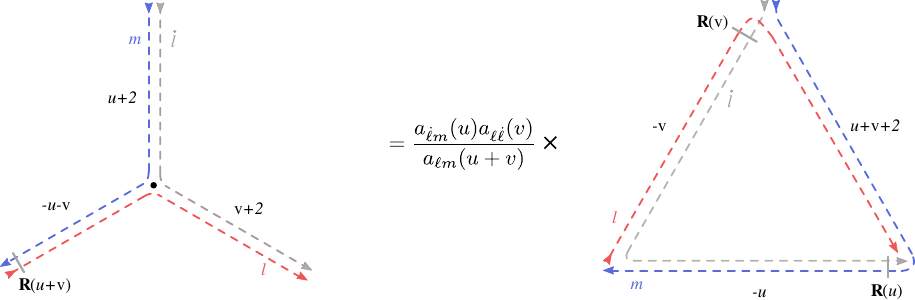}
\end{center}
\caption{Star-triangle identity \eqref{STR_ker}: on the left a cubic vertex of conformal propagators with scaling dimensions $u+2\,,v+2\,,-u-v$, and left/right spins $(
\dot \ell,m)\,,(m,\ell)\,,(\ell,\dot \ell)$ - the \emph{star}. On the right, the equivalent \emph{triangle} of conformal propagators that extend between the vertices of the star, with dual scaling dimensions $-u\,,-v\,,u+v+2$. In the general case of propagators of spinning fields we use the double-dashed line notation and different colours for different spins. Both in the star and in the triangle the spinor indices are mixed by the action of fused $SU(2)$ $\mathbf{R}$-matrices \eqref{RR_fused}. For zero spins, the identity boils down to its scalar version \cite{DEramo:1971hnd, Kazakov:1983ns, Kazakov:1984km}.}
\label{STR_1}
\end{figure}

Besides \eqref{YBE}, the operator $\rmat_{ij}(u)$ satisfies another non-trivial algebraic relation involving its adjoint \cite{Derkachov:2021rrf}, crucial in order to define diagonalizable transfer matrices
\begin{align}
 \begin{aligned}
 \label{YBE_adj}
&\rmat_{12}(u){\rmat}_{32}(u+v^*)\rmat_{13}(v)^{\dagger}\,=\,\rmat_{13}(v)^{\dagger}{\rmat}_{32}(u+v^*)\rmat_{12}(u)\,.
 \end{aligned}
 \end{align}
Making use of the integral representation \eqref{p_def}, the operator \eqref{R_mat} can be introduced in an equivalent fashion as an integral operator with the kernel
\begin{align}
\begin{aligned}
\label{R_integral_form}
&\rmat(x_1,x_2|x'_1,x'_2)=\frac{(1-u+\frac{\ell+\dot{\ell}}{2})\Gamma\left(1-u+\frac{\dot{\ell}-\ell}{2}\right)\Gamma\left(1-u+\frac{\dot{\ell}-\ell}{2}\right)}{\pi^4\Gamma\left(u+\frac{\dot{\ell}-\ell}{2}\right)\Gamma\left(u+\frac{\dot{\ell}-\ell}{2}\right)}\times \\& \times
\frac{[\mathbf{\overline{x}}_{21}]^{\dot{\ell}_1}
\mathbf{R}_{\dot{\ell}_1\ell_2}(u-\Delta_{+})
[\mathbf{x}_{21}]^{\ell_2}}{x_{12}^{2\left(-u+\Delta_{+}\right)}}
\frac{[\mathbf{\overline{x}}_{12'}]^{\ell_2}
[\mathbf{x}_{12'}]^{\ell_1}}
{x_{12'}^{2\left(u+\Delta_{-}+2\right)}}
\frac{[\mathbf{\overline{x}}_{21'}]^{\dot{\ell}_2}
[\mathbf{x}_{21'}]^{\dot{\ell}_1}}
{x_{21'}^{2\left(u-\Delta_{-}+2\right)}}
\frac{[\mathbf{\overline{x}}_{1'2'}]^{\ell_1}\mathbf{R}_{\ell_1\dot{\ell}_2}(u+\Delta_{+})
[\mathbf{x}_{1'2'}]^{\dot\ell_2}}
{x_{1'2'}^{2\left(-u-\Delta_{+}\right)}}
\,,
\end{aligned}
\end{align}
where $\Delta_+ =(\Delta_1+\Delta_2)/2-2$ and $\Delta_- =(\Delta_1-\Delta_2)/2$. 
The formula \eqref{R_integral_form} allows for a graphical representation of $\rmat$-operator via the notation of lines and dots used for Feynman integrals. Indeed the function $x^{2u}\,[\mathbf{\overline{x}}]^{\ell}
[\mathbf{x}]^{\dot{\ell}}$ is the propagator of a conformal field~\cite{Sotkov:1976xe} of scaling dimension $\Delta=-u$ and left/right spins $\ell,\dot{\ell}$ - stripped of all constants - and we represent it with a double-dashed line as in Fig.\ref{fig_propag1} (I). The mixing of spinor by the $\mathbf{R}$-matrix in $x^{2u}\,[\mathbf{\overline{x}}]^{\ell} 
\mathbf{R}_{\ell\dot \ell}(u)[\mathbf{x}]^{\dot{\ell}}$ is illustrated in Fig.\ref{fig_propag1} (II).
The notation of by Fig.\ref{fig_propag1} is ubiquitous in our computations since it allows for a compact representation of complicated integral kernels in the form of Feynman graphs, and translates tedious integral identities into elementary \emph{moves} of lines of the graph. For example, the kernel \eqref{R_integral_form} takes the form of a square of propagators depicted in Fig.\ref{R_mat_YBE}. 
\noindent
The proof of Yang-Baxter equation \eqref{YBE} is a consequence of the generalized star-triangle duality
\begin{align}
\begin{aligned}
\label{STR_operatorial0}
&\hat p^{2u}\,[\mathbf{ \overline{p}}]^{\dot{\ell}}\mathbf{R}_{m\dot{\ell}}(u)
[\mathbf{p}]^{m}\,\,
x^{2(u+v)}\,[\mathbf{ \overline{x}}]^{m}\mathbf{R}_{m \ell}(u+v)
[\mathbf{x}]^{\ell}\,\,
\hat p^{2v}\,[\mathbf{\overline{ p}}]^{\ell}\mathbf{R}_{\ell\dot{\ell}}(v)
[\mathbf{p}]^{\dot{\ell}}\, = \\
&x^{2v}\,[\mathbf{ \overline{x}}]^{\dot{\ell}}\mathbf{R}_{\ell\dot{\ell}}(v)
[\mathbf{x}]^{\ell}\,\,
\hat p^{2(u+v)}\,[\mathbf{ \overline{p}}]^{\ell}\mathbf{R}_{m\ell}(u+v)
[\mathbf{p}]^{m}\,\,
x^{2u}\,[\mathbf{ \overline{x}}]^{m}\mathbf{R}_{m \dot{\ell}}(u)
[\mathbf{x}]^{\dot{\ell}}\,\,.
\end{aligned}
\end{align}
The propagators involved in both sides of \eqref{STR_operatorial0} carry spin indices in the spaces
\begin{equation}
\text{Sym}_{m}[\mathbb{C}^2]\otimes \text{Sym}_{\ell}[\mathbb{C}^2]\otimes \text{Sym}_{\dot \ell}[\mathbb{C}^2]\,,
\end{equation}
and are pair-wise mixed by the $\mathbf R$-matrix acting on the product of $SU(2)$ symmetric tensors. Applying the both sides of \eqref{STR_operatorial0} to the delta-function $\delta^{(4)}(x-z)$ we obtain the following identity for integral kernels
\begin{multline}\label{STR_ker}
\int d^4 y\,\frac{[\mathbf{\overline{x-y}}]^{\dot{\ell}}
[\mathbf{x-y}]^{m}}{(x-y)^{2(u+2)}}
\frac{[\mathbf{ \overline{y}}]^{m}\mathbf{R}_{m \ell}(u+v)
[\mathbf{y}]^{\ell}}{y^{-2(u+v)}}
\frac{[\mathbf{\overline{y-z}}]^{\ell}
[\mathbf{y-z}]^{\dot{\ell}}}{(y-z)^{2(v+2)}} = \\ = \pi^2 \frac{a_{\dot{\ell}m}(u)a_{\ell\dot{\ell}}(v)}{a_{\ell m}(u+v)}
\frac{[\mathbf{ \overline{x}}]^{\dot{\ell}}\mathbf{R}_{\ell \dot{\ell}}(v)
[\mathbf{x}]^{\ell}}{x^{-2v}}
\frac{[\mathbf{\overline{x-z}}]^{\ell}
[\mathbf{x-z}]^{m}}{(x-z)^{2(u+v+2)}}
\frac{[\mathbf{ \overline{z}}]^{m}\mathbf{R}_{m \dot{\ell}}(u)
[\mathbf{z}]^{\dot{\ell}}}{z^{-2u}}\,,
\end{multline}
where $a_{\ell\dot{\ell}}(u)$ is a prefactor \begin{align}
a_{\ell\dot{\ell}}(u) =  \frac{(i)^{\dot \ell-\dot{\ell}}\,\Gamma\left(-u-\frac{\ell+\dot{\ell}}{2}\right)}
{\,\left(u+\frac{\ell+\dot{\ell}}{2}+1\right)\,
\Gamma\left(u+1+\frac{\ell-\dot{\ell}}{2}\right)}\,.
\end{align}
\noindent
The form \eqref{STR_ker} of the duality - obtained in \cite{Derkachov:2021rrf} - can be represented compactly in the graphical notation of propagators and vertices as in Fig.\ref{STR_1}. 
As it was pointed out in \cite{Derkachov2020,Derkachov:2020zvv,Derkachov:2021rrf}, the diagram notation defined in Fig.\ref{fig_propag1} is way more compact than analytic formulas, especially when dealing with $SU(2)$ symmetric spinors and their indices. The star-triangle identity can be cast in a few equivalent forms, obtained via conformal boosts and/or reshuffling the position of $\mathbf{R}$-matrices via unitarity and crossing symmetry, and we list some of them in the appendix \ref{app:str_amp}. 
\begin{figure}[H]
\begin{center}
\includegraphics[scale=1.0]{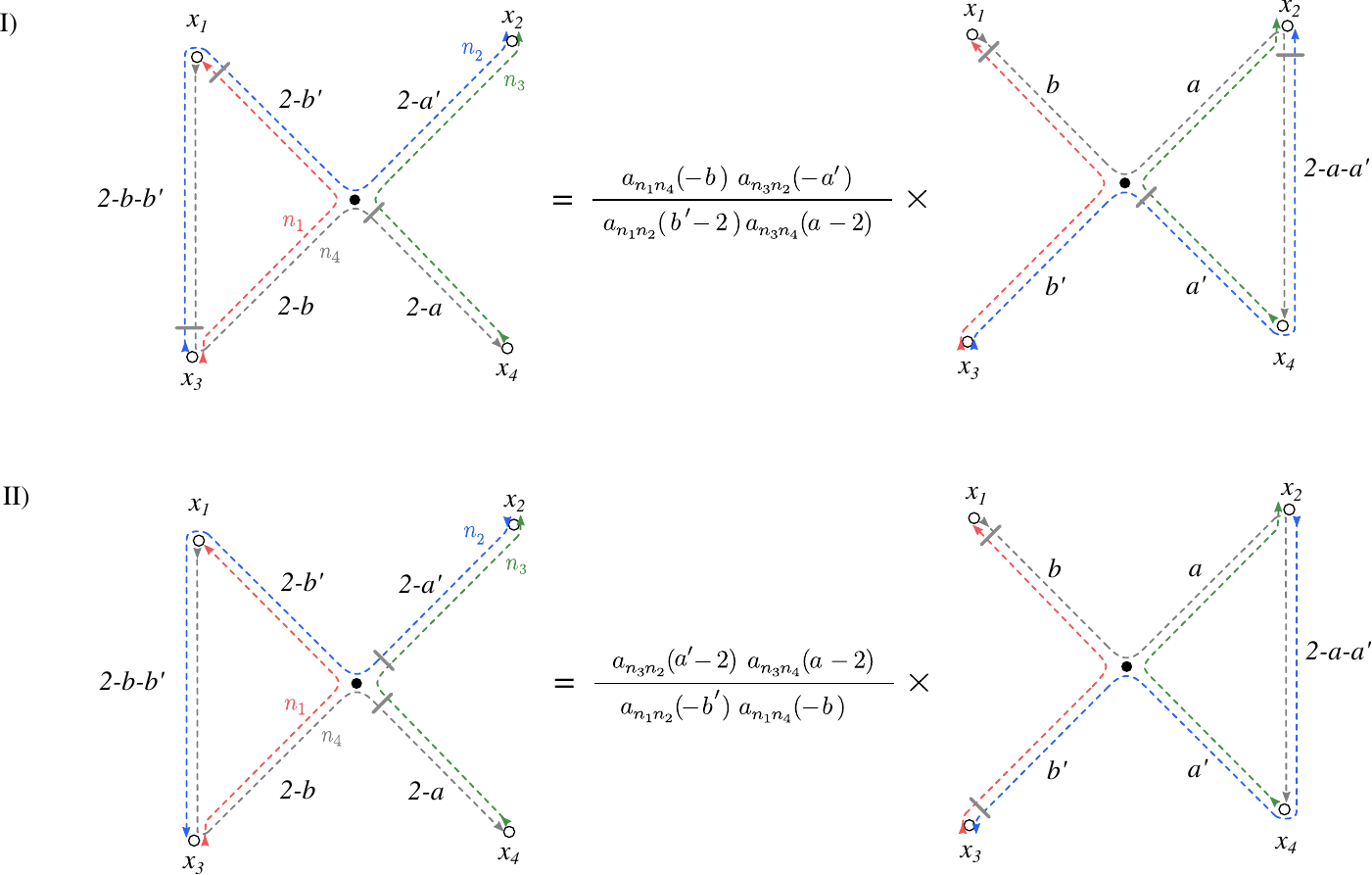}
\end{center}
\caption{Diagram of the interchange relations I and II, valid under the constraint $a+a'=4-b-b'$. The move of a vertical propagator across the quartic vertex exchanges the blue/gray spinorial structures and changes the weights of the propagators of the vertex.}
\label{interchanges}
\end{figure}
Moreover \eqref{STR_ker} is the building block of more involed identities that describe the move of a spinning propagator across a quartic vertex; two of them are listed in Fig.\ref{interchanges} and are referred to as \emph{interchange relations}. The latter are the fundamental step of several spin-chain computations in section \ref{sect:eigenf}.
\subsection{Inhomogeneous spinning Fishnet}
The solution $\rmat_{ij}(u)$ of the Yang-Baxter equation acting on any two unitary irreps of the conformal group $SO(1,5)$ is the building block of a family of integrable non-compact spin chains with $SO(1,5)$ symmetry \cite{Chicherin:2013rma,Derkachov:2021rrf}. These are systems of (quasi-)particles labelled by an index $k=1,\dots,L$, which interact only with their nearest neighbours $k+1$ and $k-1$. Each particle is carachterized by a scaling dimension $\Delta_k$ and by left/right spin numbers $\ell_k$ and $\dot{\ell}_k$, in a given representation of the principal series. Therefore, the quantum states of a spin chain with $L$ sites are wavefunctions in the Hilbert space \cite{Derkachov:2021rrf}
\begin{equation}
\label{Hilbert_chain}
\mathcal{V}=\mathbb{V}_1\otimes \mathbb{V}_2\otimes \cdots \otimes \mathbb{V}_L \,,
\end{equation}
where each $\mathbb{V}_k$ is the module of the representation $\left(\Delta_k,\ell_k,\dot{\ell}_k\right)$ - in general different at each site - and the scalar product on $\mathcal{V}$ is inherited by those on the sites $\mathbb{V}_k$ (see \eqref{scalar_product}). 
Each wavefunction depends on the position of the particles $x^{\mu}_k\,\in\,\mathbb{R}$ and carries $(\ell_k,\dot{\ell}_k)$-symmetric spinor indices $\mathbf{a}_k$, $\mathbf{\dot{a}_k}$; under a conformal change of coordinates it transforms in the tensor product of the representations defined at each site $k$
\begin{equation}
\Phi_{\mathbf{a}_1\mathbf{\dot{a}}_1\dots \mathbf{a}_L\mathbf{\dot{a}}_L}(x_1,\dots,x_L)\,,\,\,\,\,\,\mathbf{a}_k = (a_{k,1},\dots,a_{k,\ell_k})\,,\,\,\,\mathbf{\dot a}_k = (\dot a_{k,1},\dots,\dot a_{k,\dot \ell_k})\,.
\end{equation}
The Hamiltonian operator of the model with closed boundary conditions, i.e. where the $(L+1)$-th and the first particle are identified\footnote{This condition can be relaxed by the introduction of a twist at the boundary while preserving the integrability of the model, as for the model of section \ref{sect:fixed_boundary}.}, can be introduced starting from the definition of a \emph{monodromy matrix} operator
\begin{equation}
\label{monodromy_op}
\mathbb{T}_{1,\dots,L,a}(u)\,=\,\rmat_{1a}(u+\theta_1)\rmat_{2a}(u+\theta_2)\cdots \rmat_{La}(u+\theta_L)\,,
\end{equation}
which act on the physical space of system $\mathcal{V}$ and on an auxiliary space $\mathbb{V}_a$ - the module of a unitary irrep $(\Delta_a,\ell_a,\dot{\ell}_a)$. The \emph{inhomogeneity} parameters $\theta_k$ are complex numbers to be determined.
\begin{figure}[H]
\begin{center}
\includegraphics[scale=1]{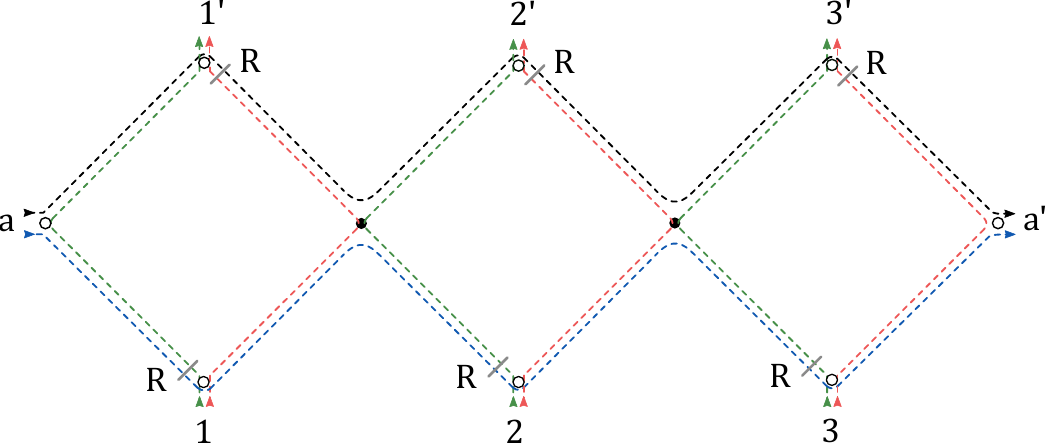}
\end{center}
\caption{Integral kernel of the monodromy matrix operator $\mathbb{T}_{1,2,3,a}(u)$ obtained from the convolution of operators $\rmat_{ka}(u+\theta_k)$ in the auxiliary space $\mathbb{V}_a$ for $k=1,2,3$. The black blobs are integrated, the circles are external coordinates. Black and blue lines denote the propagation of  auxiliary space Weyl spinors $(\ell_a,0)$ and $(0,\dot{\ell}_a)$, while the red and green lines are Weyl spinor propagators in the representations $(\ell_k,0)$ or $(0,\dot{\ell}_k)$ in the physical spaces $\mathbb{V}_k$, $k=1,\dots,L$. The mixing of spinorial indices between auxiliary (black, blue) and physical (red, green) representations is given by the $SU(2)$ R-matrices denoted with a grey line.}
\end{figure}
It follows from the YBE \eqref{YBE} that the monodromy operator satisfies the $RTT$ relation (Yangian algebra)
\begin{equation}
{\rmat}_{ab}(u-v) \mathbb{T}_{1,\dots,L,a}(u)\mathbb{T}_{1,\dots,L,b}(v)\,=\,\mathbb{T}_{1,\dots,L,a}(u) \, \mathbb{T}_{1,\dots,L,b}(v)\, {\rmat}_{ab}(u-v)\,,
\end{equation}
which in turn implies to the commutation relation of the \emph{transfer matrix} operators defined as the infinite dimensional trace of \eqref{monodromy_op} over the auxiliary spaces $\mathbb{V}_a\otimes \mathbb{V}_b$.
\begin{equation}
\label{t_comm}
t^{(a)}(u) t^{(b)}(v) =  t^{(b)}(v) t^{(a)}(u)\,,
\end{equation}
\begin{equation}
\label{t_matrix_op}
t^{(a)}(u)\equiv t^{(a)}_{1,\dots,L}(u)\,\equiv \,\text{Tr}_{\mathbb{V}_a} \left(\mathbb{T}_{1,\dots,L,a}(u)\right)\,.
\end{equation}
The conformal invariance of the model is inherited via its transfer matrices from the invariance of the $\rmat$-operator \eqref{rmat_inv}, and reads
\begin{equation}
\label{t_invariance}
t^{(a)}(u)\,\mapsto\, \left(\prod_{k=1}^L \lambda(x_k)^{\Delta_k} [U]^{\ell_k} [V]^{\dot \ell_k}\right) t^{(a)}(u)\left(\prod_{k=1}^L \lambda(x_k)^{\Delta_k} [U]^{\ell_k} [V]^{\dot \ell_k}\right)^{-1}\,.
\end{equation}
The label $(a)$ in the definition \label{t_matrix_op} distinguish the between the infinite possible representation chosen for the auxiliary space. This is a crucial fact for the complete integrability of the model, as different representation will generate - in general - independent commuting integrals of motion. As analytic functions of the parameter $u\,\in\,\mathbb{R}$, the operators $t^{(a)}(u)$ can be Taylor expanded around a chosen $u=u_0$ leading to an infinite tower of commuting operators\begin{equation}
H^{(a)}_k = \left[\frac{1}{k!}\frac{d^k}{du^k} t^{(a)}(u)\right]_{u=u_0}\,,\,\,\,\, H^{(a)}_i H^{(b)}_j\,=\,H^{(b)}_j H^{(a)}_i\,.
\end{equation}
In order to define an integrable model with commuting charges $H^{(a)}$, we should ensure that such operators are indeed diagonalizable.
In fact, we require the operator $t^{(a)}(u)$ to be normal, i.e. to commute with its hermitian conjugate\begin{equation}
\left[t^{(a)}(u),t^{(a)}(u)^{\dagger}\right]\,=\,0\,,
\end{equation}
and this constrains the inhomogeneities in the monodromy matrix to satisfy
\begin{equation}
\label{inhom_constr}
\theta_k+\theta_k^* \,=\,\theta_j+\theta_j^*\,,\,\,\,\,\,\forall j,k\,=\,1,\dots,L\,.
\end{equation}
The condition \eqref{inhom_constr} actually guarantees the stronger property
\begin{equation}
\label{t_normal}
t^{(a)}(u) \,(t^{(b)}(v))^{\dagger}\,=\,(t^{(b)}(v))^{\dagger} \,t^{(a)}(u)\,,\,\,\,\,\,\forall\,u,v\,,
\end{equation}
valid for any two auxiliary space representations $(\Delta_a,\ell_a,\dot{\ell}_a)$ and $(\Delta_b,\ell_b,\dot{\ell}_b)$. The proof of \eqref{t_normal} makes use of the R-operator property \eqref{YBE_adj}, in the same way that \eqref{t_comm} follows from the YBE. The hermitian conjugate of the monodromy operator \eqref{monodromy_op} reads
 \begin{equation}
\mathbb{T}_{1,\dots,L,a}^{\dagger}(u)\,=\,\rmat_{La}(u+\theta_L)^{\dagger}\cdots \rmat_{2a}(u+\theta_2)^{\dagger}\rmat_{1a}(u+\theta_1)^{\dagger}\,,
\end{equation}
and $\dagger$ refers to the hermitian conjugation in both the physical and auxiliary space of each $\rmat$-operator. The analogue of $RTT$ relations is a $T R T^{\dagger}$ relation
\begin{equation}
\mathbb{T}_{1,\dots,L,a}(u){\mathcal{R}}_{ba}(u+v^*+\theta+\theta^*)\,\mathbb{T}_{1,\dots,L,b}(v)^{\dagger}\,=\,\mathbb{T}_{1,\dots,L,b}(v)^{\dagger}\,{\mathcal{R}}_{ba}(u+v^*+\theta+\theta^*)\,\mathbb{T}_{1,\dots,L,a}(u)\,,
\end{equation}
where $\theta+\theta^*=\theta_k+\theta_k^*$ for any $k=1,\dots,L$.
The algebra \eqref{t_normal} follows from the trace over auxiliary spaces in the previous equation, where
\begin{equation}
(t^{(a)}(u))^{\dagger}\,=\,\text{Tr}_{\mathbb{V}_a} \left(\mathbb{T}_{1,\dots,L,a}^{\dagger}(u)\right)\,=\,\text{Tr}_{\mathbb{V}_a} \left(\rmat_{La}(u+\theta_L)^{\dagger}\cdots \rmat_{2a}(u+\theta_2)^{\dagger}\rmat_{1a}(u+\theta_1)^{\dagger}\right)\,.
\end{equation}
Following the prescription of \cite{Derkachov:2021rrf}, in this paper we realize the constraint \eqref{inhom_constr} with the choice
\begin{equation}
\theta_k= \frac{\Delta_k}{2} = 1 +i\nu_k\,,
\end{equation}
as it is suitable for studying the \emph{fishnet} reductions \cite{Gurdogan:2015csr,Gromov:2017cja,Kazakov_2019} of the operator $t^{(a)}(u)$ at special values of $u$.
Whenever \eqref{inhom_constr} is satisfied the operators $t^{(a)}(u)$ are normal operators for any choice of auxiliary space and for any $u$, hence the operator-valued coefficients of their Taylor expansions satisfy $[H_k^{(a)},H_j^{(b)\,\dagger}]=0$. In particular, this property guarantees that operators $H_k^{(a)}$ form a continuous infinity of mutually diagonalizable operators for any $k$, $\mathbb{V}_a$. The Hamiltonian of the spin chain \cite{Chicherin:2012yn,Faddeev:1996iy} is usually picked to be the local nearest-neighbor operator
\begin{equation}
\mathbb{H}\,\equiv\, H_1^{(a)}\,,\,\,\,\,u_0=0\,,
\end{equation}
therefore any other operator $H^{(a)}_k$ is a conserved charge\footnote{In this respect, any other $H^{(a)}_k$ or hermitean linear combinations thereof can be taken as the Hamiltonian of the system. This includes infinite linear combinations, for instance $t^{(a)}(u_0)$ itself.} \begin{equation}
\left[\mathbb{H}, H^{(a)}_i\right]\,=\, \frac{d}{dt} H^{(a)}_i =0\,.
\end{equation}
The complete integrability of the quantum spin-chain model is the statement that one can extract from the infinite tower of conserved charges $\{H^{(a)}_k\}$, a maximal subset of linearly independent operators.
\section{Fixed boundary conditions}
\label{sect:fixed_boundary}
\begin{figure}[H]
\begin{center}
\includegraphics[scale=0.6]{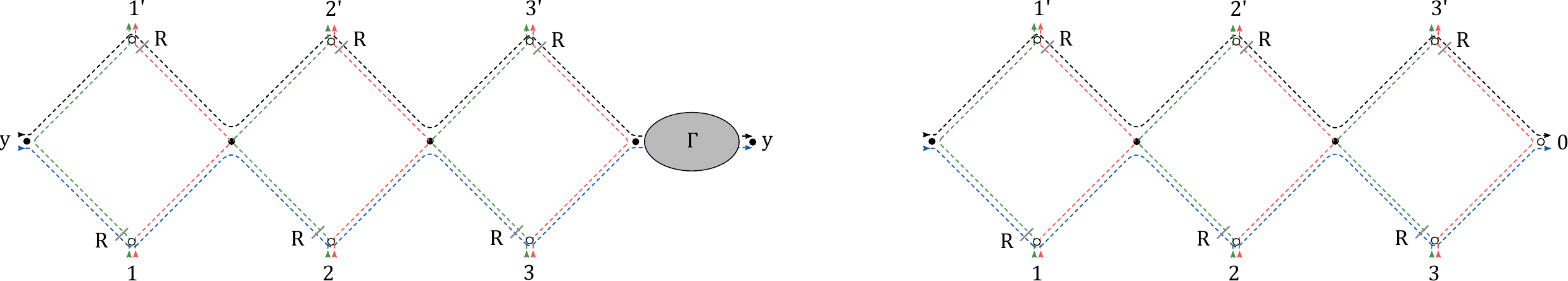}
\end{center}
\caption{\textbf{Left:} Transfer matrix for the twisted model $t^{(a)}_{\Gamma}(u)$. The auxiliary space twist is represented by the grey blob as an integral operator, and the points $y$ are the same. \textbf{Right:} Transfer matrix $\mathbf{Q}_{a}$ of the mirror chain model, resulting from the choice of integrable twist \eqref{delta_twist}. Here the spinorial structures of the auxliary space (black and blue dashed lines) are closed and traced over. Without Kronecker-$\delta$ in the spinor indices of the twist, the picture is the same but the $SU(2)$ indices of the auxiliary spaces are open between last and first sites.}
\label{Q_twist_diag}
\end{figure}

In this section we generalize the \emph{open} conformal spin chain with fixed boundaries defined in \cite{Derkachov2020,BassoFerrando}, and applied the computation of the Basso-Dixon integrals in \cite{Derkachov2020,Derkachov:2020zvv}, to the inhomogeneous and spinning setup of the previous section - which means that the $k$-th site of the chain hosts a function in the representation $(\Delta_k,\ell_k,\dot{\ell}_k)$ of the conformal group $SO(1,5)$. Besides the homogeneous scalar choice $(1,0,0)$ of \cite{Derkachov2020,BassoFerrando}, other useful reductions of the model for the computation of planar diagrams are the homogeneous chain of Weyl fermions $(3/2,1,0)$ which generates an hexagonal ``honeycomb" lattice of Yukawa vertices \cite{Pittelli2019}, and the inhomogeneous chain of mixed scalars and fermions describing the planar Feynman integrals of the doubly-scaled $\gamma$-deformation of $\mathcal{N}=4$ SYM (9) of \cite{Caetano:2016ydc,Kazakov_2019}.\\
The wavefunctions of the chain with $L$ particles/sites belong to the Hilbert space 
\begin{equation}
\label{Hilbert_sp}
\mathbb{V}_1\otimes \mathbb{V}_2 \otimes \cdots \otimes \mathbb{V}_L\,,
\end{equation}
where - otherwise than the \emph{closed} chain treated in the previous section - they are not subject to any cyclicity constraint. 
The model with fixed boundary conditions, that we may dub the \emph{mirror chain} - is defined via an integrable twist in the transfer matrix operator \eqref{t_matrix_op}, as represented in Fig.\eqref{Q_twist_diag} (left). As a general fact, the introduction of an auxiliary space twist $\Gamma_{a}:\mathbb{V}_a \to \mathbb{V}_a$, reads
\begin{equation}
\label{twist_t}
t_{\Gamma}^{(a)}(u)=\text{Tr}_{a}[\rmat_{1a}(u_1)\dots \rmat_{La}(u_L)\Gamma_a]\,,
\end{equation}
and it preserves the integrability of the model, i.e. the properties \eqref{t_comm} and  \eqref{t_normal}, under two sufficient conditions
\begin{equation}
\label{int_twist}
\left[\rmat_{ab}(u),\Gamma_{a}\Gamma_{b}\right]\,=\,0\,,\,\,\,\,\, \Gamma_{a}^{\dagger}\rmat_{ab}(u)\Gamma_{b}\,=\,\Gamma_{b}\rmat_{ab}(u)\Gamma_{a}^{\dagger}.
\end{equation}
\noindent
The integrable model we aim to is defined by breaking the auxiliary space convolution in \eqref{twist_t} with a twist $\Gamma_{x_0}$, which is an integral operator with kernel
\begin{equation}
\label{delta_twist}
\Gamma_{x_0}(x|y)_{\mathbf{a}\mathbf{c}}^{\mathbf{{b}}\mathbf{{d}}}\,\,=\, \delta_{\mathbf a}^{\mathbf{b}}  \delta_{\mathbf{{c}}}^{\mathbf{{d}}} \delta^{(4)}(x-x_0)\,,\,\,\,\,\text{and}\,\,\,\,\, \Gamma_{x_0}^{\dagger}(x|y)_{\mathbf{a}\mathbf{c}}^{\mathbf{{b}}\mathbf{{d}}}\,=\,\Gamma_{x_0}(y|x)_{\mathbf{b}\mathbf{d}}^{\mathbf{a}\mathbf{c}}\,,
\end{equation}
where the point $x_0^{\mu}$ is a parameter of the twist. We refer to this type of twisted transfer matrix with the notation $\mathbf{Q}_{a}(u)\equiv t_{\Gamma_{x_0}}^{(a)}(u)$. The first property in \eqref{int_twist} ensures the commutation of two twisted transfer matrices at any different values of $u$ and for any auxiliary space representations, and it holds as a consequence of \begin{equation}
\label{twist_comm_1}
\rmat_{ab}(u)(\Gamma_{x_0})_{a}(\Gamma_{x_0})_{b}=C(u)(\Gamma_{x_0})_{a}(\Gamma_{x_0})_{b}\mathbf{R}_{\ell_a\ell_b}\left(u+{\Delta_{ab}}\right)\otimes\mathbf{R}_{\dot \ell_a \dot \ell_b}\left(u-{\Delta_{ab}}\right)\,=\,(\Gamma_{x_0})_{a}(\Gamma_{x_0})_{b}\rmat_{ab}(u)\,,
\end{equation}
where $\Delta_{ab}=(\Delta_a-\Delta_b)/2$ and $C(u)$ is the complex number defined in \eqref{coeff_C}. The proof of \eqref{twist_comm_1} is given in detail in appendix \ref{app:comm_tw}.
The second relation in \eqref{int_twist} guarantees that the twisted operators commute with their hermitian conjugates. One can verify that such property for $\mathbf{Q}_a$ follows from the relation
\begin{equation}
\label{twist_comm_2}
(\Gamma_{x_0}^{\dagger})_{a}\rmat_{ab}(u)(\Gamma_{x_0})_{b}=C(u)(\Gamma_{x_0})_{a}(\Gamma_{x_0})_{b}\mathbf{R}_{\ell_a\ell_b}\left(u+{\Delta_{ab}}\right)\otimes \mathbf{R}_{\dot \ell_a \dot \ell_b}\left(u-{\Delta_{ab}}\right)=(\Gamma_{x_0})_{b}\rmat_{ab}(u)(\Gamma_{x_0}^{\dagger})_{a}\,
\end{equation}
The proof of \eqref{twist_comm_2} follows closely the steps of \eqref{twist_comm_1} as given in the appendix and is based on the following properties of the kernel $\rmat_{ab}(x_a,x_b|y_a,y_b)$
\begin{align}
\label{int_twist_II}
\begin{aligned}
&\int d^4 x_a  d^4y _b \, \rmat_{ab}(x_b,x_a|y_a,y_b)(u) \,=\,C(u)\,\mathbf{R}_{\ell_a\ell_b}\left(u+{\Delta_{ab}}\right)\mathbf{R}_{\dot \ell_a \dot \ell_b}\left(u-{\Delta_{ab}}\right) \,,\\
& \rmat_{ab}(x_b,x_0|y_a,x_0)\,=\, C(u)\,\mathbf{R}_{\ell_a\ell_b}\left(u+{\Delta_{ab}}\right)\mathbf{R}_{\dot \ell_a \dot \ell_b}\left(u-{\Delta_{ab}}\right)  \delta^{(4)}(x_0-x_a)\delta^{(4)}(x_0-y_b)\,.
\end{aligned}
\end{align}The twisted transfer matrix $\mathbf{Q}_{a}(u)$ has the diagrammatic representation of Fig.\ref{Q_twist_diag} (right), and can be rewritten as
\begin{align}
\label{Q_op}
\begin{aligned}
\mathbf{Q}_{a,L}(u)&=\sum_{\mathbf{a},\mathbf{\dot{a}}}\int d^4 x_a \left[\mathbb{T}_{1,\dots,L,a}(u)\,\delta^{(4)}(x_a-x_0)\right]^{\mathbf{a\dot{a}}}_{\mathbf{a\dot{a}}}\,=\\&=\,\sum_{\mathbf{a}\,\mathbf{\dot{a}}}\int d^4 x_a\, \left[\rmat_{1a}(u+\theta_1)\rmat_{2a}(u+\theta_2)\cdots \rmat_{La}(u+\theta_L)\,\delta^{(4)}(x_a-x_0)\right]^{\mathbf{a\dot{a}}}_{\mathbf{a\dot{a}}}\,,
\end{aligned}
\end{align}
namely the convolution in the space $\mathbb{V}_a$ - the auxiliary space of the spin chain - of $L$ copies of the $\rmat$-operator - each acting on a different physical space $\mathbb{V}_k$ for $k=1,\dots,L$ -  where the last auxiliary space point is fixed to $x_0$ and the first is integrated. In fact, we can regard this first integrated point to be connected by a propagator to the point at $\infty$, and eventually restore it via a conformal boost. It is useful to define explicitly a related operator with open spinor indices in auxiliary space
\begin{align}
\label{tilda_Q_op}
\begin{aligned}
\left(\mathbb{Q}_{a,L}(u)\right)_{\mathbf{a\dot{a}}}^{\mathbf{b\dot{b}}}=\int d^4 x_a \left[\mathbb{T}_{1,\dots,L,a}(u)\,\delta^{(4)}(x_a-x_0)\right]^{\mathbf{a\dot{a}}}_{\mathbf{b\dot{b}}}\,,\,\,\mathbf{Q}_{a,L}(u)= \text{Tr}_{\ell_a}\otimes \text{Tr}_{\dot \ell_a} \mathbb{Q}_{a,L}(u)\,,
\end{aligned}
\end{align}
whose reduction for scalar physical spaces $\Delta_k=1,\ell_k=\dot\ell_k=0$ and a fermion in auxiliary space $\Delta_a=3/2,\ell_a=1,\dot{\ell}_a=0$ has been studied in \cite{Derkachov:2020zvv} as graph-building operator of fishnet integrals with fermions on the disk. We point out that - differenlty than its trace in the auxiliary space spinors - the operator $\mathbb{Q}_{a,L}(u)$ is not diagonalizable.\\
The hamiltonians of the model are defined as an infinite set of mutually commuting operators given by the operator-valued coefficients of the Taylor expansion of \eqref{Q_op} around a point $u_0$
\begin{equation}
{H}^{(k)}_{a}\,\equiv\,\frac{1}{k!}\frac{d^k}{du^k}\,\mathbf{Q}_{a}(u)_{u=u_0}\,.
\end{equation}
The operators ${H}^{(k)}_{a}$ mutually commute and are moreover diagonalizable under the condition \eqref{inhom_constr}, in fact the equations \begin{equation}
\left[\mathbf{Q}_{a}(u),\mathbf{Q}_{b}(v)\right]=0\,,\,\,\,\left[\mathbf Q_{a}(u),\mathbf Q_{b}(v)^{\dagger}\right]=0\,,
\end{equation}
hold after the twisting due \eqref{int_twist}.  The constraint on inhomogeinities \eqref{inhom_constr} is satisfied setting $\theta_k =-\Delta_k/2$, as we consider unitary irreps of $SO(1,5)$ in the principal series $\Delta_k-2 = i\nu_k$.

In the next section of the paper we construct the eigenfunctions of $\mathbf Q_{a}(u)$, compute its spectrum and the overlap of the eigenfunctions, i.e.  the spectral measure. The logic we follow aims to the separation of variables (SoV) developed in \cite{Derkachov2020} (and inspired by the $SL(2,\mathbb{C})$ technique of \cite{Derkachov:2001yn}), promoted to the general setup of spinning particles and inhomogeneous spin chain. The achievement of this program relies ultimately on the identity \eqref{STR_ker} and its diagrammatic interpretation (Fig.\ref{STR_1}). 

We point out that due to lack of invariance of the twist \eqref{delta_twist}, the mirror transfer matrix $\mathbf{Q}_{a,L}(u)$ is not invariant under a conformal change of coordinates, but transforms non-trivially under conformal inversion $I(x^{\mu})=x^{\mu}/x^2$. Indeed, the result of such map reads
\begin{equation}
\label{Q_conformal_law}
\left(\prod_{k=1}^L \frac{ [\mathbf{x_k}]^{\ell_k} [\mathbf{\overline{x_k}}]^{\dot \ell_k}}{(x_k^{2})^{-\Delta_k}} \right)\left( \int d^4x_a \frac{[\mathbf{x_{0a}}]^{ \ell_0}  [\mathbf{\overline{x_{0a}}}]^{ \dot \ell_0}}{ (x_0^{2})^{4-\Delta_a}} \mathbb{T}_{a,L}(u)  \delta^{(4)}(x_a-x_0)\,\frac{[\mathbf{x_0}]^{\dot \ell_0}  [\mathbf{\overline{x_0}}]^{ \ell_0}}{ (x_0^{2})^{4-\Delta_a}} \right) \left(\prod_{k=1}^L \frac{ [\mathbf{x_k}]^{\dot \ell_k}  [\mathbf{\overline{x_k}}]^{ \ell_k}}{(x_k^{2})^{\Delta_k}}\right) \,,
\end{equation}
the quantity in the left and right brackets are the standard $SO(1,5)$ invariance in the sites of the chain, while the central bracket describes the simple modification affecting the kernel of $\mathbf{Q}_{a,L}$ with respect to the definition \eqref{Q_op}.

\begin{figure}[H]
\begin{center}
\includegraphics[scale=0.65]{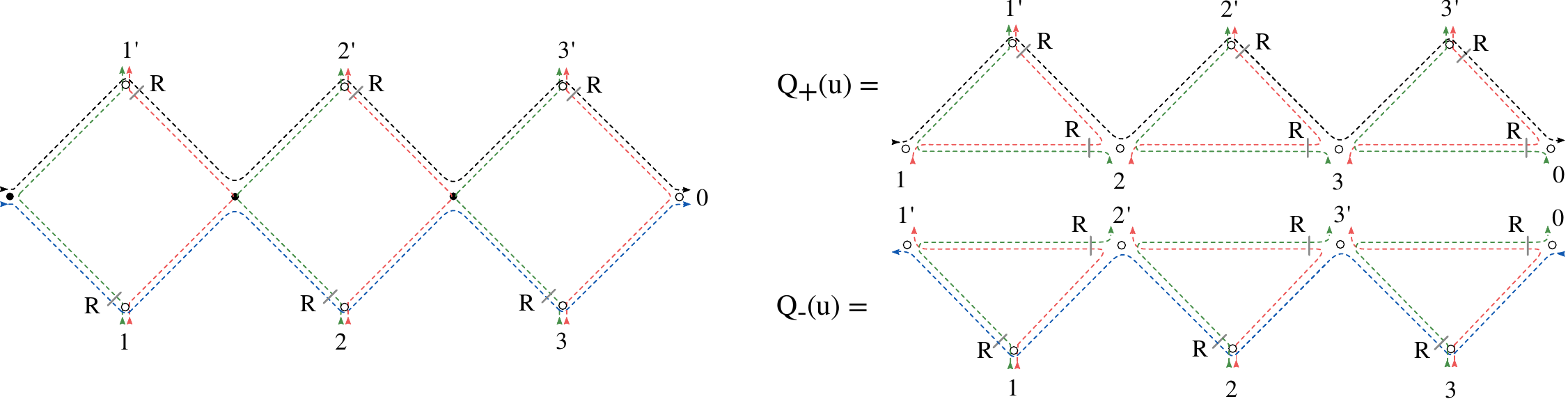}
\end{center}
\caption{Factorization of the transfer matrix $\mathbf{Q}_{a}$ via the factorization of its kernel. The squares of propagators corresponding to $\rmat$-operators are decomposed into two triangles by insertion of $1=(x_{kk+1}^2)^{2-\Delta_k}[\mathbf{x}_{kk+1}] \mathbf{R}(\Delta_k-2)[\mathbf{\bar x}_{kk+1}] \, (x_{kk+1}^2)^{\Delta_k-2}[\mathbf{x}_{kk+1}] \mathbf{R}(2-\Delta_k)[\mathbf{\bar x}_{kk+1}]$. }
\label{Q+Q-}
\end{figure}

The computation of the eigenvalues of $\mathbf{Q}_{a}(u)$ of sect.\ref{sect:eigenf} is is split into two parts following the factorization of the transfer matrix into the product of two operators $\mathbf{Q}_{a-}(u)\mathbf Q_{a+}(u)$ via the representation of its integral kernel as the convolution of two ``halves" defined by Fig.\ref{Q+Q-} (see also appendix \ref{app:R_fact} for the underlying factorization of $\rmat$-operator).
\subsection{Fishnet graph-builders}
\label{sect:graph_builder}
\begin{figure}[H]
\begin{center}
\includegraphics[scale=1]{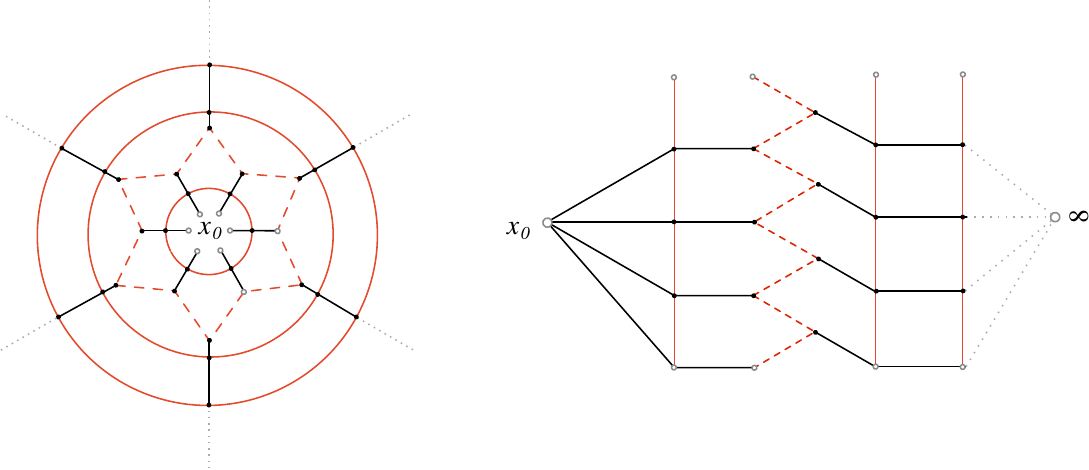}
\end{center}
\caption{Planar Feynamn diagram with mixed square-lattice and honeycomb topology, on the sphere of poles $x_0$ and $\infty$. Dotted lines are the amputated legs to infinity. On the left its view from the north-pole $x_0$: each wrapping of scalar (continuos) or fermionic (dashed) type is added by the action of a corresponding graph-building operator in radial direction. Any such operator belongs to the family of the periodic transfer matrix operator $t^{(a)}(u)$. On the right, the same graph can be obtained acting with a graph building operator of the family $\mathbf{Q}_{a}(u)$ in angular direction. Before the convolution between the last operator (up) and the first (down), the graph lies on a disk as the sphere is cut along $x_0- \infty$ direction.}
\label{mirror_direct}
\end{figure}

In the papers \cite{Gurdogan:2015csr,Gromov:2017cja,Kazakov_2019,Derkachov:2021rrf} the model with periodic boundary conditions of sect.\ref{sect:spin_chain} has been shown to reduce, for appropriate choices of the representations in the physical and auxiliary spaces, to the graph-building operator of the Feynman integrals that dominate the planar limit of bi-scalar Fishnet CFT or the more general chiral fishnet theories defined in the double-scaling limit of $\gamma$-deformed $\mathcal{N}=4$ SYM.  The choice of periodic boundary corresponds to a fishnet graph wrapped onto a cylinder (Fig.\ref{mirror_direct}, left) - hence such transfer matrices are graph-building operators $\hat B$ acting in radial direction - or \emph{direct channel} - around the insertion of a local operator in a planar correlator. This kind of diagrams can be re-summed \'a-la Bethe-Salpeter, relating the dilation operator $\hat D$ to the graph-builder 
\begin{equation}
\hat D \sim \log{\frac{1}{1-\xi^2 \hat B}}\,,
\end{equation}
hence the spectrum of the spin chain model captures the quantum correction to the scaling behaviour of local operator at finite coupling. 
On the other hand, the \emph{open} chain with fixed boundaries \eqref{Q_op} provides graph-building operators for the same type of Feynman graphs acting in angular direction - or \emph{mirror channel} - as exemplified in Fig.\ref{mirror_direct} (right).
The name \emph{direct} and \emph{mirror} channel are borrowed from the language of sigma-models on $\mathbb{R}\times S^{1}$: the interchange of the radial (non-compact) and angular (compact) directions by means of a double Wick-rotation is called \emph{mirror transformation} (see \cite{Arutyunov_2014} and references therein).

The mirror chain defined in this paper finds application to the computation of the class of fishnet integrals on the disk that describe at finite coupling single-trace correlators of point-split fields
\begin{equation}
 \langle \text{Tr}\left[\Phi_1(x_1) \Phi_2(x_2) \cdots\Phi_n(x_n) \right]\rangle\,.
\end{equation}
Due to the absence of gauge invariance in the action of Fishnet theories, these correlators are meaningful objects, and a worked-out example in this sense is the Basso-Dixon fishnet \cite{Basso:2017jwq,Derkachov2020} and its generalization with fermions \cite{Derkachov:2020zvv}. Here we list three reductions of the general spinning and inhomogeneous model that are relevant for the fishnet theories:
\begin{itemize}
\item Scalar fields in auxiliary space and physical spaces: $(\Delta_k,\ell_k,\dot{\ell}_k)=(\Delta_a,\ell_a,\dot{\ell}_a)=(1,0,0)$, giving rise to the graph-builder of the squared-lattice Basso-Dixon integrals.
\item Fermions propagating in auxiliary space and scalar fields propagating in physical space: $(\Delta_k,\ell_k,\dot{\ell}_k)=(1,0,0)$ and $(\Delta_a,\ell_a,\dot{\ell}_a)=(3/2,0,1)$,  which is the graph builder of a honeycomb of Yukawa vertices. In the formulation of the present paper, $\mathbb{Q}_{a}(u)$ correspond to the fishnet with generic fermionic indices (not diagonalizable), while $\mathbf{Q}_{a}(u)$ builds the trace over fermion indices of the graph (diagonalizable and integrable).
\item Fermions propagating along the physical space, and scalars in auxiliary space: $(\Delta_k,\ell_k,\dot{\ell}_k)=(3/2,0,1)$ and $(\Delta_a,\ell_a,\dot{\ell}_a)=(1,0,0)$, describing the Yukawa honeycomb with fermions wrapping around a cylinder - as the one describing planar integrals in $\mathcal{N}=2$ fishnet theories \cite{Pittelli2019}.
\end{itemize}
\begin{figure}[H]
\includegraphics[scale=0.51]{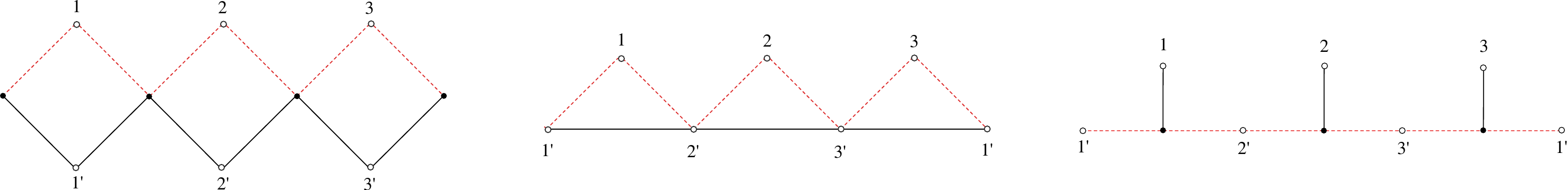}
\caption{\textbf{Left:} Integral kernel of the transfer matrix operator $\mathbf{Q}_{a,L}(u)$ for a spin chain of length $L=3$, for auxiliary space in the Weyl fermion representation $(3/2,0,1)$ and physical spaces in the the scalar representation $(1,0,0)$. \textbf{Center:} the transfer matrix kernel at $u=-5/4$ takes the shape of a triangular lattice wrapped onto a cylinder. \textbf{Right:} by means of star-triangle duality the triangular lattice can be re-written as an hexagonal lattice, with fermions propagating in angular direction and scalars along the cylinder.}
\end{figure}
\begin{figure}[H]
\includegraphics[scale=0.50]{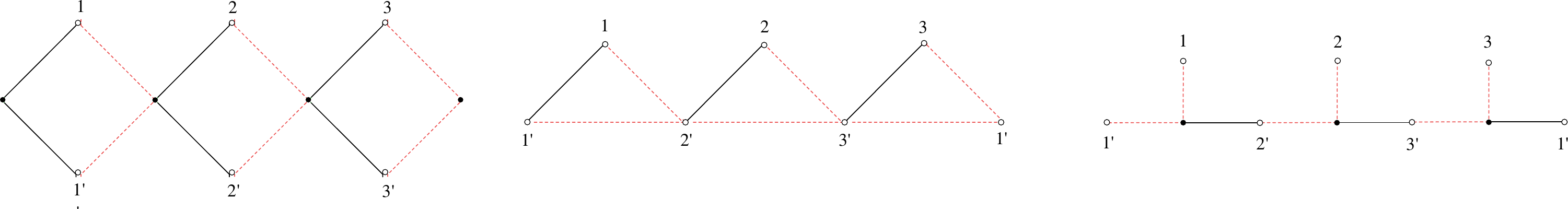}
\caption{\textbf{Left:} Integral kernel of the transfer matrix operator $\mathbf{Q}_{a,L}(u)$ for a spin chain of length $L=3$, for spinless auxiliary space $\ell_a=\dot{\ell}_a=0$ and physical spaces in the Weyl fermion representation $(3/2,0,1)$. \textbf{Center:} the transfer matrix kernel at $u=-5/4$ takes the shape of a triangular lattice wrapped onto a cylinder. \textbf{Right:} by means of star-triangle duality the triangular lattice can be re-written as an hexagonal lattice of Yukawa vertices, where fermions propagate along the cylinder and scalar around it.}
\end{figure}
Furthermore, the extension to inhomogeneous chain allows to mix scalars $(1,0,0)$ and Weyl fermions $(3/2,0,1)$ in the physical space, describing new classes of planar Feynman integrals. The transfer matrix is a graph builder that mixes the topology of square lattice and \emph{honeycomb} lattice, describing the bulk of a planar graph in the fishnet theory \eqref{chiCFT4_intro}.
\begin{figure}[H]
\includegraphics[scale=0.52]{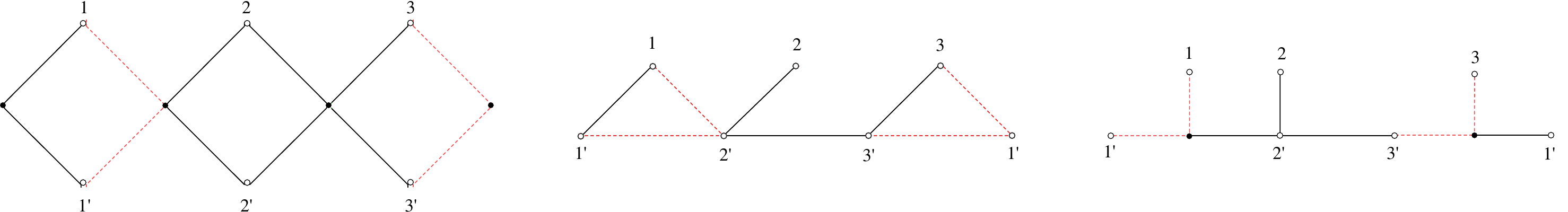}
\caption{\textbf{Left:} Integral kernel of the transfer matrix operator $\mathbf{Q}_{a,L}(u)$ for a spin chain of length $L=3$, for scalar auxiliary space $\ell_a=\dot{\ell}_a=0$ and inhomogeneous representation in mallthe physical spaces: scalar $\mathbb{V}_1=(1,0,0)$ or  Weyl fermion $\mathbb{V}_2=(3/2,0,1)$. \textbf{Center:} reduction of the kernel at the point $u_1=u+\theta_1=-1$, $u_2=u+\theta_2=-5/4$. \textbf{Right:} by star-triangle duality the reduced kernel can be written as the graph building operator for a planar Feynman diagram on the disk, whose topology mixes square-lattice and Yukawa hexagonal lattice.}
\end{figure}
\section{Eigenfunctions}
\label{sect:eigenf}
In this section we carry on the computation of eigenfunctions and spectrum of the integrable mirror chain. In order to be in closer contact with the notations and definitions of the papers \cite{Derkachov2020,Derkachov:2020zvv},  the operators $\mathbf{Q}_{a,k}(u)$ in this section are related to the definition \eqref{Q_op} by a shift $u\to -2+u, \Delta_k \to 4-\Delta_k = \Delta_k^*$, which conserves all the properties of the model. Moreover, for the sake of simplicity, we assume that any operator $\rmat_{ij}(u)$ entering the definition \eqref{Q_op} of $\mathbf{Q}_{a,L}(u)$ is given by the integral kernel \eqref{R_integral_form} stripped of all the constants with respect to the spacetime coordinates $x_i^{\mu}$.
\subsection{Iterative construction}
\label{sect:iter}
The eigenfunctions of the transfer matrix $\mathbf{Q}_a(u)$ were found in \cite{Derkachov2020} for the homogeneous spinless model, namely for a given unitary irrep $(\Delta,0,0)$ at each site of the chain, and for the special choice of spinless auxiliary space $\ell_a=\dot{\ell}_a=0$. The main result of that paper is an iterative construction of the eigenfunctions based on the equation
\begin{equation}
\label{SoVold}
\mathbf Q_{a,k}(u)\mathbf \Lambda_{k}(Y) =q_{a}(Y) \,\mathbf{\Lambda}_{k}(Y)  \mathbf{Q}_{a,k-1}(u) \,,\,\,\,\, q_{a}(Y) \in \mathbb{C}\,,
\end{equation}
where $\mathbf Q_{a,k}(u)$ is the transfer matrix acting on the space of $k$ particles $\mathbb{V}^{\otimes k}$, all in the same representation $(\Delta,0,0)$, while $q_{a}(Y)$ is a function of the quantum number $Y$ - a separated variable of the system. The quantum numbers $Y$ found in \cite{Derkachov2020} are of the form
\begin{equation}
Y = \frac{n}{2}+i\nu\,,\,\,\,\,\, Y^* = \frac{n}{2}-i\nu\,,\,\,\,\,\, \nu\in\mathbb{R}\,,\,\,\,n\in \mathbb{N}\,,
\end{equation}
and encode the principal series scaling dimension $\Delta=2+i\nu$ and the spin $n/2$ of the excitation carried by the layer $\mathbf\Lambda_k(Y)$. Each \emph{layer operator} $\mathbf{\Lambda}_k(Y)$ carries matrix indices of left/right $n$-fold symmetric spinors, as the excitation transforms under $SO(4)$ in the left/right $(\mathbf{n+1})\otimes (\mathbf{\overline{n+1}})$ representations of $SU(2)$ 
\begin{equation}
\label{left_right_action}
\mathbf{\Lambda}_k(Y)_{\mathbf{\dot a}}^{\mathbf{b}} \,\longrightarrow \, [V^{\dagger}]_{\mathbf{\dot a}}^{\mathbf{\dot b}} \mathbf{\Lambda}_k(Y)_{\mathbf{\dot b}}^{\mathbf{b}}[U]_{\mathbf{b}}^{\mathbf{a}}\,.
\end{equation}
The spinor indices of the layer are not specified in \eqref{SoVold} since they are the same in the two sides of the equation.
The {layer} operators map an eigenfunction of the chain of length $L=k-1$ sites to an eigenfunction of the chain of length $L=k$, adding the $k$-th site
\begin{equation}
\label{adding_site}
\mathbf{\Lambda}_k(Y)_{\mathbf{\dot a}}^{\mathbf{b}} : \underbrace{\mathbb{V}\otimes \cdots \otimes \mathbb{V}}_{k-1} \,\longrightarrow \underbrace{\mathbb{V}\otimes \cdots \otimes \mathbb{V}}_{k}\,.
\end{equation}
The iteration of \eqref{SoVold} reduces the spectral problem of the transfer matrix from a model of length $L$ to that of length one, of easy solution. 
In this section we extend the construction of \cite{Derkachov2020} to the most general situation in which each particle of the chain transforms in a unitary irrep $(\Delta_k,\ell_k,\dot{\ell}_k)$ of the conformal group, and also to transfer matrices with spinning auxiliary space. In particular, the spectral problem of transfer matrices with spinless auxiliary space can be solved by separation of variables, and we are going to solve the equation \eqref{SoVold} in its generalized form\footnote{The separation of variables in a classical system is achieved by a canonical transformation on the phase space that brings the Hamiltonian into a separated form - i.e. a form in which the Hamilton-Jacobi equations can be solved as a collection of 1-dimensional uncoupled systems \cite{gantmacher1975lectures}. At the quantum level the canonical transformation is replaced by a unitary transformation - i.e. an isometry of two Hilbert spaces realized by an orthonormal change of basis. In this regard, the condition \eqref{SoV} is the quantum version of a the classical separable Hamiltonian of type $\mathcal{H}(q_1,p_1,\dots,q_L,p_L)=f_L(q_L,p_L,f_{L-1}(q_{L-1},p_{L-1},\dots f_1(q_1,p_1))\dots )$.}
\begin{equation}
\label{SoV}
\mathbf{Q}_{a,k}(u)\mathbf \Lambda_{k}(Y|\eta,\bar \eta) =q_{a,k}(Y) \,\mathbf\Lambda_{k}(Y|\eta,\bar \eta)\mathbf{Q'}_{a,k-1}(u) \,,\,\,\,\, q_{a,k}(Y) \in \mathbb{C}\,.
\end{equation}
In \eqref{SoV} $\mathbf Q_{a,k}$ is the transfer matrix for a system of $k$ particles, acting on the tensor product of $k$ Hilbert spaces $\mathbb{V}_1\otimes \cdots \otimes \mathbb{V}_{k}$, where $\mathbb{V}_j=(\Delta_j,n_j,\dot{n}_j)$, while the tranfer matrix $\mathbf{Q'}_{k-1}^{(a)}(u)$ acts on $k-1$ particles $\mathbb{V}_1'\otimes \cdots \otimes  \mathbb{V}_{k-1}'$ with shifted spins $\mathbb{V}_j' = (\Delta_{j},\ell_{j},\dot \ell_{j+1})\,,\,\, j=1,\dots, k-1$. For a chain of spinless sites $\ell_j=\dot \ell_j=0$ - as in the case described by \eqref{SoVold} - the two spaces coincide $\mathbb{V}_j'\equiv \mathbb{V}_j$. The operators
\begin{align}
\begin{aligned}
\label{layer_def}
\mathbf\Lambda_{k}(Y|\eta,\bar \eta): \mathbb{V}_1'\otimes \cdots \otimes  \mathbb{V}_{k-1}'\longrightarrow \mathbb{V}_1\otimes \cdots  \otimes \mathbb{V}_k  \,,
\end{aligned}
\end{align}
define a linear map on the Hilbert space of $k-1$ particles $
\mathbb{V}_j'$ to the space of $k$ particles $\mathbb{V}_j$ parametrized by a quantum number $Y$ and two spinors $\bar \eta_{\mathbf{\dot c}}$, $ \eta^{\mathbf{\dot c}}$ in the $\dot\ell_1-$fold and $\ell_k$-fold symmetric representations of $SU(2)$. The spinor structure of a layer is that of a matrix acting on $k$ dotted and $k$ undotted symmetric spinors belonging respectively to the representations $(\ell_1,\dot \ell_1),\dots ,(\ell_k,\dot \ell_k)$ of the physical space, in addition to left and right $n$-fold symmetric indices of the excitation \eqref{left_right_action}, that we keep implicit
\begin{equation}
\label{layer_mat}
\mathbf{\Lambda}_k(Y|\eta,\bar \eta)_{\mathbf{a_1\dots a_{k} \dot a_1\dots \dot a_{k}}}^{\mathbf{ b_1\dots  b_{k-1}\dot b_2\dots \dot b_{k}}}  = \mathbf{\Lambda}_k(Y)_{\mathbf{a_1\dots a_{k}\dot a_1\dots \dot a_{k}}}^{\mathbf{ b_1\dots  b_{k}\dot b_1\dots \dot b_{k}}}  \bar \eta_{\mathbf{\dot b}_1}  \eta_{\mathbf{b}_k}\,.
\end{equation}
In the following we'll often use the compact notation
\begin{equation}
 \mathbf{\Lambda}_k(Y|\eta,\bar \eta) = \mathbf{\Lambda}_k(Y)|\eta\rangle |\bar\eta\rangle\,.
\end{equation}
Assuming the knowledge of layer operators at length $k>1$, the equations \eqref{SoV} and \eqref{1_p_sol} determine a solution for the model of $L$-particles, and the eigenfunctions take an iterative structure
\begin{align}
\begin{aligned}
\label{full_eigen_gen}
\Psi(\mathbf{Y}|\mathbf{x},\boldsymbol{\eta},\boldsymbol{\bar \eta})& = \mathbf{\Lambda}_L (Y_L) \mathbf{\Lambda}_{L-1} (Y_{L-1}) \cdots \mathbf{\Lambda}_2 (Y_2) \mathbf{\Lambda}_1(Y_1)|\boldsymbol{\eta}\rangle |\boldsymbol{\eta'}\rangle =\\&= 
 \mathbf{\Lambda}_L (Y_L|\eta_L  ,\bar \eta_1)   \cdots \mathbf{\Lambda}_2 (Y_2|\eta_{2},\bar \eta_{L-1}) \mathbf{\Lambda}_1(Y_1|\eta_1 ,\bar \eta_L)\,,
\end{aligned}
\end{align}
where we used the compact notation for the quantum numbers $\mathbf{Y}=(Y_1,\dots,Y_L)$, the coordinates $\mathbf{x}=(x_1,\dots,x_L)$ and the symmetric spinors $\boldsymbol{\eta}= (\eta_1,\dots,\eta_L)$, $\bar{\boldsymbol{\eta}}= (\bar \eta_1,\dots,\bar \eta_L)$. The layer at depth $h$ in the eigenfunction acts on the tensor product of $L-h$  Hilbert spaces
\begin{equation}
\label{Hilbert_shift}
\mathbb{V}_1^{(h)}\otimes \cdots \otimes \mathbb{V}_h^{(h)}\,,\,\,\,\, \mathbb{V}^{(h)}_j =(\Delta_j,\ell_j,\dot{\ell}_j+h)\,,\,\,\,j=1,\dots,L-h\,.
\end{equation}
\begin{figure}[H]
\begin{center}
\includegraphics[scale=0.7]{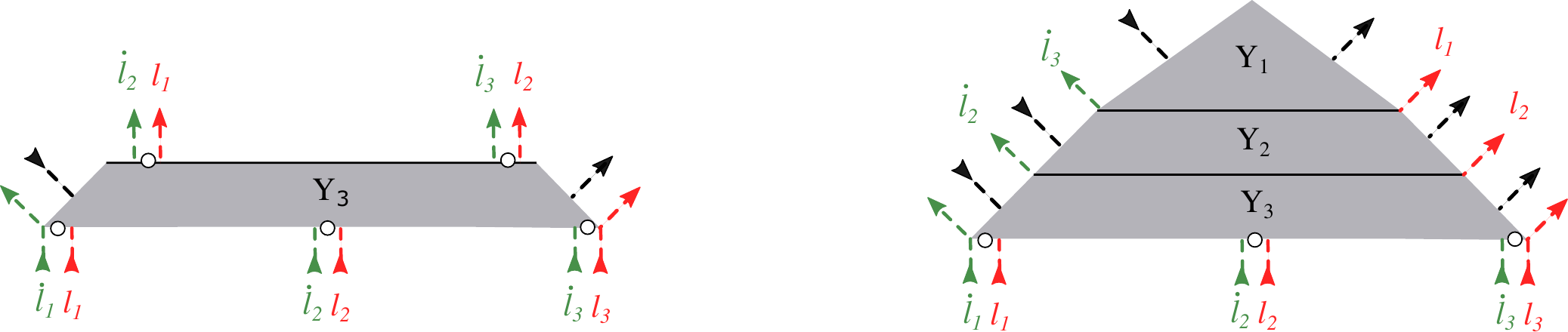}
\end{center}
\caption{\textbf{Left:} Scheme of the layer $\mathbf{\Lambda}_3(Y_3)$ with the red/green lines being the left/right part of physical space particles, and the black line standing for the excitation $Y_3$ carried by the layer. \textbf{Right:} The eigenfunction \eqref{full_eigen_gen} has a pyramidal structure of layers of decreasing length (here $L=3$). The first right (green) spin and the last left (red) spin of the chain gets factored out at each level. Therefore, the spins of the physical spaces at depth $h$ (layer of length $L-h$) are shifted according to \eqref{Hilbert_shift}. }
\label{layer_gen_structure}
\end{figure}
For clarity, let's write explicitly the contractions between spinor indices belonging to different layers for $L=2$, $L=3$
\begin{align}
\begin{aligned}
&\Psi(\mathbf{Y}|\mathbf{x},\boldsymbol{\eta},\bar{\boldsymbol \eta})_{\mathbf{a}_1 \mathbf{a}_2 \mathbf{\dot a}_1
\mathbf{\dot a}_2}= \Lambda_{2}(Y_2|\eta_2,\bar \eta_1)_{\mathbf{a}_1 \mathbf{a}_2\mathbf{\dot a}_1
\mathbf{\dot a}_2}^{\mathbf{c}_1 \mathbf{\dot c}_2} \Lambda_{1}(Y_1|\eta_1,\bar\eta_{2})_{\mathbf{c}_1\mathbf{\dot{c}}_{2}}\,,\\\\
&\Psi(\mathbf{Y}|\mathbf{x},\boldsymbol{\eta},\bar{\boldsymbol \eta})_{\mathbf{a}_1 \mathbf{a}_2\mathbf{a}_3 \mathbf{\dot a}_1
\mathbf{\dot a}_2\mathbf{a}_3}= \Lambda_{3}(Y_3|\eta_3,\bar \eta_1)_{\mathbf{a}_1 \mathbf{a}_2 \mathbf{a}_3 \mathbf{\dot a}_1
\mathbf{\dot a}_2 \mathbf{\dot a}_3 }^{\mathbf{b}_1\mathbf{b}_2 \mathbf{\dot b}_2\mathbf{\dot b}_3 } \Lambda_{2}(Y_2|\eta_2,\bar \eta_2)_{\mathbf{b}_1 \mathbf{b}_2\mathbf{\dot b}_2
\mathbf{\dot b}_3 }^{\mathbf{c}_1 \mathbf{\dot c}_3 } \Lambda_{1}(Y_1|\eta_1,\bar\eta_{3})_{\mathbf{c}_1\mathbf{\dot{c}}_{3}}\,.
\end{aligned}
\end{align}
The eigenvalue corresponding to \eqref{full_eigen_gen} is factorized into $L$ contributions (to be determined) according to \eqref{SoV}, each depending on one of the $Y_j$
\begin{equation}
\label{spectrum gen}
\prod_{k=1}^L q_{a,k}(Y_k)\,,
\end{equation}
and this is a statement of separation of variables of the model. 
Indeed, the iteration of \eqref{SoV} defines a linear transformation via the $\mathbf{\Lambda}_L(Y_L)\cdots \mathbf \Lambda_1(Y_1)$ from the physical spaces of the chain ${\mathbb{V}}_j$ to the spaces $\widetilde{\mathbb{V}}_j$ of functions over the excitation quantum numbers $Y_j$ and spinor indices
\begin{align}
\begin{aligned}
\label{sov_trans}
\mathcal{U}: &\mathbb{V}_1\otimes \cdots \otimes \mathbb{V}_L  \longrightarrow \widetilde{\mathbb{V}}_1\otimes \cdots \otimes \widetilde{\mathbb{V}}_L\,,
 \\&f(x_1,\dots,x_L) \,\mapsto\, \langle f, \Psi(Y_1,\dots,Y_L) \rangle_{\mathbb{V}_1\otimes\cdots \otimes \mathbb{V}_L}\,,
\end{aligned}
\end{align}
With respect to such change of basis the transfer matrices are factorized, i.e. the variables are separated
\begin{equation}
\label{eigenf_sov}
\mathcal{U}\,\mathbf Q_{a,L}(u)\, \mathcal{U} ^{-1}= q_{a,1}(Y_1)\cdots q_{a,L}(Y_L)\,,\end{equation}
and the excitation numbers $Y_j$ are the quantum separated variables (or to be more correct, the eigenvalues of the s.v.).
The scheme outlined in \eqref{sov_trans} and \eqref{eigenf_sov} should be made rigorous by the definition of the space of functions of $Y_j$ as an Hilbert space. For one layer only, i.e. for the model of one site $L=1$, such an Hilbert space has the standard scalar product inherited from its definiton
\begin{equation}
\label{V_tilda_def}
 \widetilde{\mathbb{V}} = L^{2}(\nu,d\nu)\otimes \left(\bigoplus_{n=0}^{\infty}  \text{Sym}_{n}[\mathbb{C}^2]^{*} \otimes \text{Sym}_{n}[\mathbb{C}^2]\right)\,, 
\end{equation}
where the star $*$ is the standard notation for linear functionals (dual space). On the other hand, for $L>1$ we need to equip the tensor product of spaces $\widetilde{\mathbb{V}}_j$ with a measure $\rho(Y_1,\dots,Y_L)$ that defines a structure of Hilbert space
\begin{equation}
\label{dual_Hilbert}
\widetilde{\mathcal{V}} = \widetilde{\mathbb V}_1 \otimes \cdots \otimes \widetilde{\mathbb{V}}_L\,,
\end{equation}
In order for \eqref{sov_trans} to be an isometry, the measure amounts to the overlap of eigenfunctions, and it is computed in the next subsections. Moreover we notice that in \eqref{sov_trans} the physical spin vectors $\eta, \bar \eta$ are not involved in the transformation and they are omitted. This fact holds true for the transfer matrices with spinless auxiliary space, for which $\eta, \bar \eta$ variables are just spectators. 

\subsection{$L=1$ eigenfunctions}
\label{sec:1p}
Let us assume to know a solution  $\mathbf{\Lambda}_k$ of \eqref{SoV}: after $L$ iterations the spectral problem reduces to the single-particle model
\begin{equation}
\label{1_p_prob}
\mathbf{Q}_{a,1}(u)\mathbf \Lambda_1(Y_1|\eta,\bar \eta) =\text{Tr}_a\left[\rmat_{1a}(u+\theta_1)\Gamma_{x_0}\right]  \mathbf \Lambda_1(Y|\eta,\bar \eta) = q_{a,1}(Y) \mathbf\Lambda_1(Y|\eta,\bar \eta) \,,
\end{equation}
that is, with explicit spinor indices:
\begin{align}
\begin{aligned}
\label{1_p_spectral}
\left[\mathbf{Q}_{a,1} (u)\right]_{\mathbf{a}_1\mathbf{\dot a}_1}^{\mathbf{b}_1\mathbf{\dot b}_1} \mathbf\Lambda_1(Y_1|\eta,\bar \eta)_{\mathbf{b}_1\mathbf{\dot b}_1} = q_{a,1}(Y_1)\mathbf\Lambda_1(Y_1|\eta,\bar \eta)_{\mathbf{a}_1\mathbf{\dot a}_1} \,,
\end{aligned}
\end{align}
where the only particle in of this one-site chain transforms in the representation $(\Delta_1,\ell_1,\dot{\ell}_{L-1})$, thus $\eta_{\mathbf{a}}$, $\bar{\eta}_{\mathbf{\dot{a}}}$ are symmetric spinors of degree $\ell_1,\dot{\ell}_{L-1}$, according to \eqref{SoV},\eqref{Hilbert_shift}. 
As in this section we care only about the $1$-site spectral problem, we drop indices and use the lighter notation $(\Delta_1,\ell,\dot{\ell})$. The solution of \eqref{1_p_spectral} is constrained by the conformal invariance of the $\mathcal R$-operator \eqref{R_mat} and it is just a function with spinor indices (for $k=1$  the layer \eqref{layer_def} is acts from no site to one site, \eqref{adding_site})
\begin{equation}
\label{1_p_sol}
\Psi(Y|x,\eta,\bar{\eta})_{\mathbf{a\dot{a}}} = \frac{[\mathbf{\overline{x-x_0}}]^{n}}{(x-x_0)^{2(2-\Delta_1-i \nu)}} [\mathbf{x-x_0}]_{\mathbf{ a}}^{\mathbf{\dot c}} \eta_{\mathbf{\dot c}}  \bar \eta_{\mathbf{\dot a}}\,,
\end{equation}
where
\begin{equation}
Y= \frac{n}{2}+i\nu\,,\, \nu \in \mathbb{R}\,,\,n \in \mathbb{N}\,,\,\,\,\, \mathbf{a} =(a_1,\dots,a_{\ell})\,,\,  \mathbf{\dot c} =(\dot c_1,\dots, \dot{c}_{\dot \ell})\,.
\end{equation}
The eigenvalue $q_{a,1}(u,Y)$ corresponding to \eqref{1_p_sol} reads
\begin{equation}
\label{1_p_eigen}
\pi^4 \frac{\Gamma\left(-u-\frac{\Delta_a}{2} +\frac{\ell}{2} \right) \Gamma\left(4-u-\frac{\Delta_a}{2}+\frac{\ell}{2}  \right)}{\Gamma\left(u+2+\frac{\Delta_a}{2} +\frac{\ell}{2} \right) \Gamma\left(-u-2+\frac{\Delta_a}{2} +\frac{\ell}{2}\right)}  \frac{\Gamma\left(-2-u+\frac{\Delta_a}{2} -i\nu+\frac{n}{2} \right) \Gamma\left(u+\frac{\Delta_a}{2}+i\nu +\frac{n}{2}\right)}{\Gamma\left(u+4-\frac{\Delta_a}{2} +i\nu+\frac{n}{2} \right) \Gamma\left(2-u-\frac{\Delta_a}{2} -i\nu+\frac{n}{2} \right)}.
\end{equation}
The computation of \eqref{1_p_eigen} follows from two applications of the star triangle identity in its amputated form (Fig.\ref{STR_amp_app} in appendix \ref{app:str_amp}), and is illustrated in Fig.\ref{diag_1_p}.
\begin{figure}[H]
\begin{center}
\includegraphics[scale=0.95]{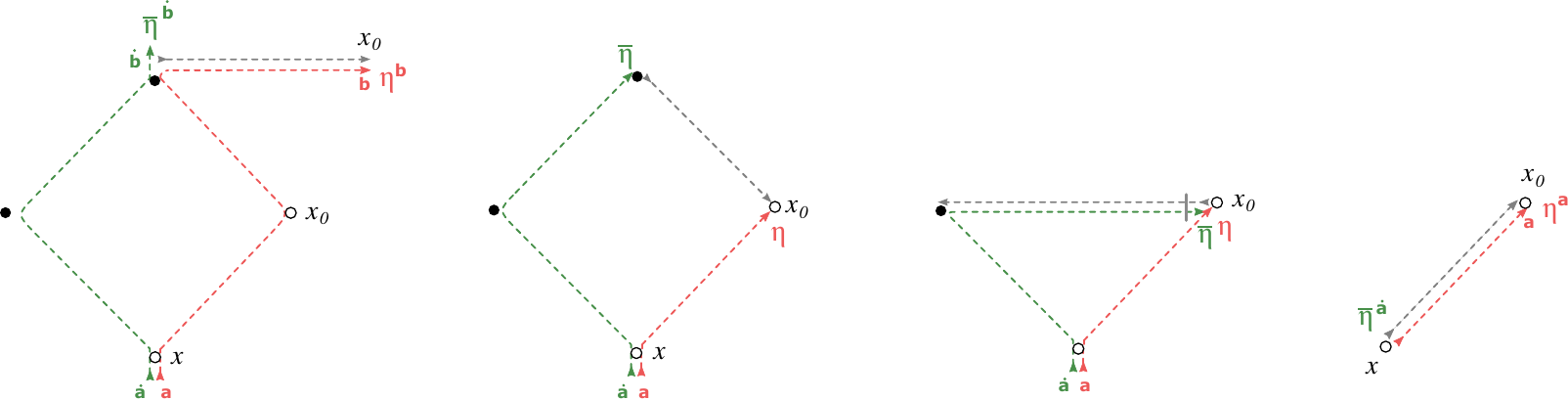}
\end{center}
\caption{\textbf{Left to right:} Left hand side of equation \eqref{1_p_spectral}; identification of $x_0$ and simplification of the spinorial structure via identity $[\mathbf{x\overline{x}}]=\mathbbm{1}$; (amputated) star-triangle integration of the top blob; the (amputated) star-triangle integration of the left blob delivers the initial function \eqref{1_p_sol}.}
\label{diag_1_p}
\end{figure}
We focus the reader's attention on the two sources of spinor indices in the function \eqref{1_p_sol}: one is the physical space spins of the chain $(\ell_1,\dot{\ell}_1)$, explicitly expressed in \eqref{1_p_eigen} (red and green in Fig.\ref{diag_1_p}), the other is carried by $[\mathbf{\overline{x-x_0}}]^{n}$ and it depends on the quantum number $n$ of the excitation (grey in Fig.\ref{diag_1_p}). It may be convenient to pair the latter with some auxiliary spinors $\alpha_{\dot a}, \beta^{a}$
\begin{equation}
\label{aux_spins}
[\mathbf{\overline{x-x_0}}]^{\ell} \to(\alpha^*)^{\mathbf{\dot c}} [\mathbf{\overline{x-x_0}}]_{\mathbf{\dot c}}^{\mathbf{c}}  \,\beta_{\mathbf c} \equiv \langle  \alpha |\overline{\mathbf{x-x_0}}|\beta \rangle^{\ell}\,.
\end{equation}
When dealing with $SO(4)$ invariant charges $(\ell_a=\dot{\ell}_a =0)$ the auxiliary spinors \eqref{aux_spins} are left untouched by the action of $\mathbf Q_{a,k}(u)$, and the spectrum is degenerate respect to spacetime rotations, hence we will often omit them from formulae. In fact, we will restore the notation \eqref{aux_spins} when dealing with charges generated by a choice of spinning auxiliary space $(\ell_a,\dot{\ell}_a)\neq (0,0)$, where the action of $\mathbf Q_{a,k}(u)$ on layer operators rotates the $n$-spinor indices, breaking the degeneracy.
The exponent $(2-\Delta_1+i \nu)$ in the denominator of \eqref{1_p_eigen} is fixed requiring of orthogonality and completeness.
In the notation of auxiliary spinors \eqref{aux_spins} we can rewrite \eqref{1_p_sol} as
\begin{equation}
\langle \beta |\Psi(Y|x,\eta,\bar \eta)|\alpha \rangle \,=\, \frac{\langle\beta|\mathbf x|\alpha\rangle^{n}\, }{(x^2)^{2-\Delta+i\nu}}[\mathbf{\overline{x}}]\eta \rangle^{\ell}| \bar \eta\rangle^{\dot \ell}\,.
\end{equation}
We define the conjugate eigenfunction by complex conjugation followed by a shift $\nu \to \nu+2i$ (convenient choice that simplifies the orthogonality relation)
\begin{equation}
\langle \alpha |\bar{\Psi}(Y|x,\eta,\bar \eta)|\beta \rangle \,=\, \frac{\langle\alpha|\bar{\mathbf x}|\beta \rangle^{n}\, }{(x^2)^{\Delta-i\nu}}\langle \eta  [\mathbf{x}]^{\ell} \langle \bar \eta|^{\dot \ell}\,.
\end{equation}
The overlap of two such functions reads
\begin{align}
\begin{aligned}
\langle\Psi(Y'),\Psi(Y)\rangle_{\mathbb{V}}&=\,\int d^4x\, \frac{\langle\alpha'|\mathbf{\overline x}|\beta'\rangle^{n'}\langle\beta|\mathbf{x}|\alpha\rangle^{n}}{(x^2)^{2+i(\nu-\nu')}}\, \langle \eta' |[\mathbf{\overline{x}}][\mathbf{x}]|\eta\rangle^{\ell} \langle\bar \eta '| \bar \eta \rangle^{\dot \ell} =\\&=   \langle \eta'|  \eta \rangle^{ \ell}  \langle\bar \eta '| \bar \eta \rangle^{\dot \ell}  \int d^4x\,\frac{\langle\alpha'|\mathbf{\overline x}|\beta'\rangle^{n'}\langle\beta|\mathbf x|\alpha\rangle^{n}}{(x^2)^{2+i(\nu-\nu')}}\,,
\end{aligned}
\end{align}
thus the integration reduces to the scalar case $\ell=\dot{\ell}=0$, and the result is
\begin{equation}
\label{ortho_1_p}
\langle\Psi(Y'),\Psi(Y) \rangle_{\mathbb{V}}\,=\,\frac{2\pi^3}{(n+1)}\delta(\nu-\nu') \delta_{n,n'}\,\langle \beta|\beta'\rangle^{n} \langle \alpha '|\alpha\rangle^{n}\, \langle \eta'|  \eta \rangle^{ \ell}  \langle\bar \eta '| \bar \eta\rangle^{\dot \ell}  \,.
\end{equation}
The same result without auxiliary spinors \eqref{aux_spins} reads
\begin{equation}
\label{clean_1_p}
\langle\Psi(Y'),\Psi(Y) \rangle_{\mathbb{V}} = \bar{\mathbf \Lambda}_1(Y'|\eta',\bar\eta') {\mathbf \Lambda}_1(Y|\eta',\bar\eta') = \,\frac{2\pi^3}{(n+1)}\delta(Y-Y')\, \mathbb{P}_{11'} \langle \eta'|  \eta \rangle^{ \ell}  \langle\bar \eta '| \bar \eta\rangle^{\dot \ell} \,,
\end{equation}
where we introduced the compact notation $\delta(Y-Y')=\delta_{n,n'}\delta(\nu-\nu')$ and $\mathbb{P}_{11'}$ exchanges the spinor indices of the two excitations.
The terms dependent on spinors in \eqref{ortho_1_p} are actually $\delta$-functions in the symmetric spinor space
\begin{equation}
\int D\eta \,\langle \eta'|  \eta \rangle^{ \ell}  \phi(\eta) =\int d^2\eta_1 d^2\eta_2 \,e^{- \eta^* \cdot \eta} \,\langle \eta'|  \eta \rangle^{ \ell}  \phi(\eta) = \phi(\eta')\,,
\end{equation}
therefore the orthogonality (and completeness) of the eigenfunctions at $L=1$ is a straightforward extension of the spinless case. Taking into account the normalization in \eqref{clean_1_p} and writing explicitly the excitation's spinor indices $\mathbf{a},\mathbf{\dot{a}}$, the completeness relation follows as
\begin{align}
\begin{aligned}
\label{comp_1_p}
\pi^3 \sum_{n=0}^{\infty} \int_{\mathbb{R}} \frac{d\nu}{n+1} \,\Psi(Y|x,\eta,\bar \eta)_{\mathbf{a \dot a}}  \left(\bar \Psi(Y|y,\eta,\bar \eta)^* \right)^{\mathbf{a\dot a}} = \delta^{(4)}(x-y) \left(|\eta\rangle \langle \eta |\right)  \otimes  \left(|\bar\eta \rangle \langle \bar \eta|\right)\,.
\end{aligned}
\end{align}
\noindent
 
\subsection{Spinless inhomogeneous model}
\label{scalar_inhom}
\begin{figure}[H]
\begin{center}
\includegraphics[scale=1.0]{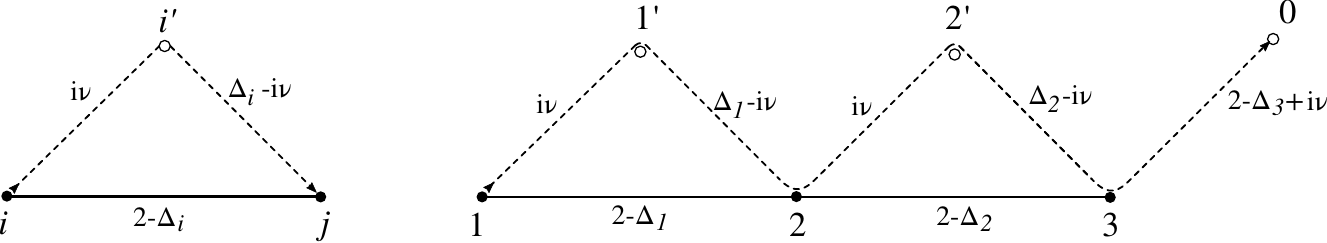}
\end{center}
\caption{\textbf{Left:} Kernel of the operator $\mathbb{R}^{(n)}_{ij}(\Delta_1/2-i\nu)$; as usual the dashed lines stand for $SU(2)$ matrices that in a product are alternatively defined as $[{\mathbf{x}}]=\boldsymbol{\sigma}_{\mu} \hat x^{\mu}$ and $[\overline{\mathbf{x}}]=\overline{\boldsymbol{\sigma}}_{\mu} \hat x^{\mu}$, to power $n$ of the symmetric representation. \textbf{Right:} Product of $\mathbb{R}$-operators defining the kernel of $\mathbf{\Lambda}_3(Y)$ according to \eqref{Layer_inhom}: the $SU(2)$ indices flow from the point $x_1$ to the point $x_0$ according to the arrows on dashed lines.}
\label{r_scalar_fig}
\end{figure}

Let us start by considering the model with spinless particles $\Delta_k=2+i \lambda_k$ and $\ell_k=\dot{\ell}_k=0$ at each site. We study the spectral problem of a transfer matrix defined by a spinless auxiliary space, that is $\ell_a=\dot{\ell}_a=0$. In the setup $\Delta_i\neq \Delta_j$ the solution of \eqref{SoV} is a generalization of the homogeneous one treated in \cite{Derkachov2020}, and reads
\begin{align}
\begin{aligned}
\label{Layer_inhom}
&\mathbf{\Lambda}_1(Y)=\Psi_1(Y|x) = \frac{[\mathbf{(x-x_0)}]^{n}}{(x-x_0)^{2\left(2-{\Delta_1} +i\nu \right)}}\,,\\&
\mathbf \Lambda_k(Y)\equiv \mathbf \Lambda_k(n,\nu) =\mathbb{R}^{(n)}_{12}\left(\frac{\Delta_1}{2}-i\nu\right)\mathbb{R}^{(n)}_{23}\left(\frac{\Delta_2}{2}-i\nu\right)\cdots \mathbb{R}^{(n)}_{k-1k}\left(\frac{\Delta_{k-1}}{2}-i\nu\right)\frac{[\mathbf{(x_k-x_0)}]^{n}}{(x_k-x_0)^{2\left(2-{\Delta_k}+i\nu \right)}}\,,\end{aligned}
\end{align}
where the operators $\mathbb{R}^{(n)}_{ij}(u)$ are defined by their integral kernel
\begin{align}
\begin{aligned}
&[\mathbb{R}^{(n)}_{ij}(u)] \Phi(x_i,x_j) = \int d^4 y \,{R}^{(n)}_u (x_i,x_j|y)  \Phi(y,x_j) \\
&{R}^{(n)}_u (x_i,x_j|y) = \frac{[\mathbf{(x_1-y)(\overline{y-x_2})}]^n}{(x_i-x_j)^{2(2-\Delta_i)}(x_i-y)^{2\left(-u +\frac{\Delta_i}{2} \right)}(y-x_j)^{2\left(u+\frac{\Delta_i}{2}\right)}} \,,
\end{aligned}
\end{align}
represented in Fig.\ref{r_scalar_fig} together with the layer operator \eqref{Layer_inhom}.
\noindent
Following the very same steps of the homogeneous model $\Delta_i=\Delta_j$ treated in detail in \cite{Derkachov:2020zvv}, one can check the iterative relation \eqref{SoV}
and an exhaustive proof can also be read out of the more involved spinning model treated in section \ref{spinning_inhom}.
According to \eqref{SoV} the action of the layer $\mathbf{\Lambda}_k(Y)$ singles out the space of the $k$-th particle $\mathbb{V}_k$ from the model, leaving behind a transfer matrix of length $k-1$ and factor
\begin{align}
\begin{aligned}
\label{eigenvalue_k}
q_{a}(u,Y)=  \pi^4   \frac{\Gamma\left(-u-\frac{\Delta_a}{2} \right) \Gamma\left(4-u-\frac{\Delta_a}{2} \right)}{\Gamma\left(u+2+\frac{\Delta_a}{2} \right) \Gamma\left(-u-2+\frac{\Delta_a}{2}\right)} \frac{\Gamma\left(-2-u+\frac{\Delta_a}{2} +Y^* \right) \Gamma\left(u+\frac{\Delta_a}{2}+Y\right)}{\Gamma\left(u+4-\frac{\Delta_a}{2} +Y \right) \Gamma\left(2-u-\frac{\Delta_a}{2} +Y^* \right)} \,,
\end{aligned}
\end{align}
The first fraction appearing in \eqref{eigenvalue_k} is in fact independent from $Y_j$ and can be included into the definition of $\mathbf{Q}_{a,k}(u)$ by a suitable normalization, and the  spectrum is invariant w.r.t. any permutation of the quantum numbers $(Y_1,\dots,Y_L)$. In agreement with such symmetry, permuting the numbers $Y_j$ between the layers of the eigenfunctions amounts just to a rotation of the $SU(2)$ indices of the function
\begin{equation}
\label{symm_scalar}
 \mathbf \Lambda_{k+1}(Y)\cdot  \mathbf \Lambda_{k}(Y') = \frac{r_{k}(Y')}{r_k(Y)}\times  \mathbf{R}(Y',Y) \mathbf \Lambda_{k+1}(Y')\cdot  \mathbf \Lambda_{k}(Y)\mathbf{R}(Y,Y')\,
\end{equation}
where we use the compact notation $\mathbf{R}(Y,Y') \equiv \mathbf{R}_{n,n'}(i\nu-i\nu')$, and the coefficients $r_j(Y)=r_j(n,\nu)$ are given by
\begin{equation}
\label{r_coeff}
r_j(n,\nu) = \frac{\Gamma\left(i\nu+\frac{n}{2} \right)\Gamma\left(2-\Delta_{j}+i\nu+\frac{n}{2} \right)}{\Gamma\left(2-i\nu+\frac{n}{2} \right)\Gamma\left(\Delta_j-i\nu+\frac{n}{2}\right)}\,.
\end{equation}
\begin{figure}[H]
\begin{center}
\includegraphics[scale=0.62]{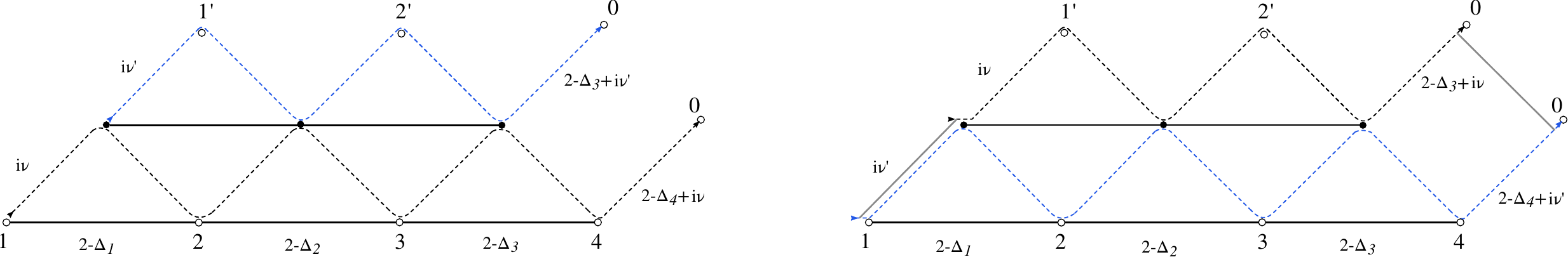}
\end{center}
\caption{L.h.s and r.h.s of the equation \eqref{symm_scalar} for a layer operators of length $k+1=4$.}
\label{symm_scalar_fig}
\end{figure}
\noindent
We define the eigenfunctions of $\mathbf{Q}_{a,L}(u)$ with a suitable normalization
\begin{equation}
\label{eig_inhom_scalar}
\Psi(\mathbf{Y}|\mathbf{x}) =\mathbf  \Lambda_L(Y_L)\cdots  \mathbf \Lambda_2(Y_2) \mathbf \Lambda_1(Y_1)  \prod_{k=1}^{L} r_k(Y)^{k-1} \,.
\end{equation}
For a general permutation $\pi$ of quantum numbers $(Y_1,\dots,Y_L)$ in the eigenfunction \eqref{eig_inhom_scalar}, the equation \eqref{symm_scalar} implies that there is an associated product of matrices $\mathbf{R}_{\pi}$ according to the rule
\begin{align}
\begin{aligned}
\label{perm_R}
(\dots,Y_k,Y_{k+1},\dots) \to (\dots,Y_{k+1},Y_{k},\dots) \,&\Longrightarrow \, \mathbf{R}(Y_k,Y_{k+1})\,\\
(Y_{1},\dots, Y_{L}) \to (Y_{\pi(1)},\dots, Y_{\pi(L)}) \,&\Longrightarrow \, \mathbf{R}_{\pi}(\mathbf{Y})\,,
\end{aligned}
\end{align}
and the matrices $\mathbf{R}(Y_k,Y_{k+1})$ define a representation of the symmetric group $\mathbb{S}_L$ over the spaces of $SU(2)$ spinors, where unitarity \eqref{R_unit} and YBE \eqref{Rmat_YBE} are  the Coxeter relations. 
The straightforward consequence of \eqref{symm_scalar} on the eigenfunctions is
\begin{equation}
\Psi(Y_{\pi(1)},\dots, Y_{\pi(L)}) = \mathbf{R}_{\pi}(\mathbf{Y})^{\dagger}\, \Psi(Y_1,\dots, Y_L) \,\dot{\mathbf{R}}_{\pi}(\mathbf{Y})\,.
\end{equation}
The orthogonality of eigenfunctions at $L=1$ was established in the most general situation in last section, while the overlap of eigenfunctions of length $L>1$ is analyzed starting from the convolution of two layer operators at $\nu\neq \nu'$
\begin{align}
\begin{aligned}
\label{ortho_scalar}
 \mathbf{\bar \Lambda}_{k+1}(Y') \cdot  \mathbf  \Lambda_{k+1}(Y) =\frac{\pi^4}{\mu(Y,Y')}\frac{ r_k(Y)}{r_k(Y')}\times \left[ \mathbf{R}(Y',Y)^{t'} \mathbf  \Lambda_{k}(Y)\cdot  \mathbf{\bar \Lambda}_{k} (Y')^{t'}\mathbf{R}(Y,Y')^{t'}\right]^{t'}\,,
\end{aligned}
\end{align}
where $t'$ is the transposition in the spinor indices of the excitation $Y'$, and
\begin{equation}
\mu(Y,Y')= \left| i(\nu-\nu')+\frac{n-n'}{2}\right|^2\left|1+ i(\nu-\nu')+\frac{n+n'}{2}\right|^2\,.
\end{equation}
In the r.h.s. of \eqref{ortho_scalar} the layer operators appear with length diminished by one respect to the l.h.s., thus the iterative application of \label{ortho_scalar} reduces the overlap of two eigenfunctions of lenght $L$ to the product of $L$ overlaps of eigenfunctions of length-$1$, computed in \eqref{clean_1_p}.
\begin{figure}[H]
\begin{center}
\includegraphics[scale=0.65]{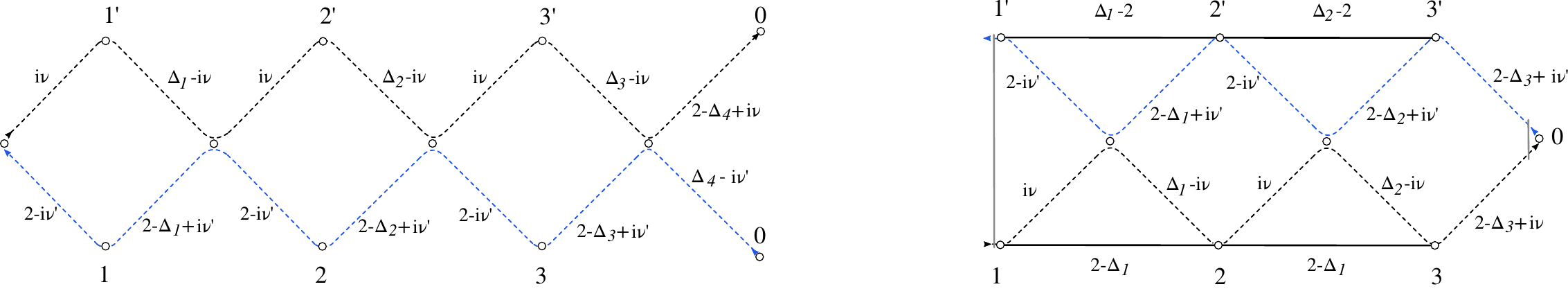}
\end{center}
\caption{L.h.s and r.h.s of the equation \eqref{ortho_scalar} for a layer operators of length $k+1=4$.}
\label{ortho_scalar_fig}
\end{figure}
\noindent
The first non-trivial example is the overlap at $L=2$ can be comptued via \eqref{ortho_scalar} as
\begin{align}
\begin{aligned}
\label{overlap_2}
&\mathbf{\bar \Lambda}_1(Y_1')\mathbf{\bar \Lambda}_2(Y_2')\mathbf{ \Lambda}_2(Y_2)\mathbf{ \Lambda}_1(Y_1) = \\&= \frac{\pi^4}{\mu(Y_1,Y_2)}\frac{r(Y_2)}{r(Y_2')} \times  \left[\mathbf{R}(Y_2',Y_2)^{t_2'}\mathbf{\bar \Lambda}_1(Y_1')\mathbf{ \Lambda}_1(Y_2)\mathbf{ \bar \Lambda}_1(Y_2')^{t_2'}\mathbf{ \Lambda}_1(Y_1) \mathbf{R}(Y_2,Y_2')^{t_2'}
\right]^{t_2'}= \\& =\frac{4\pi^{10}}{(n_1+1)(n_2+1)\mu(Y_1,Y_2)} \frac{r(Y_2)}{r(Y_1)}  {\delta(Y_1-Y_2') \delta(Y_2-Y_1')} \times \mathbf{R}(Y_2,Y_2')\otimes  \mathbf{R}(Y_1,Y_1') \mathbb{P}_{1'2} \mathbb{P}_{12'}\,.
\end{aligned}
\end{align}
The operators $\mathbb{P}_{1'2}$ and $\mathbb{P}_{12'}$ are permutations in the tensor product of symmetric spinors
\begin{align}
\begin{aligned}
\mathbb{P}_{ij'}: \,&\text{Sym}_{n_i}[\mathbb{C}]\otimes  \text{Sym}_{n_j'}[\mathbb{C}] \rightarrow \text{Sym}_{n_i}[\mathbb{C}]\otimes  \text{Sym}_{n_j'}[\mathbb{C}]\\ &\quad \qquad\qquad |\chi\rangle \otimes  |\eta \rangle \,\mapsto \,  |\chi\rangle \otimes  |\eta \rangle\,,
\end{aligned}
\end{align}
and the definition is well posed since the condition $n_i =n_j'$ is ensured by $\delta(Y_i-Y_j')$ in \eqref{overlap_2}.
The complete result of the overlap must be symmetric respect to the permutations of $Y$-s between layers, therefore for $L=2$ we repeat \eqref{overlap_2} after the exchange $Y_1 \leftrightarrow Y_2$
\begin{align}
\begin{aligned}
&\mathbf{\bar \Lambda}_1(Y_1')\mathbf{\bar \Lambda}_2(Y_2')\mathbf{ \Lambda}_2(Y_2)\mathbf{ \Lambda}_1(Y_1) = \\&= \frac{r(Y_1)}{r(Y_2)} \times  \mathbf{R}(Y_1,Y_2) \mathbf{\bar \Lambda}_1(Y_1') \mathbf{ \bar \Lambda}_2(Y_2') \mathbf{ \Lambda}_2(Y_1) \mathbf{ \Lambda}_1(Y_2)  \mathbf{R}(Y_2,Y_1)= \\&=\frac{4\pi^{10} {\delta(Y_1-Y_1') \delta(Y_2-Y_2')} }{(n_1+1)(n_2+1)\mu(Y_1,Y_2)} \times  \mathbf{R}(Y_1,Y_2)  \mathbb{P}_{1'1}  \mathbf{R}(Y_1,Y_2')  \mathbf{R}(Y_2,Y_1') \mathbb{P}_{22'}\mathbf{R}(Y_2,Y_1)= \\& =\frac{4\pi^{10}}{(n_1+1)(n_2+1)\mu(Y_1,Y_2)} {\delta(Y_1-Y_1') \delta(Y_2-Y_2')} \times \mathbb{P}_{1'1} \mathbb{P}_{22'}\,,
\end{aligned}
\end{align}
and it follows that
\begin{align}
\begin{aligned}
&\langle \Psi_2(Y_1',Y_2') ,\Psi_2(Y_1,Y_2)\rangle =\frac{1}{\, \mu(Y_1,Y_2)} \frac{2\pi^{10}}{(n_1+1)(n_2+1)}\times \\&\times \left[  \delta(Y_1-Y_1') \delta(Y_2-Y_2') \mathbb{P}_{11'} \mathbb{P}_{22'}+ \delta(Y_1-Y_2') \delta(Y_2-Y_1') \mathbf{R}(Y_2,Y_1)\otimes \dot{\mathbf{R}}(Y_1,Y_2) \mathbb{P}_{12'} \mathbb{P}_{1'2}   \right]\,.
\end{aligned}
\end{align}
Similarly, for higher length $L$ the overlap is a sum of $\delta$-functions over the permutations of the quantum numbers $(Y_1,\dots,Y_L)$, each associated with the corresponding product of $\mathbf{R}$ according to \eqref{perm_R}. The scalar product reads
\begin{equation}
\langle \Psi_L(\mathbf{Y}') ,\Psi_L(\mathbf{Y})\rangle  = \frac{1}{\rho(\mathbf{Y}) L!}\sum_{\pi}  \mathbf{R} _{\pi}( \mathbf{Y}) \otimes \dot{\mathbf{R}}_{\pi} (\mathbf{Y}) \mathbb{P}_{\pi}  \prod_{j=1}^L \delta (Y_j-Y_{\pi(j)})\,,
\end{equation}
where
\begin{equation}
\mathbb{P}_{\pi} = \mathbb{P}_{1\pi(1)'}\otimes \mathbb{P}_{2\pi(2)'}\otimes\cdots \otimes\mathbb{P}_{L\pi(L)'}\,,
\end{equation}
and the compact notation $\rho(\mathbf{Y})$ stands for
\begin{equation}
\label{rho_def}
\rho(Y_1,\dots,Y_L) =  \prod_{j=1}^L \frac{(n_j+1)}{2\pi^{(2L+1)}} \prod_{k
\neq j}^L {\mu(Y_j,Y_k)}\,,
\end{equation}
and defines the measure over the space of separated variables $Y_j$, i.e. the Hilbert space structure in \eqref{dual_Hilbert}. Hence, the spectral transformation $\mathcal{U}$ of \eqref{sov_trans} is extended straightforwardly from the homogeneous case of \cite{Derkachov2020} to the present one as an isometry between Hilbert spaces $\mathcal{V}$ and $\widetilde{\mathcal{V}}$
\begin{align}
\begin{aligned}
\label{SoV_transform}
&\mathcal{U}: \Phi(x_1,\dots,x_L) \,\mapsto \, \left(\mathcal{U}\Phi\right) (\mathbf{Y}) =\int d^4 x_1\cdots d^4 x_L \Psi(\mathbf{x}|\mathbf{Y})^* \Phi(\mathbf{x}) = \langle \Psi(\mathbf{Y}) , \Phi \rangle_{\mathcal{V}}\,,\\
&\mathcal{U}^{-1}: \widetilde{\Phi}(Y_1,\dots,Y_L) \,\mapsto \, \left(\mathcal{U}^{-1}\widetilde{\Phi}\right) (\mathbf{x}) =\sum_{n_1,\dots,n_L=0}^{\infty} \int d \nu_1\cdots d \nu_L\, \rho(\mathbf{Y}) \widetilde{\Phi}(\mathbf{Y}) \Psi(\mathbf{x}|\mathbf{Y})=\langle \widetilde \Phi,\Psi(\mathbf{x})  \rangle_{\widetilde{\mathcal{V}}}\,.
\end{aligned}
\end{align}

The properties \eqref{symm_scalar} and \eqref{ortho_scalar} generalize the analogues ones proved in \cite{Derkachov:2020zvv} for the homogeneous model $\Delta_i=\Delta_j$, and their proofs follows the very same steps. Alternatively,  \eqref{symm_scalar} and \eqref{ortho_scalar} can be recovered as a special case of the more general algebraic relations holding for a spinning model, treated in the section \ref{spinning_inhom}. The only difference respect to the homogeneous model lies in the normalization coefficients $r_k(Y)$  and in the spectrum of the transfer matrix, since in the present case both depend on the scaling behaviour of the sites $\Delta_k$.
\subsubsection{Spinning charges}
\label{sect:spinning}
The generic conserved charge of the twisted chain with fixed boundaries is generated by the a transfer matrix \eqref{Q_op} whose auxiliary space transforms in the representation with non-zero spins $(\ell_a,\dot{\ell}_a)\neq (0,0)$. In this case, \eqref{full_eigen_gen} does not define eigenfunctions of $\mathbf{Q}_{a,L}(u)$ since the exchange of the layer $\mathbf \Lambda_k(Y)$ with the operator $\mathbf{Q}_{a,k}(u)$ is not simple as in \eqref{SoV}. Indeed, the transfer matrix acts non-trivially on a layer's excitation spinor indices, mixing them with the ones of the auxiliary space.
The equation \eqref{SoV} gets modified as follows (derived in \cite{Derkachov:2020zvv} for $\Delta_i =\Delta_j$)
\begin{align}
\begin{aligned}
\label{non_SoV}
\mathbb{Q}_{a,k} (u) \mathbf{\Lambda}_k(Y)= q_{a}(Y)  \times \mathbf{R}_{\dot \ell_a,n}\left(2+u+i\nu-\frac{\Delta_a}{2}\right)   \mathbf{\Lambda}_k(Y)\mathbb{Q}_{a,k}(u) \mathbf{R}_{\ell_a,n}\left(-u-i\nu-\frac{\Delta_a}{2}\right)  \,,
\end{aligned}
\end{align}
and by $\mathbb{Q}_{a,k}(u)$ we denote the operator \eqref{tilda_Q_op}, obtained from $\mathbf{Q}_{k,a}(u)$ by opening the auxiliary space trace over $\ell_a$ and $\dot \ell_a$ spinor indices between site $k$ and site $1$ 
\begin{align}
\begin{aligned}
\mathbf{Q}_{a,k}(u)= \text{Tr}_{\ell_a } \otimes \text{Tr}_{\dot \ell_a }\,\mathbb{Q}_{a,k} (u)\,.
\end{aligned}
\end{align}
The function $q_{a}(Y)$ is given by
\begin{align}
\begin{aligned}
q_{a}(Y) = \pi^4 &\frac{\Gamma\left(2-i\nu -u+\frac{\Delta_a}{2}+\frac{n-\dot \ell_a}{2}\right)}{\Gamma\left(3+i\nu+u-\frac{\Delta_a}{2}+\frac{n-\dot \ell_a}{2}\right)\left(3+i\nu+u-\frac{\Delta_a}{2}+\frac{n+\dot \ell_a}{2}\right)}\times \\& \times\frac{\Gamma\left(u+i\nu+\frac{\Delta_a}{2}+\frac{n-\ell_a}{2}\right)}{\Gamma\left(1-u-i\nu-\frac{\Delta_a}{2}+\frac{n- \ell_a}{2}\right)\left(1-u-i\nu-\frac{\Delta_a}{2}+\frac{n+\ell_a}{2}\right)}\times \\& \times\frac{\Gamma\left(-u-\frac{\Delta_a}{2}+\frac{ \ell_a}{2}\right)\Gamma\left(4+u-\frac{\Delta_a}{2}+\frac{\dot \ell_a}{2}\right)}{\Gamma\left(u+2+\frac{\Delta_a}{2}+\frac{\ell_a}{2}\right)\Gamma\left(-u-2+\frac{\Delta_a}{2}+\frac{\dot \ell_a}{2}\right)}\,,
\end{aligned}
\end{align}
and for $\ell_a=\dot \ell_a =0$ reduces to \eqref{eigenvalue_k}.
The iteration of \eqref{non_SoV} and the trace over auxiliary spinor indices delivers the action of $\mathbf{Q}_{a,k}(u)$ over the functions \eqref{eig_inhom_scalar}
\begin{align}
\begin{aligned}
\label{Q_mix_t}
\mathbf{Q}_{a,L} (u) \Psi(\mathbf{Y}|\mathbf{x}) = q_{a}(Y_1) \cdots q_{a}(Y_L)  \times \mathbf{\bar t}_{a,L\dots 1}(u|\mathbf{Y})  \, \Psi(\mathbf{Y}|\mathbf{x}) \, \mathbf{  t}_{a,1\dots L}(u|\mathbf{Y}) \,,
\end{aligned}
\end{align}
where
\begin{align}
\begin{aligned}
\label{t_t_bar}
&\mathbf{t}_{a,1\dots L}(u|\mathbf{Y}) = \text{Tr}_{\ell_a}\left[\mathbf{R}_{a}(u|Y_1)\mathbf{R}_{a}(u|Y_2)\cdots \mathbf{R}_{a}(u|Y_L)\right]\,,\\
&\mathbf{ \bar t}_{a,L\dots 1}(u|\mathbf{Y}) = \text{Tr}_{\dot \ell_a}\left[\mathbf{R}_{\dot a}(u|Y_L)\cdots \mathbf{R}_{\dot a}(u|Y_2) \mathbf{R}_{\dot a}(u|Y_1)\right]\,,
\end{aligned}
\end{align}
and the $\mathbf{R}$-matrices inside the traces in \eqref{t_t_bar} are a compact notation for
\begin{equation}
\label{RR_XXX}
 \mathbf{R}_{\dot a}(u|Y)=\mathbf{R}_{\dot \ell_a,n}\left(2+u+i\nu-\frac{\Delta_a}{2}\right)\,,\, \mathbf{R}_{a}(u|Y)=\mathbf{R}_{\ell_a,n}\left(-u-i\nu-\frac{\Delta_a}{2}\right)\,.
\end{equation}
The equation \eqref{t_t_bar} defines the transfer matrices of two generalized $SU(2)$ Heisenberg magnet where the $j$-th site is in the $\mathbf{(n_j+1)}$ irreducible representation
\begin{equation}
\text{Sym}_{n_1}[\mathbb{C}^2] \otimes \cdots \otimes \text{Sym}_{n_k}[\mathbb{C}^2] \,.
\end{equation}
and the quantum number $\nu_j$ is the inhomogeneity parameter on the $j$-th site.

In order to determine the eigenfunctions of the $\mathbf{Q}_{a,L}(u)$ we need to pair each layer with auxiliary spinors \eqref{aux_spins}, and diagonalize the $SU(2)$ transfer matrices acting on them. The equation \eqref{Q_mix_t} can be rewritten - modulo normalization - as
\begin{align}
\begin{aligned}
& \mathbf{Q}_{a,L}(u) \langle \alpha_L | \mathbf\Lambda_L(Y_L)|\beta_L\rangle \cdots \langle \alpha_2| \mathbf \Lambda_2(Y_2)|\beta_2\rangle\langle \alpha_1| \mathbf \Lambda_1(Y_1)|\beta_1\rangle=\\&=q_{a}(Y_1) \cdots q_{a}(Y_L)  \langle \alpha_L,\dots,\alpha_1|\mathbf{\overline t}_{a ,L\dots 1}(u|\mathbf{Y})  \,\mathbf \Lambda_L(Y_L)\cdots \mathbf \Lambda_1(Y_1) \,\mathbf{t}_{a,1\dots L}(u|\mathbf{Y}) |\beta_L, \dots ,\beta_1\rangle\,,
\end{aligned}
\end{align}
The diagonalization of \eqref{t_t_bar} can be achieved by Bethe ansatz technique, which ultimately relies on the Yang-Baxter equation for the fused $SU(2)$ matrices
\begin{align}
\begin{aligned}
&\mathbf{R}_{\ell_a,\ell_b}(u-v)  \left(\mathbf{R}_{a}(u|Y_1)\cdots \mathbf{R}_{a}(u|Y_L) \right) \left( \mathbf{R}_{b}(v|Y_1)\cdots \mathbf{R}_{b}(v|Y_L)\right)= \\&\quad \quad\qquad \,\,\,=\left( \mathbf{R}_{b}(v|Y_1)\cdots \mathbf{R}_{b}(v|Y_L)\right) \ \left(\mathbf{R}_{a}(u|Y_1)\cdots \mathbf{R}_{a}(u|Y_L) \right) \mathbf{R}_{\ell_a,\ell_b}(u-v)\,,
\end{aligned}
\end{align}
and therefore
\begin{equation}
\label{SU_2_integr}
\left[ \mathbf{t}_{a,1\dots L}(u|\mathbf{Y}),\mathbf{t}_{b,1\dots L}(v|\mathbf{Y}) \right] = 0\,,\,\,\,\,\, \left[ \mathbf{\overline t}_{a,L\dots 1}(u|\mathbf{Y}),\mathbf{\overline t}_{b,L\dots 1}(v|\mathbf{Y}) \right] = 0\,.
\end{equation}
Since commuting operators \eqref{SU_2_integr} share a basis of eigenfunctions, it is enough to solve the spectral problem for $\mathbf{t}^{(a)}_{1\dots L}(u|\mathbf{Y})$ in the fundamental representation $\ell_a=1$, that is by the choice of a spin-$\frac{1}{2}$ auxiliary space in the spin chain. This latter class of transfer matrices includes several known models: when all the excitations have the same spin number $n_1=\dots = n_L=2s $ we recover the transfer matrix of the XXX$_s$ Heiseberg model \cite{BABUJIAN1982479}, while for $n_j=2s'$ and $n_{i\neq j}=2s$ we recover the magnet with an impurity of the Kondo problem \cite{PhysRevLett.45.379, Schlottmann_1991}. Here we deal with the general inhomogeneous case $n_i \neq n_j$, whose solution has been studied by O.~Castro Alvaredo and J.~Maillet in \cite{Castro-Alvaredo:2007psp} via the algebraic Bethe ansatz \cite{Faddeev:1996iy,korepin_bogoliubov_izergin_1993}.

The eigenvectors and eigenvalues of the transfer matrix $\mathbf{t}_{a,1,\dots,L}(u|\mathbf{Y})$ depend on a set of Bethe roots $\{w_1,\dots,w_M \}$, complex numbers that solve a set of finite difference equations 
\begin{equation}
\label{BAE}
\prod_{k=1}^{L}\frac{w_k -i\nu_k -\frac{n_j-1}{2}i}{w_k-i\nu_k +\frac{n_j+1}{2}i} = \prod_{k=1}^{M} \frac{w_k-w_j +i}{w_k-w_j - i}\,,\,\,\,\,\,\, j=1,\dots,M\,.
\end{equation}
Each eigenvector is the (on-shell) Bethe state corresponding to a set of Bethe roots, according to the recipe of quantum inverse scattering method \cite{Faddeev:1996iy}. 
For a state
\begin{equation}
|\mathbf{w}\rangle = |w_1,w_2,\dots,w_M\rangle\,\in\, \text{Sym}_{n_1}[\mathbb{C}^2] \otimes \cdots \otimes \text{Sym}_{n_L}[\mathbb{C}^2]\,,
\end{equation}
 the spectral equation reads
\begin{align}
\begin{aligned}
&\mathbf t_{a,1\dots,L}(u|\mathbf{Y}) |w_1,\dots w_L\rangle = \tau_a(u|\mathbf{Y},w_1,\dots,w_M) |w_1,\dots w_L\rangle\,,\\
\end{aligned}
\end{align}
and the expression with $\ell_a=1$ is an adaptation from (2.16) of \cite{Castro-Alvaredo:2007psp}
\begin{equation}
\label{eigen_bethe}
\tau_{\ell_a=1}(u|\mathbf{Y},w_1,\dots,w_M)  = \prod_{j=1}^M \frac{w_j+u+\frac{\Delta_a}{2}+i}{w_j+u+\frac{\Delta_a}{2}}+\prod_{k=1}^{L} \frac{u+\frac{\Delta_a}{2}-i\nu_k+\frac{n_j-1}{2}i}{u+\frac{\Delta_a}{2}-i\nu_k-\frac{n_j+1}{2}i}\prod_{j=1}^M\frac{w_j+u+\frac{\Delta_a}{2}}{w_j+u+\frac{\Delta_a}{2}-i}\,.
\end{equation}
The Bethe states for the transfer matrix $\mathbf{\bar t}_{L\dots 1}(u|\mathbf{Y})$ and the formula for its eigenvalues follow from \eqref{eigen_bethe}, with the change of spectral parameter and inhomogeneities \eqref{RR_XXX}
\begin{equation}
u \,\to \,2-u\,,\,\,\, i\nu_j\,\to\,- i\nu_j\,. 
\end{equation}
For a choice of $\ell_a,\dot \ell_a \neq 1$  we the actual eigenvalue of the higher spin transfer matrices $\mathbf{t}_{a,1\dots L}(u|\mathbf{Y})$ and $\mathbf{\bar t}_{a,L\dots 1}(u|\mathbf{Y})$. Its general expression was computed by \cite{Castro-Alvaredo:2007psp}, and for the matrix $\mathbf{t}_{a,1\dots,L}(u|\mathbf{Y})$ with $M$ Bethe roots it reads
\begin{align}
\begin{aligned}
&\tau_{\ell_a}(u|\mathbf{Y},w_1,\dots,w_M) = \sum_{r=1}^{\ell_a} \mathcal{P}^{(\ell_a)}_{r}(u)\prod_{j=1}^M \frac{\left(u+\frac{\Delta_a}{2}-i\nu_j-i\frac{\ell_{a}+1}{2}\right)\left(u+\frac{\Delta_a}{2}-i\nu_j-i\frac{\ell_{a}+1}{2}\right)}{\left(u+\frac{\Delta_a}{2}-i\nu_j-i\frac{\ell_{a}-r-1}{2}\right)\left(u+\frac{\Delta_a}{2}-i\nu_j+i\frac{\ell_{a}+1-r}{2}\right)}\,,\\
&\mathcal{P}^{(\ell_a)}_{r}(u)=\prod_{h=r}^{\ell_a-1} \frac{\left(u+\frac{\Delta_a}{2}-i\nu_j+i\frac{2 h-\ell_{a}+1}{2}\right)}{\left(u+\frac{\Delta_a}{2}-i\nu_j+i\frac{2h-\ell_{a}-\ell_j-1}{2}\right)}\,,\,\,\, \mathcal{P}^{(1)}_{r}(u)=1\,.
\end{aligned}
\end{align}
Finally, the eigenfunctions of the model \eqref{Q_op} - for general auxiliary space and spinless physical spaces - are labeled by the rapidity and spin of the layer excitations $Y_j$, and also by $M,\bar M$ Bethe roots $w_j$ and $\bar w_j$
\begin{equation}
\label{eigenf_spinn}
\Psi(\mathbf{Y,w,\bar w}|\mathbf{x}) = \langle \bar w_1,\dots, \bar w_{\bar M}| \mathbf \Lambda_L(Y_L) \cdots \mathbf \Lambda_1(Y_1) | w_1,\dots, w_M \rangle\,. 
\end{equation}
 and therefore depend at most on of $2L$ quantum numbers, namley $\{\nu_k,n_k\}$ plus the two sets of Bethe roots $\{w_1,\dots,w_M\}$ and $\{\bar w_1,\dots \bar w_{\bar M}\}$. The eigenvalue corresponding to \eqref{eigenf_spinn} follows from \eqref{Q_mix_t}
 \begin{equation}
\tau_{a,1\dots,L}(u|\mathbf{Y},\mathbf{w})\, \bar{\tau}^*_{a,L\dots,1}(u|\mathbf{Y},\mathbf{\bar w}) \prod_{k=1}^L q_{a,k}(u)
 \end{equation}
The property \eqref{symm_scalar} for the exchange of the excitations $Y_{k+1},Y_{k}$ between adjacent layers implies
\begin{equation}
\Psi(\mathbf{Y,w,\bar w}|\mathbf{x}) = \langle \mathbf{\bar w}| \mathbf{R}(Y_{k-1},Y_{k}) \mathbf \Lambda_L(Y_L) \cdots  \mathbf\Lambda_{k+1}(Y_k) \mathbf\Lambda_{k}(Y_{k+1}) \cdots \mathbf \Lambda_1(Y_1) \mathbf{R}(Y_{k+1},Y_k)|\mathbf{w}\rangle\,,
\end{equation}
mapping one-to-one the Bethe states $|\mathbf w\rangle$ of the transfer matrix $\mathbf{t}_{a,1\dots k,k+1\dots L}(u|\dots Y_k,Y_{k+1}\dots)$ to those of the model with swapped spins and inhomogeneities between sites $k,k+1$. The map between the two models is established via the Yang-Baxter equation as follows
\begin{equation}
\label{reshuffle_spinn}
\mathbf{R}(Y_{k+1},Y_k)\mathbf{t}_{a,1\dots k,k+1\dots L}(u|\mathbf{Y})  =  \mathbf{t}_{a,1\dots k+1,k\dots L}(u|Y_1\dots Y_{k+1},Y_k\dots Y_L) \mathbf{R}(Y_{k+1},Y_k)\,,
\end{equation}
and leaves the eigenvalues unchanged. The very same considerations hold for the transfer matrix $\mathbf{\bar t}_{a,L\dots,1}(u|\mathbf Y)$ acting on the left vectors $\langle \mathbf{\bar w}|$ and re-shuffled by $\mathbf{R}(Y_k,Y_{k+1})$.
Because of the pairing of layer operators with Bethe states in \eqref{eigenf_spinn}, the overlap of eigenfunctions reads
\begin{align}
\begin{aligned}
\label{overlap_spin_aux}
\langle \Psi_L(\mathbf{Y',w',\bar w'}) ,\Psi_L(\mathbf{Y,w,\bar w})\rangle = \frac{1}{\rho(\mathbf{Y}) L!}\sum_{\pi}  \langle \mathbf{\bar w}|\langle \mathbf{w'}|{\mathbf{R} }_{\pi}( \mathbf{Y}) &\otimes \dot{\mathbf{R}}_{\pi} (\mathbf{Y}) \mathbb{P}_{\pi}|\mathbf{w}\rangle |\mathbf{\bar w'}\rangle \,.\end{aligned}
\end{align}
The eigenfunctions \eqref{eigenf_spinn} are related to \eqref{eig_inhom_scalar} by the pairing of the spinor indices of excitations with the left/right Bethe vectors, therefore we may use the notation of auxiliary spinors \eqref{aux_spins} and write
\begin{equation}
\Psi(\mathbf{Y,w,\bar w}) = \langle \mathbf{\bar w}| \Psi(\mathbf{Y})|\mathbf{w}\rangle\,.
\end{equation}
Let's analyze the overlap in the space of spin vectors. First, the operator $\mathbb{P}_{\pi}$ is charachterized by its action on decomposable tensors $|\alpha_1,\dots, \alpha_L\rangle=|\alpha_1\rangle \otimes \cdots \otimes| \alpha_L\rangle$
\begin{equation}
 \mathbb{P}_{\pi}|\alpha_1,\dots,\alpha_L\rangle\otimes |\alpha'_1,\dots,\alpha'_L\rangle  = |\alpha'_{\pi(1)},\dots,\alpha'_{\pi(L)}\rangle\otimes |\alpha_{\pi(1)},\dots,\alpha_{\pi(L)}\rangle\,,
\end{equation}
which is well defined because $n_j = n'_{\pi(j)}$ by virtue of $\delta(Y_j-Y'_{\pi(j)})$ (in the r.h.s. of \eqref{overlap_spin_aux}). Secondly, the action of $\mathbf{R}_{\pi}(\mathbf{Y})$-matrices maps the Bethe states to the chain with permuted sites, reshuffling the sites according to \eqref{reshuffle_spinn}
 \begin{align}
 \begin{aligned}
 \mathbf{t}_{a,1'\dots L'}(u|\mathbf{Y}') \,{\mathbf{R}}_{\pi}(\mathbf{Y}') &= {\mathbf{R}}_{\pi}(\mathbf{Y'})\,  \mathbf{t}_{a,\pi(1')\dots \pi(L')}(u|{Y}'_{\pi(1)}\dots Y'_{\pi(L)})\,,\\
  \mathbf{\bar t}_{a,L\dots 1}(u|\mathbf{Y})\,\mathbf{R}_{\pi}(\mathbf{Y}) & =   \mathbf{R}_{\pi}(\mathbf{Y})  \,\mathbf{\bar t}_{a,\pi(L)\dots \pi(1) }(u|{Y}_{\pi(L)}\dots Y_{\pi(1)})\,,
 \end{aligned}
 \end{align}
 that is the states
 \begin{equation}
 \langle \mathbf{w}'| \langle \mathbf{\bar w} |{\mathbf{R} }_{\pi}( \mathbf{Y}) \otimes \dot{\mathbf{R}}_{\pi} (\mathbf{Y})=  \langle \mathbf{\bar w} | {\mathbf{R} }_{\pi}( \mathbf{Y}) \otimes \langle \mathbf{w}'| \dot{\mathbf{R}}_{\pi} (\mathbf{Y})\,,
  \end{equation}
  are the Bethe states of roots $\mathbf{\bar w}$ and $\mathbf{w}'$ belonging respectively to the (dual of) spaces
 \begin{equation}
 \text{Sym}_{n_{\pi(1)}}[\mathbb{C}^2] \otimes \cdots \otimes \text{Sym}_{n_{\pi(L)}}[\mathbb{C}^2]\,,\,\text{and}\,\,\text{Sym}_{n'_{\pi(1)}}[\mathbb{C}^2] \otimes \cdots \otimes \text{Sym}_{{n'_{\pi(L)}}}[\mathbb{C}^2]\,,
  \end{equation}
 and for every permutation $\pi$ the paring reduces to two spin-chain scalar products
 \begin{equation}
 \langle \mathbf{\bar w}|\langle \mathbf{w'}|{\mathbf{R} }_{\pi}( \mathbf{Y})\otimes \dot{\mathbf{R}}_{\pi} (\mathbf{Y}) \mathbb{P}_{\pi}|\mathbf{w}\rangle |\mathbf{\bar w'}\rangle \,=\, \langle \mathbf{w'}|\mathbf{w}\rangle\langle \mathbf{\bar w} |\mathbf{\bar w'}\rangle\,.
 \end{equation}
 The resulting expression reads \cite{Korepin_1982}
\begin{align}
\begin{aligned}
 &\langle \mathbf{\bar w}|\mathbf{\bar w'}\rangle   \langle \mathbf{w'}|\mathbf{ w}\rangle = \delta_{M,M'}\delta_{\bar M,\bar M'} \frac{\det {H}(\mathbf{w,w'})}{\prod_{i\neq j}^M(w_i-w_j) (w'_i-w'_j)}\frac{\det {H}(\mathbf{\bar w,\bar w'})}{\prod_{i\neq j}^{\bar M}(\bar w_i-\bar w_j) (\bar w'_i-\bar w'_j)}\,,
 \end{aligned}
 \end{align}
 with the $M\times M$ matrix $H(\mathbf{w},\mathbf{w}')$
 \begin{equation}
  H(\mathbf{w,w'})_{i,j} =\frac{i}{w_i-w'_j} \left[\prod_{k\neq i}^{ M} (w_k-w'_j+i)+\prod_{h=1}^L \frac{w_h+i \nu_h-\frac{n_j+1}{2}i}{w_h+i \nu_h+\frac{n_j-1}{2}i} \prod_{k\neq i}^M (w_k-w'_j+i) \right]\,.
 \end{equation}
The measure \eqref{rho_def} can be cast in the form of a Vandermonde determinant (see section 2.2 of \cite{Basso:2021omx})
\begin{equation}
\rho(\mathbf{Y}) = \prod_{k\neq h}^{2L} \left(\xi_k-\xi_h\right)\,;\,\,\,\,\,\,\,  \xi_{2h+1} = \frac{1}{2}+Y_h\,,\, \,\xi_{2h} = Y_h^* -\frac{1}{2}\,,
\end{equation}
thus the scalar product of eigenfunctions \eqref{eigenf_spinn} is a product of determinants - i.e. a determinant. The eigenfunctions \eqref{eigenf_spinn} are a basis of eigenfunctions common to any transfer matrix \eqref{Q_op} with spinless particles $(\Delta_k,0,0)$. The eigenvalues depend on the choice of auxiliary space representation $(\Delta_a,\ell_a,\dot{\ell}_a)$ in an explicit fashion, and the degeneracy of the spectrum respect to $SO(4)$ is removed as the eigenvalues depend on the Bethe roots.
\subsection{Spinning inhomogeneous model}
\label{spinning_inhom}
In the most general setup the $k$-th particle of the chain transforms in a unitary representations with non-zero spins $(\ell_k,\dot{\ell}_k)\neq (0,0)$. The spectral problem of transfer matrices with spinless auxiliary space $\ell_a=\dot \ell_a=0$ can be solved by separation of variables similarly to the scalar case of section \ref{scalar_inhom}. For the sake of applications to Feynman integrals, this class of transfer matrices includes the graph-building operator for the most general topology of planar graphs of the fishnet theory \eqref{chiCFT4_intro}, that is a mix of hexagonal ``honeycomb" and squared ``fishnet" as that of Fig.\ref{disk_chi}.

\begin{figure}[H]
\begin{center}
\includegraphics[scale=1]{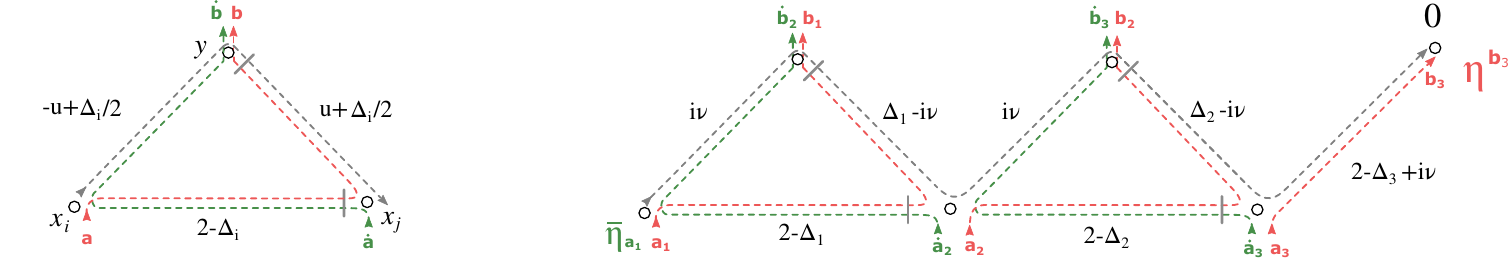}
\end{center}
\caption{\textbf{Left:} Diagram of the integral kernel ${R}^{(n)}_u (x_i,x_j|y)_{\mathbf{a \dot a}}^{\mathbf{b \dot b}}$. The dashed line starting in $x_i$ and ending in $x_j$ carries $n$-symmetric spinor indices. The green/red dashed lines carry spinor indices of the particles of the chain. \textbf{Right:} Diagram corresponding to the kernel of the layer operator $\mathbf \Lambda_3(Y|\eta,\bar \eta)_{\mathbf{\underline{a},\underline{\dot {a}}}}^{\mathbf{\underline{b},\underline{\dot{b}}}}$. The various operators $\mathbf{R}_{i,i+1}^{(n)}(u)$ are convoluted in the $n$-symmetric spinor indices, as represented by the continuous grey dashed line starting.}
\label{R_layer_sp}
\end{figure}

\begin{figure}[H] 
\begin{center}
\includegraphics[scale=0.6]{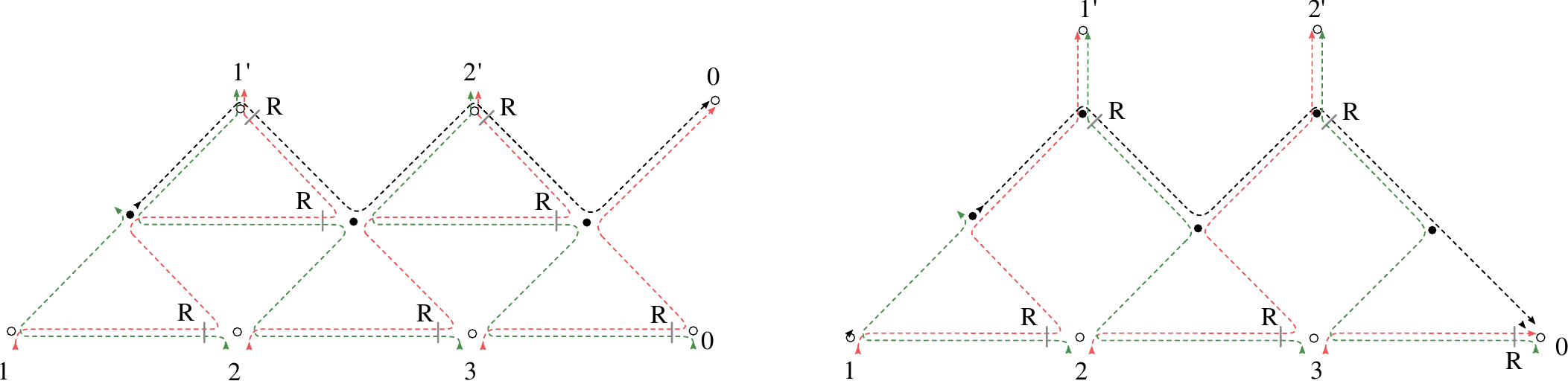}\\ \vspace{8mm}
\includegraphics[scale=0.6]{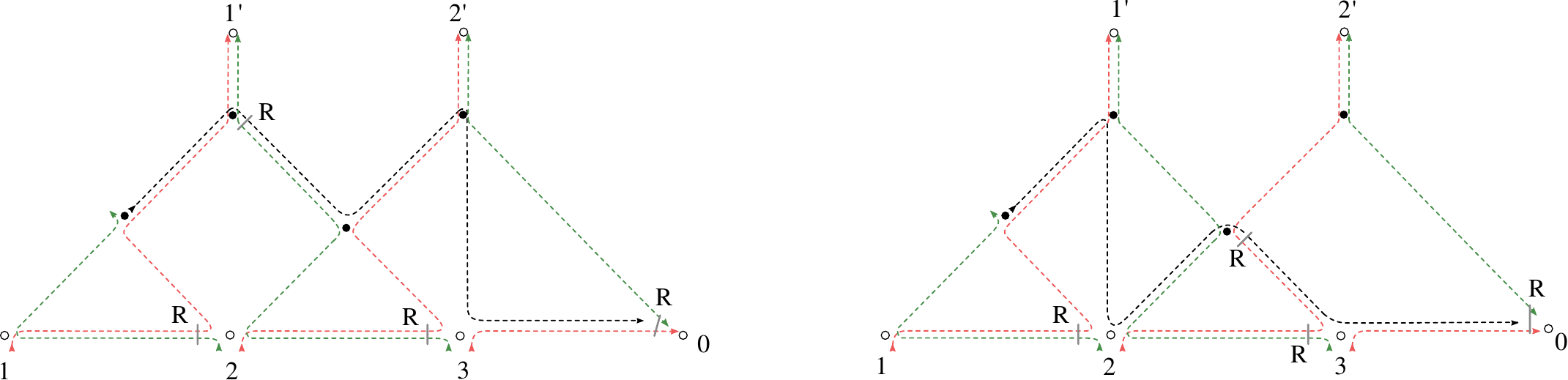}\\ \vspace{8mm}
\includegraphics[scale=0.6]{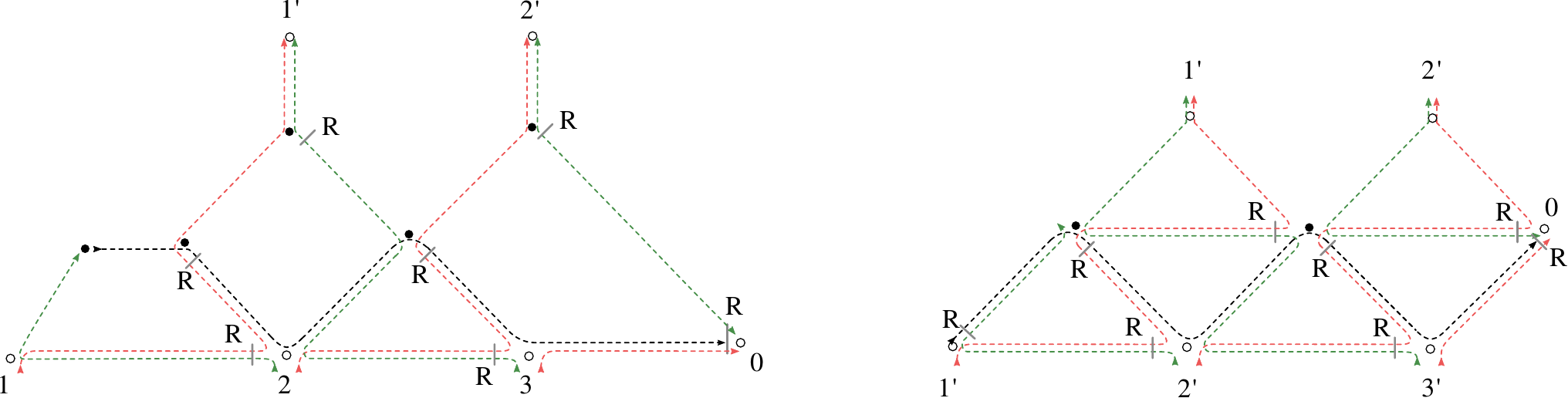}
\end{center}
\caption{\textbf{Up left:} diagram representation of $\mathbf{Q}_{a+}(u) \mathbf{\Lambda}_k(Y)$. \textbf{Up right:} transformation of triangles into star integrals in the layer $\mathbf{\Lambda}_k$, and identification of $x_0$ points. \textbf{Middle left:} star-triangle integration of the rightmost blob. \textbf{Middle right:} the interchange relation I of Fig.\eqref{interchanges} moves the vertical black line from right to left in the diagram. \textbf{Down left:} the triangle with basis the vertical black line is replaced by a star integral. \textbf{Down right:} result of the integration of the leftmost blob.}
\label{SOV_spin_1}
\end{figure}

\begin{figure}[H]
\begin{center}
\includegraphics[scale=0.6]{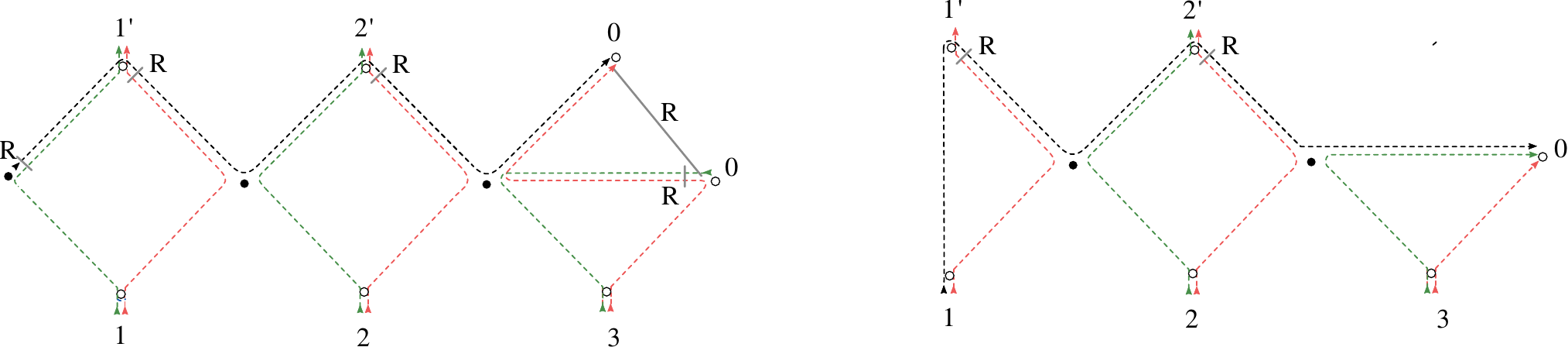}\\ \vspace{8mm}
\includegraphics[scale=0.6]{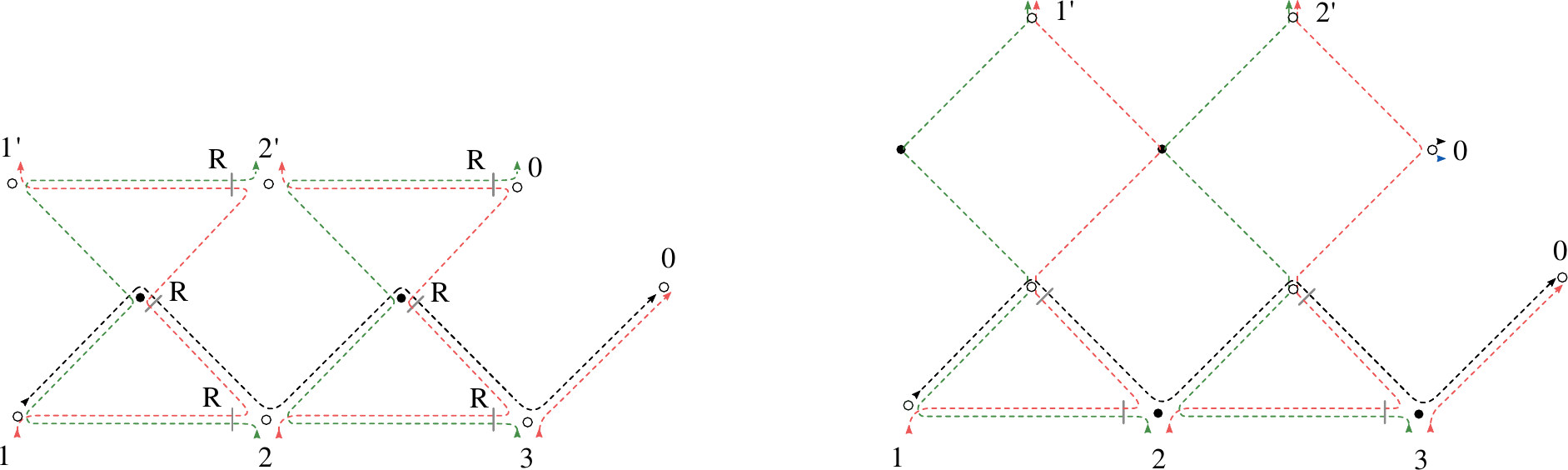}
\end{center}
\caption{\textbf{Up left:} $\mathbf{Q}_{a-}(u)$ is applied to the first row of triangles of the down-right picture in Fig.\ref{SOV_spin_1}; the horizontal edges of triangles get canceled. \textbf{Up right:}  identification of $x_0$ points and (amputated) star-triangle integration of the leftmost blob. \textbf{Down left:} twice the application of the interchange relation I of Fig.\ref{interchanges} moves the vertical black line from $x_{11'}$ to $x_{30}$, with the effect of adding horizontal edges to form two rows of triangles. \textbf{Down right:} the upper row of triangles in down-right of Fig.\ref{SOV_spin_1} is glued back, obtaining the diagram of $\mathbf{\Lambda}_k(Y)\mathbf{Q'}_{a,k-1}(u)$, and completing the proof of \eqref{SOV_spinning}.}
\label{SOV_spin_11}
\end{figure}

The definition of layer operators is generalized to the case of spinning particles by the substitution of the operator $\mathbb{R}_{ij}(u)$ with a spinning version. We adopt the notation of explicit $SU(2)$ indices for the chain's spinning particles, while we leave them implicit for the excitation carried by the layers
\begin{align}
\label{Lambda_spin_inhom}
&\mathbf\Lambda_k(Y|\eta,\bar\eta)_{\mathbf{\underline{a},\underline{\dot {a}}}}^{\mathbf{\underline{b},\underline{\dot{b}}}} =\mathbf \Lambda_k(Y)_{\mathbf{a_1\dots  a_{k},\dot{a}_1\dots  \dot{a}_k}}^{\mathbf{b_1\dots b_{k},\dot{b}_1\dots \dot{b}_{k}}} \,\eta_{\mathbf{ b_k} }\bar{\eta}_{\mathbf{\dot b_1}}\,,
\\\notag \\
&\notag \mathbf\Lambda_k(Y)_{\mathbf{\underline{a},\underline{\dot {a}}}}^{\mathbf{\underline{b},\underline{\dot{b}}}} =\mathbf \Lambda_k(Y)_{\mathbf{a_1\dots  a_{k},\dot{a}_1\dots  \dot{a}_k}}^{\mathbf{b_1\dots b_{k},\dot{b}_1\dots \dot{b}_{k}}} =\\&= \delta_{\mathbf{\dot a_1}} ^{\mathbf{\dot b_1}}  \left[\mathbb{R}^{(n)}_{12}\left( \frac{\Delta_1}{2}-i\nu\right)\right]_{\mathbf{a_1 \dot{a}_2}}^{\mathbf{b_1 \dot{b}_1}} \cdots \left[\mathbb{R}^{(n)}_{k-1k}\left(\frac{\Delta_{k-1}}{2} -i\nu\right)\right]_{\mathbf{a_{k-1} \dot{a}_k}}^{\mathbf{b_{k-1} \dot{b}_{k}}}\cdot \frac{[\mathbf{x_k-x_0}]^{n} [\mathbf{\overline{x_k-x_0}}]_{\mathbf{a_{k}}}^{\mathbf{ b_k}}  }{(x_k-x_0)^{2\left(2-{\Delta_k}+i\nu \right)}} \,,
\end{align}
where the operators $\mathbb{R}^{(n)}_{12}(u)$ are defined by their integral kernel
\begin{align}
\begin{aligned}
\label{R_bb_op}
&\mathbb{R}^{(n)}_{ij}(u)_{\mathbf{a \dot a}}^{\mathbf{b \dot b}} \Phi(x_i,x_j)_{\mathbf{b\dot{b}}} = \int d^4 y \,{R}^{(n)}_u (x_i,x_j|y)_{\mathbf{a \dot a}}^{\mathbf{b \dot b}}   \,\Phi(y,x_j)_{\mathbf{b \dot b}} \, \\
&{R}^{(n)}_u (x_i,x_j|y) =\\&= \frac{\left[\mathbf{(x_1-y)(\overline{y-x_2})}\right]^n\left[\,\mathbf{\overline{x_{1}-x_{2}}}\,\right]^{\ell_i} \mathbf{R}_{\dot{\ell}_j,\ell_i}(\Delta_i-2) \mathbf{R}_{n,\ell_i}\left(-u-\frac{\Delta_i}{2}\right)  [\mathbf{x_2-y}]^{\ell_i}[\mathbf{({x_{2}-x_{1}})(\overline{x_1-y})}]^{\dot \ell_j}}{(x_i-x_j)^{2(2-\Delta_i)}(x_i-y)^{2\left(-u+\frac{\Delta_i}{2} \right)}(y-x_j)^{2\left(u +\frac{\Delta_i}{2}\right)}} \,,
\end{aligned}
\end{align}
represented in Fig.\ref{R_layer_sp} together with their product \eqref{Lambda_spin_inhom}. 
\noindent
The remarkable properties \eqref{symm_scalar} and \eqref{ortho_scalar} of operators $\mathbf \Lambda_k(Y)$ hold in the spinning case with some modification, but the proof of \cite{Derkachov:2020zvv} - despite its logic and steps are still valid - deserves to be generalized explicitly as the structure of layers \eqref{Lambda_spin_inhom} involves also $\mathbf{R}$-matrices between physical spaces and excitations. We dedicate to such properties the sections \ref{exch_spin_inhom} and \ref{orto_spin_sect}. In the next paragraphs we rather focus on the proof of the SoV equation \eqref{SoV} and the computation of the spectrum. The operators \eqref{Lambda_spin_inhom} solve the separation of variables condition, that reads
\begin{align}
\begin{aligned}
\label{SOV_spinning}
\mathbf Q_{a,k}(u)_{\mathbf{\underline{a},\underline{\dot {a}}}}^{\mathbf{\underline{c},\underline{\dot{c}}}}\,\mathbf \Lambda_k(Y|\eta,\bar \eta)_{\mathbf{\underline{c},\underline{\dot {c}}}}^{\mathbf{\underline{b},\underline{\dot{b}}}}\,= \,q_{a,k}(u,Y)\,\mathbf \Lambda_{k}(Y|\eta,\bar \eta)_{\mathbf{\underline{a},\underline{\dot {a}}}}^{\mathbf{\underline{c},\underline{\dot{c}}}}\, \mathbf Q_{a,k-1}(u)_{\mathbf{\underline{c},\underline{\dot {c}}}}^{\mathbf{\underline{b},\underline{\dot{b}}}}\,,
\end{aligned}
\end{align}
where $\underline{\mathbf{a}}= (\mathbf{a_1,\dots, a_k})$, $\underline{\mathbf{b}}= (\mathbf{b_1,\dots, b_{k-1}})$, and repeated spinor indices are contracted. In a more explicit fashion we can write \begin{align}
\begin{aligned}
\mathbf Q_{a,k}(u)_{\mathbf{a_1\dots a_k,\dot a_1 \dots \dot a_k}}^{\mathbf{c_1\dots c_k,\dot c_1 \dots \dot c_k}}& \mathbf\Lambda_k(Y|\eta,\bar \eta)_{\mathbf{c_1\dots \dots c_k,\dot c_1 \dots\dots  \dot c_k}}^{\mathbf{b_1\dots b_{k-1},\dot b_1 \dots \dot b_{k-1}}}=\\&= q_{a,k}(u,Y)\,\mathbf\Lambda_{k}(Y|\eta,\bar \eta)_{\mathbf{a_1\dots \dots a_k,\dot a_1 \dots\dots  \dot a_k}}^{\mathbf{c_1\dots c_{k-1},\dot c_1 \dots \dot c_{k-1}}}\,\mathbf Q_{a,k-1}(u)_{\mathbf{c_1\dots c_{k-1},\dot c_1 \dots \dot c_{k-1}}}^{\mathbf{b_1\dots b_{k-1},\dot b_1 \dots \dot b_{k-1}}}
\,.
\end{aligned}
\end{align}
The proof of \eqref{SOV_spinning} follows from a few application of the star-triangle duality \eqref{STR_ker} and is split into two parts according to the factorization of the transfer matrix $\mathbf{Q}_{a}(u)= \mathbf Q_{a-}(u)\mathbf Q_{a+}(u)$ of Fig.\ref{Q+Q-}. The action of $\mathbf Q_{a+}(u)$ on a layer operator is showed in Fig.\ref{SOV_spin_1}, while the action of $\mathbf Q_{a-}(u)$ on what results is illustrated in Fig.\ref{SOV_spin_11}. The function $q_{a,k}(u,Y)$ is defined as
\begin{align}
\begin{aligned}
\label{eigenvalue_spin}
q_{a,k}(u,Y)&= \frac{\Gamma\left(-u-\frac{\Delta_a}{2} +\frac{\ell_k}{2} \right) \Gamma\left(4-u-\frac{\Delta_a}{2}+\frac{\ell_k}{2}  \right)}{\Gamma\left(u+2+\frac{\Delta_a}{2} +\frac{\ell_k}{2} \right) \Gamma\left(-u-2+\frac{\Delta_a}{2} +\frac{\ell_k}{2}\right)}\times \\&\times \pi^4   \frac{\Gamma\left(-2-u+\frac{\Delta_a}{2} -i\nu+\frac{n}{2} \right) \Gamma\left(u+\frac{\Delta_a}{2}+i\nu +\frac{n}{2}\right)}{\Gamma\left(u+4-\frac{\Delta_a}{2} +i\nu+\frac{n}{2} \right) \Gamma\left(2-u-\frac{\Delta_a}{2} -i\nu+\frac{n}{2} \right)} \,.
\end{aligned}
\end{align}
The spectrum of the transfer matrix $\mathbf{Q}_{a,k}(u)$ is symmetric with respect to any permutation of quantum numbers $(Y_1,\dots,Y_L)$. The dependence of $\eqref{eigenvalue_spin}$ on the site $k$ of the chain is factorized respect to the dependence over the $Y_j$ of the layer, and can be stripped out and included into the  transfer matrix normalization.
\subsection{Exchange symmetry}
\label{exch_spin_inhom}
\begin{figure}[H]
\begin{center}
\includegraphics[scale=0.6]{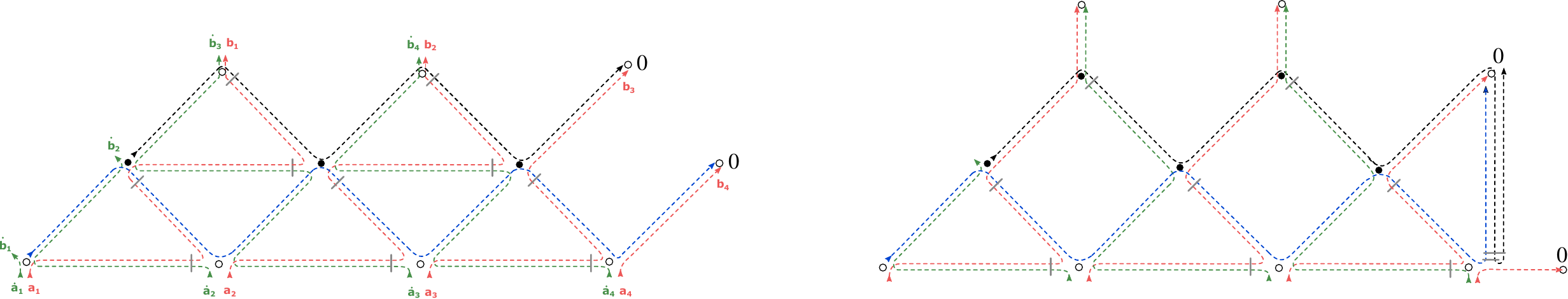}\\\vspace{5mm}
\includegraphics[scale=0.65]{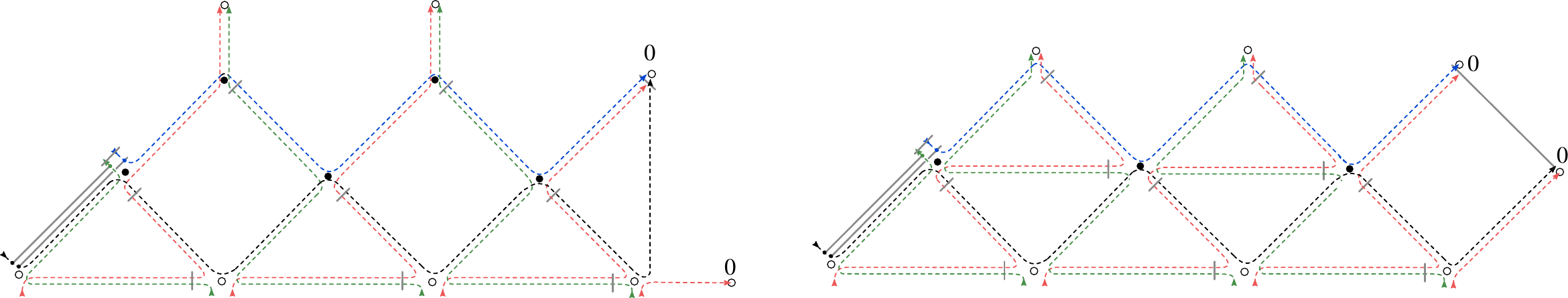}
\end{center}
\caption{Proof of the exchange relation \eqref{exchange_gen}. \textbf{Up left:} Diagram of the l.h.s. of \eqref{exchange_gen}.  \textbf{Up right:} triangles in the layer $\mathbf{\Lambda}_{k-1}$ are rewritten as star integrals and two vertical lines $[\mathbf{x}\overline{\mathbf{x}}]^{n'}\mathbf{R}(Y,Y')\mathbf{R}(Y',Y) = \mathbbm{1}$ are added on the left of the diagram.
\textbf{Down left:} thrice the application of the interchange relation (I) of Fig.\ref{interchanges} moves the vertical lines from left to right - and the last interchange make the lines disappear due to amputation of the left-most vertex. \textbf{Down right:} the upper layer star integrals are replaced by triangles, obtaining the diagram of the r.h.s. of equation \eqref{exchange_gen}.}
\label{exch_proof}
\end{figure}

The layer operators \eqref{Lambda_spin_inhom} satisfy a simple algebra respect to the exchange of two quantum numbers $Y=(n,\nu)$ and $Y'=(n',\nu')$ between two consecutive  operators $\mathbf \Lambda_{k+1}(Y)\mathbf \Lambda_{k}(Y')$. If we use the shorthand notation for $\mathbf{R}$-matrices between excitations and physical particles
\begin{equation}
\mathbf{R}(0,Y') \equiv \mathbf{R}_{n',\dot \ell_1}(-i\nu')\,,\,\mathbf{R}(Y,-2i) \equiv \mathbf{R}_{n,\dot \ell_1}(i \nu -2)\,,
\end{equation}
 the compact form of the exchange of two adjacent excitations reads
 \begin{align}
\begin{aligned}
\label{exchange_gen}
&\mathbf\Lambda_{k}(Y) \cdot \mathbf\Lambda_{k-1}(Y') \,=\, \frac{r_k(Y')}{r_{k}(Y)}\times \,\hat{\mathbf{R}}_1(Y',Y)\mathbf\Lambda_k(Y')\cdot \mathbf\Lambda_{k-1}(Y) \mathbf{R}(Y,Y') \,,
\end{aligned}
\end{align}
where the coefficients $r_k(Y) = r_k(n,\nu)$ extending the definition \eqref{r_coeff} by the introduction of physical space spins $\ell_1,\dot{\ell}_k$
\begin{equation}
r_k(n,\nu) = \frac{\Gamma\left(i\nu-\Delta_k+\frac{n-\dot \ell_k}{2} \right)}{\Gamma\left(i\nu -\Delta_k-1+\frac{n-\dot \ell_k}{2} \right) \left(i\nu-\Delta_k-1+\frac{n+\dot \ell_k}{2} \right)} \frac{\Gamma \left(i\nu +\frac{n- \ell_1}{2} \right)}{\Gamma\left(2-i\nu+\frac{n- \ell_1}{2} \right)}\,.
\end{equation}
The matrix $\hat{\mathbf{R}}_1$ is a product of three $\mathbf{R}$-matrices accounting for the exchange of excitations across the left component $(\ell_1,0)$ of the particle in site $1$
\begin{equation}
\label{hatRR_def}
\hat{\mathbf{R}}_1(Y,Y')\equiv \frac{\left(1-i\nu' +\frac{n'- \ell_1}{2} \right)}{\left(1-i\nu'+\frac{n'+ \ell_1}{2} \right)} \frac{ \left(i\nu-1 +\frac{n- \ell_1}{2} \right)}{ \left(i\nu-1 +\frac{n+ \ell_1}{2} \right)} \left[ \mathbf{R}(0,Y')^{t_1} \mathbf{R}(Y',Y) \mathbf{R}(Y,-2i)^{t_1} \right]^{t_1}\,,
\end{equation}
and $t_1$ denotes the transposition in the space of dotted spinors $\dot \ell_1$.
The equation \eqref{hatRR_def} can be eventually re-written with the order of $\mathbf{R}$-matrices reshuffled via the Yang-Baxter equation and crossing symmetry:
 \begin{equation}
 \label{YBE_trans}
\mathbf{R}(0,Y)^{t_1} \mathbf{R}(Y',Y) \mathbf{R}(Y',-2i)^{t_1}
= \mathbf{R}(Y',-2i)^{t_1}\mathbf{R}(Y',Y) \mathbf{R}(0,Y)^{t_1}\,.
 \end{equation}
Being based on the star-triangle identity \eqref{STR_ker} and the interchange relation II of Fig.\ref{interchanges}, the proof of \eqref{exchange_gen} can be delivered - avoiding tedious computations - in the diagrammatic form of Fig.\ref{exch_proof}.
The definition of an eigenfunction with a convenient normalization is
\begin{align}
\begin{aligned}
\label{eigen_spin_genL}
\Psi(\mathbf{Y}|\mathbf{x},\boldsymbol{\eta},\boldsymbol{\bar \eta})& = 
 \mathbf{\Lambda}_L (Y_L,\eta_L ,\bar \eta_1)   \cdots \mathbf{\Lambda}_2 (Y_2,\eta_{2},\bar \eta_{L-1})  \mathbf{\Lambda}_1(Y_1,\eta_{1},\bar \eta_{L}) \prod_{k=1}^L r_k(Y_k)^{k-1} \,.
\end{aligned}
\end{align}
The formula \eqref{exchange_gen} spells the exchange of two \emph{adjacent} quantum numbers as a conjugation of the layers by a few $\mathbf{R}$-matrices
\begin{align}
\begin{aligned}
\label{exchange_Y_func}
\Psi(\mathbf{x}|Y_1,\dots,Y_{k+1},Y_{k},\dots,Y_L)=\hat{\mathbf{R}}_k(Y_{k},Y_{k+1}) \Psi(\mathbf{x}|Y_1,\dots,Y_{k},Y_{k+1},\dots,Y_L)\,\mathbf{R}(Y_{k+1},Y_{k})\,.
\end{aligned}
\end{align}
Since any permutation can be decomposed in elementary exchanges between adjacent elements $Y_k,Y_{k+1}$, it follows from that the only effect of a permutation of the set $(Y_1,\dots,Y_L)$ in the eigenfunctions is a rotation in the $SU(2)$ indices of the symmetric spinors carried by the layers' excitations. The action of a transfer matrix $\mathbf{Q}_{a,L}(u)$ with spinless auxiliary space is insensitive to the spinorial indices of layers, therefore the formula \eqref{exchange_Y_func} agrees with the invariance of the spectrum respect to permutations of quantum numbers. 
 
\begin{figure}[H]
\begin{center}
\includegraphics[scale=0.58]{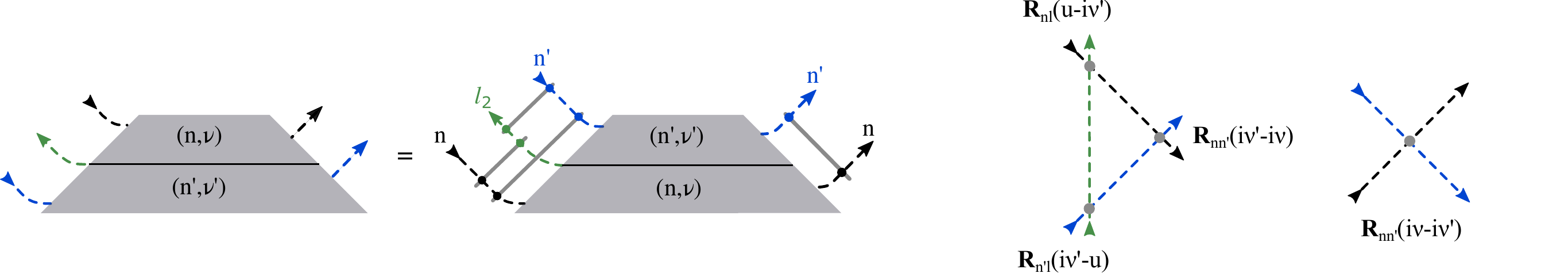}
\end{center}
\caption{\text{Left:} Scheme of the exchange relation \eqref{exchange_gen}: interchanging the two layers of quantum numbers $(n,\nu)$ and $(n',\nu')$ determines a mixing of spinor indices via the $\mathbf{R}$-matrices as showed on the r.h.s. of the equal. \text{Right:} Scattering picture of the spinor mixing, happening on the left and on the right of the two layers respectively: the two mirror magnons associated with the two layer scatter between each-other and also with a physical spinning particle. The scattering is integrable as a consequence of the Yang-Baxter property \eqref{YBE}.}
\label{schem_exch1}
\end{figure}

In addition to the mixing of indices between the $n$- and $n'$-fold symmetric spinors of the two layers, the permutation of quantum numbers in the present case involves a mixing of spinors indices between layers and physical spaces of the chain, and we need to check that the order of the exchanges leading to the same permutation is irrelevant. This fact is a consequence of the YBE \eqref{Rmat_YBE} and unitarity \eqref{R_unit} and of the analogues properties for the matrix $\hat{\mathbf {R}}_k$ (proof in appendix \ref{app:Rhat})
\begin{align}
\begin{aligned}
\label{hatR_properties}
&\hat{\mathbf{R}}_k(Y,Y')\hat{\mathbf{R}}_k(Y',Y)= \mathbbm{1}\,,\\
&\hat{\mathbf{R}}_k (Y,Y')\hat{\mathbf{R}}_{k+1}(Y,Y'')\hat{\mathbf{R}}_k (Y',Y'')=\hat{\mathbf{R}}_{k+1} (Y',Y'')\hat{\mathbf{R}}_{k}(Y,Y'')\hat{\mathbf{R}}_{k+1} (Y,Y')\,.
\end{aligned}
\end{align}
Let us analyze \eqref{exchange_gen}: as for a model of zero-spins particles, the exchange of layers spinors gives the corresponding matrix $\mathbf{R}(Y,Y')$ which accounts for the exchange of the two layer excitations. Such an exchange requires the layer spinors to \emph{scatter} through the right component of the physical particle in site $x_1$, adding to the picture the matrices $\mathbf{R}(Y',-2i)^{t_1}$ and $\mathbf{R}(0,Y)^{t_1}$, schematically showed in Fig.\ref{schem_exch1}. The transposition in the physical space is due to the fact that such scattering is of type particle-antiparticle, i.e. opposite orientation of the physical (blue, black)  and layer (green) arrows in Fig.\ref{schem_exch1} (left). The ratio $r_k(Y)/r_{k}(Y')$ correctly reproduces the normalization of the eigenfunction after the exchange $Y\leftrightarrow Y'$.
The formula \eqref{exchange_Y_func} establish the general rule for any permutation of the quantum numbers $(Y_1,\dots,Y_L)$ in the eigenfunctions of the model. The Yang-Baxter property \eqref{Rmat_YBE} and the parity \eqref{R_unit} of the matrix $\mathbf{R}(Y,Y') = \mathbf{R}_{nn'}(i\nu-i\nu')$ guarantee that it is well defines a representation of the symmetric group of the $L$ quantum numbers $Y_j$. Such a simple behaviour is identical for any choice of the physical space $\mathcal{V}$ of the spin chain, and reproduces \eqref{symm_scalar} for the spin-less particles.
\subsection{Scalar product}
\label{orto_spin_sect}
In this section we work out the overlap between eigenfunctions \eqref{eigen_spin_genL}. First, we recall that the eigenfunctions transform under a unitary irreps of the conformal group which can belong to the principal series or to the complementary series - the second describing real scaling dimensions $\Delta \in [0,4]$, relevant for applications to Feynman integrals. Here we'll limit ourselves to the principal series $\Delta=2+i\nu$ and the results for the complementary series can be recovered a posteriori by an analytic continuation of $\nu$ to the imaginary strip $[-2i,2i]$.
\subsubsection{$L>1$}
For the model of $L>1$ particles the computation of the overlap of eigenfunctions \eqref{eigen_spin_genL} is based on the algebra satisfied by a layer operator $\mathbf \Lambda_k(Y)$ with the conjugate $\mathbf{\bar \Lambda}_k(Y')$. The conjugate operator is defined respect to the scalar product \eqref{scalar_product} and with an additional shift in the quantum number $\nu\to \nu+i 2$, as established for $L=1$ eigenfunctions. The diagram of the kernel defining the conjugate layer operator is given in the first row of Fig.\ref{ortho_exch} together with the convolution $\mathbf{\bar \Lambda}_{k+1}(Y')\mathbf{\Lambda}_{k+1}(Y)$.
The remarkable property \eqref{ortho_scalar}, valid for particles of spins zero, continues to hold for the chain of spinning particles with an additional mixing of spinor indices between physical space and layer excitation
\begin{align}
\label{exch_orto}
\begin{aligned}
\mathbf{\bar \Lambda}_{k+1}(Y') \cdot \mathbf \Lambda_{k+1}(Y)\, &= \,\frac{\pi^4}{ \mu(Y,Y')}{r_k(Y)\bar r_k(Y')}\times
\\& \times\langle \eta' _k|\eta_k\rangle^{\ell_k}  \langle\bar  \eta' _1|\bar \eta_1\rangle^{\dot \ell_1}\, \left[ \hat{\mathbf{R}}(Y',Y)^{t'} \mathbf  \Lambda_{k}(Y) \cdot \mathbf{\bar \Lambda}_{k}(Y')^{t'}  \mathbf{R}(Y,Y')^{t'}\right]^{t'}\,,
\end{aligned}
\end{align}
where $t'$ is the transposition in the indices of excitation $Y'$ and $\bar r_k(Y)$ is defined as complex conjugate with shift $\nu \to \nu +2i$ according to the prescription for $\bar{\mathbf \Lambda}_k(Y)$.
\begin{figure}[H]
\begin{center}
\includegraphics[scale=0.63]{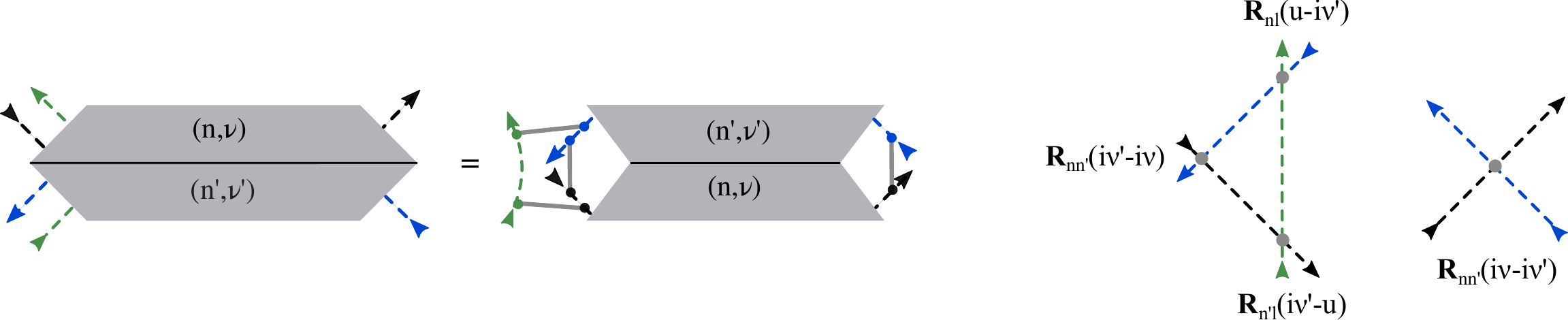}
\end{center}
\caption{\textbf{Left:} General scheme of the overlap $\bar{\mathbf{\Lambda}}_{k+1}(Y') \mathbf{\Lambda}_{k+1}(Y)$. The two layers get exchanged, i.e. and two excitations $Y,Y'$ cross each other exchanging their order, while the layers reduce their length by one. The re-ordering of $Y,Y'$ requires the excitations to cross with the right (green) component of site-$2$. \textbf{Right:} The exchange of order of two excitations corresponds to the appearence of an $\mathbf{R}$-matrix.}
\label{ortho_exch}
\end{figure}
The proof of \eqref{exch_orto} is delivered with diagrams in Fig.\ref{ortho_exch}, using the star-triangle duality and the consequent interchange relation II of Fig. \ref{interchanges}.
\begin{figure}[H]
\begin{center}
\includegraphics[scale=0.75]{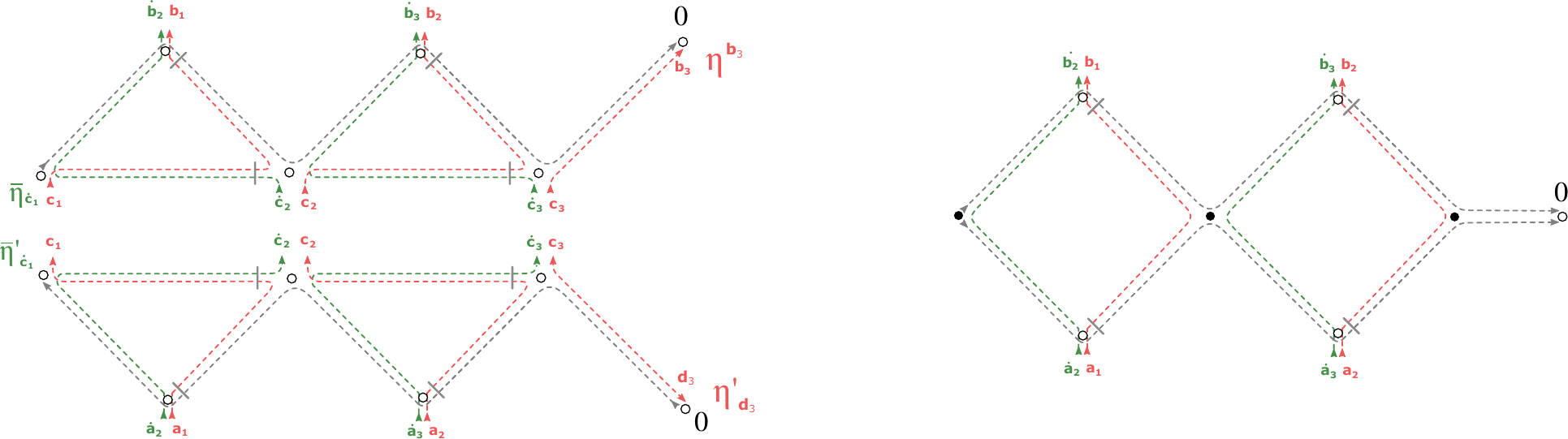}\\
\includegraphics[scale=0.75]{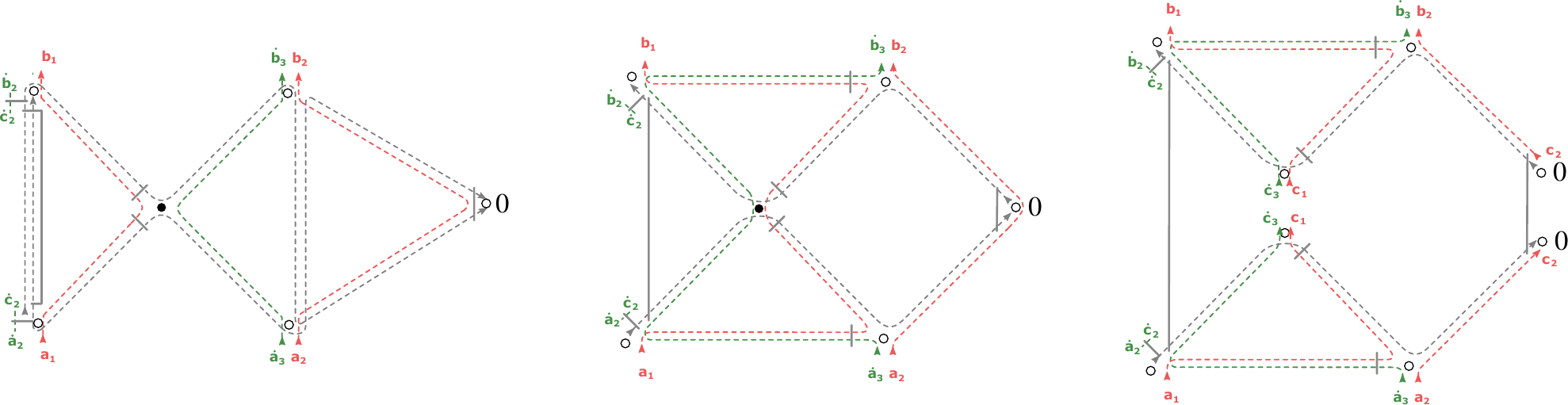}
\end{center}
\caption{\textbf{Up left:} Diagrams of layer operators $\mathbf{\bar \Lambda}_3(Y')$ and $\mathbf \Lambda_3(Y)$. \textbf{Up right:} in the convolution $\mathbf{\bar \Lambda}_3(Y') \mathbf \Lambda_3(Y)$ the basis of triangles cancel each other. The first dotted and last undotted physical spinors are contracted and factored out as $\langle \bar\eta_1| \bar\eta'_1\rangle^{\dot{\ell}_1}$ and $\langle \eta'_3|\eta_3\rangle^{\ell_3}$. \textbf{Down left:} Result of the integration by (amputated) star-triangle identity in the first and last black dots, respectively. \textbf{Down center:} the application of the interchange relation II of Fig.\ref{interchanges} delivers as a result the convolution $\mathbf \Lambda_2(Y)\mathbf {\bar {\Lambda}}_2(Y')$ together with a mixing of the layers spinor indices by the $\mathbf{R}$-matrices on the left and on the right of the diagram. \textbf{Down right:} Explicit factorization of the results in $\mathbf{\Lambda}_2(Y)$ and $\mathbf{\bar{\Lambda}}_2(Y')$ plus spinor mixing.}
\label{ortho_exch}
\end{figure}
The formula \eqref{exch_orto} strongly resembles \eqref{symm_scalar} for models of spinless particles: the physical space space spinors $\eta,\bar \eta$ get factored out of the convolution, and the only generalization sits in the depedandence of $r_k$ factors on the spins of the $k$-th of the spin chain. By means of \eqref{exch_orto} the overlap of length-$L$ eigenfunctions is reduced to the overlap of functions of length $L-1$, and this iteration eventually maps the computation to the product of $L$ overlaps of length-one functions \eqref{clean_1_p}. For example, for a length-two eigenfunction
\begin{align}
\begin{aligned}
&\langle \Psi_2(Y_1',Y_2') ,\Psi_2(Y_1,Y_2)\rangle =\frac{2\pi^{10}}{\mu(Y_1,Y_2) (n_1+1)(n_2+1)} \langle \eta'_1|\eta_1\rangle^{\ell_1}   \langle \bar \eta_2|\bar \eta'_2\rangle^{\ell_2}   \langle \bar \eta_1|\bar \eta'_1\rangle^{\dot \ell_1}    \langle \eta'_2|\eta_2\rangle^{\dot \ell_2}  \times  \\&\times \left[  \delta(Y_1-Y_1') \delta(Y_2-Y_2') \mathbb{P}_{11'} \mathbb{P}_{22'}+\delta(Y_1-Y_2') \delta(Y_2-Y_1') \mathbf{\hat R}(Y_2,Y_1)\otimes \dot{\mathbf{R}}(Y_1,Y_2)   \mathbb{P}_{1'2} \mathbb{P}_{12'} \right]\,,
\end{aligned}
\end{align}
where we recall that $\hat{\mathbf{R}}(Y,Y')$ is defined in \eqref{hatR_def}, with the obvious reduction $\hat{\mathbf{R}}(Y,Y') = {\mathbf{R}}(Y,Y')$ for the spinless site $\ell_1=0$.
Similarly, for higher length $L$ the overlap is a sum of $\delta$-functions over the permutations of the quantum numbers $(Y_1,\dots,Y_L)$, each associated with the corresponding products of matrices $\mathbf{R}$ or $\hat{\mathbf{R}}$. The scalar product reads
\begin{equation}
\langle \Psi_L(\mathbf{Y}') ,\Psi_L(\mathbf{Y})\rangle  = \frac{1}{\rho(\mathbf{Y})L!}\sum_{\pi}  \hat{\mathbf{R} }_{\pi}( \mathbf{Y}) \otimes \dot{\mathbf{R}}_{\pi} (\mathbf{Y}) \mathbb{P}_{\pi(i)}  \prod_{j=1}^L \delta (Y_j-Y_{\pi(j)}) \langle \bar \eta_j|\bar \eta'_j\rangle^{\ell_j}   \langle \bar \eta_j|\bar \eta'_j\rangle^{\dot \ell_j}  \,,
\end{equation}
where
\begin{equation}
\mathbb{P}_{\pi} = \mathbb{P}_{1\pi(1)'}\mathbb{P}_{2\pi(2)'}\cdots \mathbb{P}_{L\pi(L)'}\,.
\end{equation}\section{Scattering of mirror excitations}
\label{sect:scatter}
The eigenfunctions of the transfer matrices $\mathbf{Q}_{a,L}(u)$ presented in the last section reveal the features of wave-functions of a multi-particle state in a $(1+1)d$ integrable model. The quantum numbers $Y_j$ define the excited modes of the \emph{mirror channel} of the fishnet - i.e. of the conformal spin chain with fixed boundaries. In this respect, the properties \eqref{symm_scalar}-\eqref{exch_orto} express the exchange of  excitations by an appropriate ``scattering" matrix $\mathbf{R}(Y,Y')$, that satsifies YBE, unitarity and a crossing equation \eqref{R_cross}, and they are reminiscent of a Zamolodchikov-Faddeev algebra \cite{Faddeev_algebra,ZAMOLODCHIKOV1979253,Smirnov_book} with creation/annihilation operators $\bar{\mathbf{\Lambda}}_k(Y)$, ${\mathbf{\Lambda}}_k(Y)$. In this section we propose a scattering interpretation for the general spinning fishnet lattice, its eigenfunctions and - ultimately - for the star-triangle identity \eqref{STR_ker}.
 \subsection{Star-Triangle as a scattering process}
We claim that general $4d$ star-triangle relation \eqref{STR_ker} describes the $(1+1)d$ scattering of particles propagating in the radial direction of $\mathbb{R}^4$, and with an internal symmetry $SO(4)\sim SU(2) \times SU(2)$. Accordingly, the particles wave-functions live on the manifold $\mathbb{R} \times S^3$ and, and are defined by a momentum $p(\nu)=2\nu$ and by a degree-$n$ irreducible (i.e. symmetric and traceless) tensor of $SO(4)$
\begin{equation} 
f_{\nu,n}(y,\hat x)= e^{i\nu y} C_n(\hat x)\,,\,\,\,\, C_n(\hat x)=\hat x_{\mu_1}\cdots \hat x_{\mu_n} t^{(\mu_1\dots \mu_{n})} \,,\,\,\, y\in \mathbb{R}\,,\,\hat x \in S^3\,.\end{equation}
The quantum number $\nu$ is the \emph{rapidity} of the excitation; given its simple relation with the momentum we will use the latter word to refer to the rapidity in the rest of the section\footnote{In contrast to that, for the well-known models of $2d$ massive scattering the rapidity is related to the momentum and energy of the particle via hyperbolic functions $p(\nu)=m\sinh(\nu)\,,\, E(\nu)=m\cosh(\nu)$, e.g. \cite{Smirnov_book}.}.
We can express the symmetric tensor with $SU(2)$ spinors $|\beta \rangle$ in the representation $\mathbf{2}$ and $\langle \alpha|$ in the representation $\bar{\mathbf{2}}$, thanks to the Clifford algebra of matrices $\boldsymbol{\sigma}^{\mu},\, \boldsymbol{\bar \sigma}^{\nu}$ (see \eqref{cliff})
\begin{equation}
C^{(\mu_1\dots\mu_L)}_n(\hat x)=\langle\alpha|\boldsymbol{\sigma}^{\mu_1}|\beta\rangle \cdots \langle\alpha|\boldsymbol{\sigma}^{\mu_L}|\beta\rangle \Longrightarrow C_n(\hat x) = \langle \alpha| \mathbf{x}|\beta \rangle^n\,.
\end{equation}
The wave-function can be rewritten also without auxiliary spinors $\alpha, \beta$ in favour of the notation of symmetric indices \eqref{spin_notation}
\begin{equation}
\label{wave_f}
f_{\nu,n}(y,\hat x) =  {[\mathbf{x}]^{n}} e^{-2i\nu y}\,,
\end{equation}
and they can be regarded as propagating along the radial direction of $\mathbb{R}^4$, that amounts to the transformation
\begin{equation}
\label{eigenf_radial}
 (y,\hat x)\to (r=e^{y},\hat x)  \in \mathbb{R}^4 \,\Longrightarrow \, f_{\nu,n}(y,\hat x) \to \frac{[\mathbf{x}]^{n}}{r^{2i\nu}}\,,
\end{equation}
and the conjugate wave-function in the canonical coordinates of $\mathbb{R}^4$ is given, as the $4d$ measure is $d^4x = dr r^3 d\Omega$ and the change of variable $dy = dr/r$, by
\begin{equation}
\label{eigenf_radial_conj}
{f_{\nu,n}^*(y,\hat x)} \to \frac{[\mathbf{\bar x}]^{n}}{r^{2(2-i\nu)}}\,.
\end{equation}
We can already make two observations. 
First, the functions written as \eqref{eigenf_radial} are of the same type as \eqref{1_p_sol}, that is for $L=1$. Secondly, the rule for complex conjugation \eqref{eigenf_radial_conj} - i.e. the conjugation as radial plane waves - matches with the the prescription found in sect.\ref{sec:1p} for eigenfunctions' orthogonality.

The wave-functions \eqref{wave_f} carry $n$-fold symmetric left/right spinor indices in the numerator $[\mathbf{x}]_{\mathbf{a}}^{\mathbf{\dot a}}$, hence they can be paired with auxiliary spinor variables in $\langle \alpha| \otimes |\beta \rangle\, \in\, \text{Sym}_n^* \otimes \text{Sym}_n$; under complex conjugation the spinors get exchanged and the momentum $\nu$ flips the sign
\begin{equation}
\label{aux_scatt}
\frac{\langle \alpha| \mathbf{x}|\beta \rangle^{n}}{r^{2(i\nu)}} \,=\, \left(\frac{\langle \beta| \mathbf{\bar x}|\alpha \rangle^{n}}{r^{2(-i\nu)}}\right)^*\,.
\end{equation}
Therefore, each wave function can be regarded as having a left component of spin $(n,0)$ with momentum $\nu$, and a dual right component $(0,n)$ with momentum $-\nu$. The overlap between two wave functions with momentum $\nu$ and $\nu'$ located at different points on $\mathbb{R}\times S^3$, say $x$ and $x-z$ is nothing but an amputated form of the star-triangle identity \eqref{STR_ker} for $\ell=0$ (see Fig. \ref{STR_amp_app} of appendix \ref{app:str_amp}) 
\begin{align}
\begin{aligned}
\langle f_{m,u} , e^{i z \cdot \hat p}  f_{n,v}\rangle = \int d^4 y \frac{[\mathbf{\overline{y}}]^{m}}{(y)^{2(2-iu)}} \frac{[\mathbf{z-y}]^{n}}{(z-y)^{2(iv)}}= a_{n,m}(v,u)\frac{[\mathbf{z}]\mathbf{R}_{n,n'}(i(u-v)) [\mathbf{\overline{z}}]}{(z)^{2(i(v-u))}}\,.
\end{aligned}
\end{align}
\begin{figure}[H]
\begin{center}
\includegraphics[scale=1]{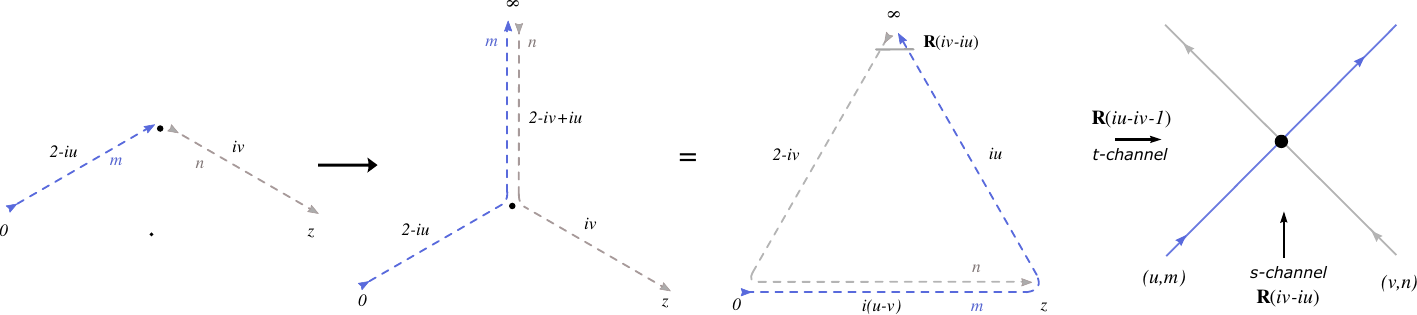}
\end{center}
\caption{\textbf{Left:} The diagram of the overlap \eqref{scatter_FF} of two wavefunctions of momenta $v,u$ and bound-state index $n+1$ and,$m+1$. The point at $\infty$ can be restored by a conformal transformation, and the diagram takes the form of a star integral. The latter is computed by star-triangle identity. \textbf{Right:} by a choice of displacement $z=(1,0,0,0)$ the triangle reduces to the $\mathbf{R}_{n,m}(iv-iu)$ (and a dynamical factor $a_{n,m}(u,v)$. We regard this matrix as the result of a scattering process, and specify the $s$- and $t$-channels.}
\end{figure}
With the pairing \eqref{aux_scatt}, we can read out clearly the spinor structure in the last formula
\begin{align}
\begin{aligned}
\label{scatter_FF}
& f_{n,v}(x) \to \langle \alpha| f_{n,v} (x)|\beta\rangle\,,\,\,\,  f_{m,u}(x) \to \langle \alpha'| f_{m,u} (x)|\beta'\rangle\,,\\&
\langle \langle \alpha'|f_{u,m}|\beta'\rangle ,e^{i z \cdot \hat p} \langle \alpha| f_{v,n}|\beta\rangle \rangle \propto \frac{\langle \alpha[\mathbf{z}] |\langle \beta'| \mathbf{R}_{m,n}(i(v-u))|\beta\rangle|[\mathbf{\overline{z}}]\alpha '\rangle}{(z)^{2(i(u-v))}}\,,
\end{aligned}
\end{align}
and reads as an integrable scattering between the two incoming left component $|\beta\rangle \otimes [\mathbf{\bar z}]|\alpha\rangle$ with momenta $\nu,\nu'$ into two outcoming left components $\langle \alpha[\mathbf{z}] \otimes \langle \beta' |$ with exchanged momenta. The $SU(2)$ matrices $[\mathbf{z}]\,,\,[\mathbf{\bar z}]$ rotate the spinors as a result of the displacement $e^{i z \cdot \hat p}$, that is also responsible for the rescaling $(z^2)^{i\nu'-i\nu}$. For the choice of a displacement only in first component $z^{\mu}=(z,0,0,0)$ the spins are not rotated, indeed
\begin{equation}
\mathbf{z} = \hat z_{\mu}\boldsymbol{\sigma}_{\mu} = \boldsymbol{\sigma}_{0}  =\mathbbm{1}\,,
\end{equation}
and furthermore setting $z=1$, the r.h.s. of \eqref{scatter_FF} is reduced to the product of an $\mathbf{R}$-matrix element between two left symmetric spinors in $(\mathbf{n+1)}\otimes (\mathbf{m+1})$ with momenta $v,u$ and two right ones in the conjugated representation, with a dynamical pre-factor 
\begin{equation}
\label{S_emergent}
a_{n,m}(iv-iu) a_{n,0}(2-iv) a_{0,m}(iu) \times {\langle \alpha |\langle \beta'| \mathbf{R}_{n,m}(iv-iu)|\beta\rangle|\alpha '\rangle}\,.
\end{equation}
The simple result \eqref{S_emergent} for the overlap of excitations at the points $(0,0,0,0)$ and $(1,0,0,0)$ allows to interpret the spin numbers $n/2$ as bound-state indices $a=n+1$ of the particles on the radial direction, because of the poles in the momenta of $a_{n,m}(iv-iu)$, in line with the language of \cite{Basso_2019}.

\begin{figure}[H]
\begin{center}
\includegraphics[scale=1.3]{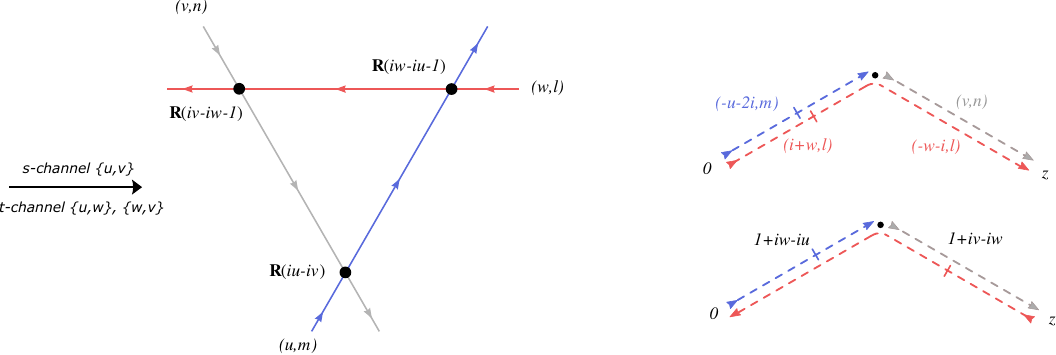}
\end{center}
\caption{\textbf{Left:} Scattering of three particles  (left-components) of momenta $\{u,v,w\}$ and spins $\{m/2,n/2,l/2\}$. The direction of arrows fixes the order of $\mathbf{R}$-matrices in the product, that is equal to \eqref{scattR_t}. \textbf{Right:} diagram of the overlap of eigenfunctions with a displacement $z$, \eqref{overlap_FF_3}. Here we distinguish $SU(2)$ matrices of type $[\mathbf{x}]$ and $[\mathbf{\bar x}]$ by the use of a bar. In the up picture barred/unbarred lines are paired, as it can be read out of \eqref{overlap_FF_3}. Downstairs we depict the result of charge-conjugation in the spinor space $\text{Sym}_{l}$, bringing the diagram in the form of a (amputated) star-triangle, where any two paired dashed lines are always of type bar/unbar (r.h.s. of \eqref{overlap_FF_3}).}
\label{injection}
\end{figure}

The logic of the last observation can be repeated also for $\ell\neq 0$ in the star-triangle \eqref{STR_ker}, that is for $3$ \emph{incoming} left spinors and $3$ \emph{outcoming} right spinors of symmetric degrees $m,l,n$. In this case we deal with a scattering of particles with momenta $u,v,w$ and bound-state indices $a=n+1,b=m+1,c=l+1$. Assuming the scattering to be factorized, we expect the $S$-matrix to be
\begin{equation}
\label{scatt3}
 \mathbf{R}_{l,m}(iw-iu) \mathbf{R}_{n,m}(iu-iv)  \mathbf{R}_{l,n}(iw-iv)\,.
\end{equation}
We represent this process in Fig.\ref{injection} and read out from the picture the momenta to inject in the star-triangle relation. 
If we look at Fig.\ref{injection} from left towards right, we see the $s$-channel for particles $u,v$, therefore a scattering matrix $\mathbf{R}_{nm}(iu-iv)$. The third particle $w$ scatters with the other two in the $t$-channel, thus the $s$-channel exchanged momenta $u-w$ and $w-v$, are analytically continued to $w-u+i$ and $v-w+i$ (see \cite{Bombardelli_2016} and references therein). The spinor indices contractions are fixed by the arrows in Fig.\ref{injection}, and the full expression of the scattering is
\begin{equation}
\label{scattR_t}
\left[\mathbf{R}_{l,m}^{t_l}(iu-iw-1) \mathbf{R}_{n,m}(iu-iv)  \mathbf{R}_{l,n}^{t_l}(iv-iw-1)\right]^{t_l}\,,
\end{equation}
and can be cast into the form \eqref{scatt3} (with additional charge conjugation and crossing factor), by the crossing equation \eqref{R_cross}.
The (amputated) star-triangle is in fact an overlap of four wave-functions, in analogy to \eqref{scatter_FF}, with displacement $z$ between the two couples of functions \begin{align}
\begin{aligned}
\label{overlap_FF_3}
&\langle f_{m,u} {f_{l,-3i-w}} , e^{i z \cdot \hat p}  f_{n,v} { f_{l,-i-w}} \rangle = \int d^4 y \frac{[\mathbf{\overline{y}}]^{m}[\mathbf{\overline{y}}]^{l}}{(y)^{2(1+i(w-u))}} \frac{[\mathbf{{z-y}}]^{l}[\mathbf{z-y}]^{n}}{(z-y)^{2(i(v-w)+1)}} = \\& =[\boldsymbol{\sigma}_2]^l\left( \int d^4 y \frac{[\mathbf{\overline{y}}]^{m}[\mathbf{{z-y}}]^{l}}{(y)^{2(1+i(w-u))}} \frac{[\mathbf{\overline{y}}]^{l}[\mathbf{z-y}]^{n}}{(z-y)^{2(i(v-w)+1)}}\right)^{t_l} [\boldsymbol{\sigma}_2]^l \,.
\end{aligned}
\end{align}
The expression in the r.h.s. between brackets is amputated star-triangle Fig.\ref{STR_amp_app} of appendix \ref{app:str_amp}, with momenta $v$ and $-i-w$ in one leg, conjugate momenta $2-u$ and $w+i$ in the other, and external points $0$ and $z$. As already observed for two excitations, a displacement $z=(1,0,0,0)$ does not lead to any dependence over $z$ in the r.h.s. of the (amputated) star-triangle, and \eqref{overlap_FF_3} is equal to 
\begin{align}
\begin{aligned}
\label{triple_FF}
\thinmuskip=-0.2mu
&a_{nm}(iu-iv)a_{nl}(iv-iw-1)a_{lm}(iw-1-iu)\,\times\\&\times [\boldsymbol{\sigma}_2]^l \mathbf{R}_{l,m}^{t_l}(iu-iw-1) \mathbf{R}_{n,m}(iu-iv)  \mathbf{R}_{l,n}^{t_l}(iv-iw-1)[\boldsymbol{\sigma}_2]^l =\\&=\frac{a_{nm}(iu-iv)a_{nl}(iv-iw-1)a_{lm}(iw-1-iu)}{r_{lm}(iu-iw)r_{ln}(iw-iv)} \times \mathbf{R}_{l,m}(iw-iu) \mathbf{R}_{n,m}(iu-iv)  \mathbf{R}_{l,n}(iw-iv) \,.
\end{aligned}
\end{align}
\begin{figure}[H]
\begin{center}
\includegraphics[scale=1.2]{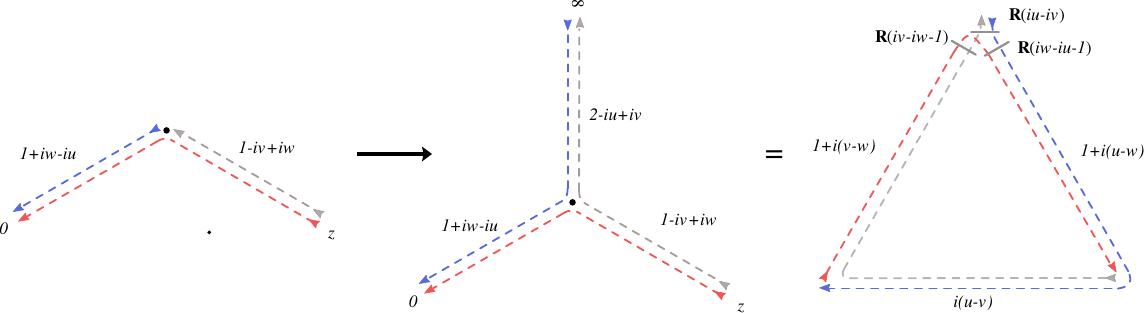}
\end{center}
\caption{Star-triangle computation of the overlap. The integral inside the brackets in the r.h.s.\eqref{triple_FF} is conformally equivalent to a star-integral, via a map that restores the point at $\infty$. A choice of vertices of the star $(0,0,0,0),(1,0,0,0),(\infty,0,0,0)$ eliminates any space-time dependence in the triangle, replacing the dashed lines $[\mathbf{x}]$ with the identity in the space of spinors.}
\label{STR_3_scattering}
\end{figure}
The result \eqref{triple_FF} is conformally equivalent to the more general overlap
\begin{align}
\begin{aligned}
\langle f_{m,u} {f_{l,-3i-w}} e^{i x_{32} \cdot \hat p}  f_{n,v}&, f_{m,u} e^{i x_{12} \cdot \hat p}  f_{n,v} { f_{l,-w-i}} \rangle =\\&= \int d^4 y \frac{[\mathbf{\overline{x_3-y}}]^{l}[\mathbf{\overline{x_3-y}}]^{m}}{(x_3-y)^{2(i(w-u)+1)}}  \frac{[\mathbf{{y-x_2}}]^{m}[\mathbf{y-x_1}]^{n}}{(y-x_2)^{2(i(u-v))}}  \frac{[\mathbf{{y-x_1}}]^{l} [\mathbf{\overline{x_2-y}}]^{n}}{(y-x_1)^{2(i(v-w)+1)}}\,,
\end{aligned}
\end{align}
which is related to the star-triangle identity of Fig. \ref{STR_3_scattering} (for more details, see the appendix \ref{app:str_amp}) by a charge conjugation in the $l$-fold symmetric spinors, just as in \eqref{overlap_FF_3}
\begin{align}
\begin{aligned}
 {[\boldsymbol{\sigma}_2]^l }\left(\int d^4 y \frac{[\mathbf{\overline{x_1-y}}]^{l}[\mathbf{\overline{x_3-y}}]^{m}}{(x_3-y)^{2(i(w-u)+1)}}  \frac{[\mathbf{{y-x_2}}]^{m}[\mathbf{y-x_1}]^{n}}{(y-x_2)^{2(i(u-v))}}  \frac{[\mathbf{{y-x_3}}]^{l} [\mathbf{\overline{x_2-y}}]^{n}}{(y-x_1)^{2(i(v-w)+1)}}\right)^{t_l} [\boldsymbol{\sigma}_2]^l \, .
\end{aligned}
\end{align}
Sending the points on the line by means of a conformal change of coordinates, $x_3\to (\infty,0,0,0)$, $x_1 \to (0,0,0,0)$ and $x_2 \to (1,0,0,0)$, all the $SU(2)$ matrices of type $[\mathbf{x_{ij}}]\,,\,[\mathbf{\overline{x_{ij}}}]$ that would rotate spinors in the r.h.s. of the star-triangle \eqref{STR_ker} are reduced to the identity, and the denominators in the r.h.s. disappear, leaving behind the result \eqref{triple_FF}.
In practise, the required conformal transform can be realized by conjugation of both sides of the star-triangle with conformal propagators, as it can be read from the comparison between Fig.\ref{STR_app} and  Fig.\ref{Finite_dim_R} in the appendix \ref{app:str_amp}.

The scattering picture for the general star-triangle duality can be applied to the interpretation of the equivalent identities obtained moving $\mathbf{R}$-matrices from the r.h.s. to the l.h.s. of the equation, or for the more involved scattering processes of the interchange relations of Fig.\ref{interchanges}. 
\subsection{Mirror excitations on the lattice}
\begin{figure}[H]
\begin{center}
\includegraphics[scale=0.9]{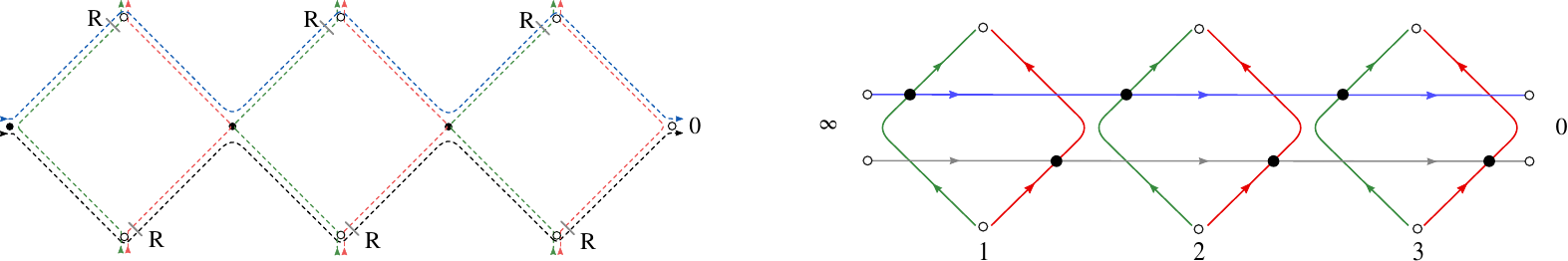}
\end{center}
\caption{Square lattice build by the transfer matrix $\mathbf{Q}_{a,3}(u)$ and its scattering interpretation. Colours of the dashes lines $[\mathbf{x}]$ and $[\mathbf{\overline{x}}]$ correspond to the colour of the lines in the scattering picture. The scattering matrix $\mathbf{R}$ is represented by black blobs. }
\label{lattice_scatter}
\end{figure}

\begin{figure}[H]
\begin{center}
\includegraphics[scale=0.8]{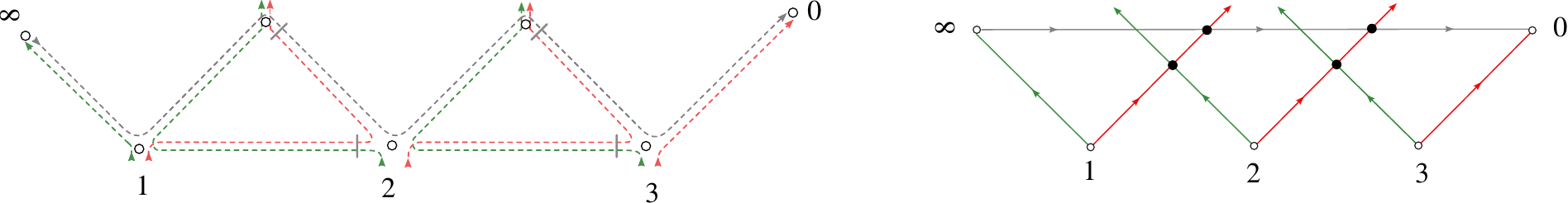}
\end{center}
\caption{Layer operator $\mathbf{ \Lambda}_3(Y)$ in the Feynman diagram notation, after an inversion that restores the point at $\infty$, and its picture as a scattering between the chain's modes and the mirror excitation. Left/right modes of the chain move towards $0$ and $\infty$ respectively. Black blobs denote scatterings, i.e. $\mathbf{R}$-matrices.}
\label{scatter_lay}
\end{figure}

\begin{figure}[H]
\begin{center}
\includegraphics[scale=0.8]{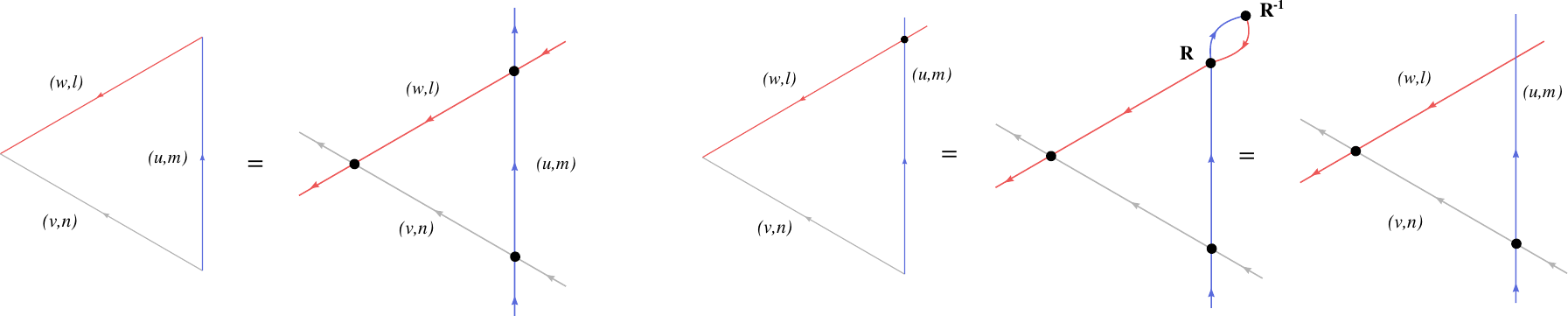}
\end{center}
\caption{\textbf{Left:} The star-triangle identity of Fig.\ref{STR_3_scattering} is represented as a \emph{move} of the lattice: the overlap of three wave-functions displaced between three points is equal to the particles crossing each-other and scattering (black blob =  (dynamical factor) $\times$ $\mathbf{R}$-matrix). \textbf{Right:} The lattice move corresponding to the star-triangle of Fig.\ref{STR_1}. In this case two overlapping wave-functions (red,blue lines) are already scattered through each-other. After crossing the trajectories (as in the left panel), the two scattering matrices between red-blue lines cancel because of opposite transferred momentum $\mathbf{R}(p)\mathbf{R}(-p)=\mathbbm{1}$.}
\label{lat_moves}
\end{figure}

Having established the identification of the \emph{triangle} of \eqref{STR_ker} with the scattering of $(1+1)d$ particles, we apply this picture to the transfer matrix $\mathbf{Q}_{a,L}(u)$ and and to the layer operators $\mathbf{\Lambda}_k(Y)$, as well as to the interpretation of their algebra \eqref{symm_scalar},\eqref{exchange_gen},\eqref{ortho_scalar},\eqref{exch_orto}.
The transfer matrix $\mathbf{Q}_{a,k}(u)$ is the graph bulding of a square lattice of spinning conformal propagators; according to \eqref{1_p_sol} the propagators are wave-functions of radial excitations and the lattice, and as they cross into vertices they give rise to a scattering process between auxiliary space excitations with physical space particles. The auxiliary space excitation $(\Delta_a,\ell_a,\dot{\ell}_a)$ is split into left/right components with momentum $(\Delta_{a}-2)/2$ and spins $\ell_a/2$ and $\dot \ell_a/2$ respectively (grey and blue in Fig.\ref{lattice_scatter}). Each physical space  $(\Delta_k,\ell_k,\dot{\ell}_k)$ contributes with left/right components as well; the momentum $2-\Delta_k$ is carried by on the left (undotted) mode, while the right (dotted) ones have momentum $0$ (red and green lines respectively in Fig.\ref{lattice_scatter})\footnote{This way the physical momenta are split is essential to integrability for an inhomogeneous model, and corresponds to the condition \eqref{inhom_constr}. In fact, for a homogeneous model the momentum can be split in any other way.}.  Auxiliary excitations traverse the lattice and scatter with physical ones, among left-left and right-right ones as in the right panel of Fig.\ref{lattice_scatter}.
\noindent
A similar picture holds for the layer operators ${\mathbf{\Lambda}}_k(Y)$: each layer is associated with an excitation $Y$ of momentum $\nu$ and bound state index $n+1$, propagating from $\infty$ to $0$ along the layer. Since the layer's kernels are composed of triangles of the type in r.h.s. of \eqref{STR_ker}, the scattering picture is straightforwardly illustrated by Fig.\ref{scatter_lay}.
According to Fig.\ref{scatter_lay}, in the general inhomogeneous and spinning model, the mirror excitation carried by a layer scatters with the left modes of physical space (red lines), that move from $x_1,\dots,x_k$ towards $x^{\mu}=0$, while it is transparent to the right modes that point towards $\infty$ (green lines).
In the simplified setup of a homogeneous chain with $\Delta_k=1$ and spinless particles $\ell_k=\dot{\ell}_k=0$, there is no scattering between mirror excitations and chain's particles. Such a choice coincide with the bi-scalar fishnet reduction of the lattice, and therefore describes the mirror excitations of the theory \eqref{biscalar_intro}. 
The $(1+1)d$ scattering picture for the graph-builder of the lattice (Fig.\ref{lattice_scatter}) and for the layer operators (Fig.\ref{scatter_lay}), can be extended to the equation \eqref{SoV}. In general, the moves of lines in the Feynman diagrams by star-triangle and interchange identities, become moves (crossing/uncrossing) of particle trajectories in the scattering picture, as explained in Fig.\ref{lat_moves}.

\begin{figure}[H]
\begin{center}
\includegraphics[scale=0.67]{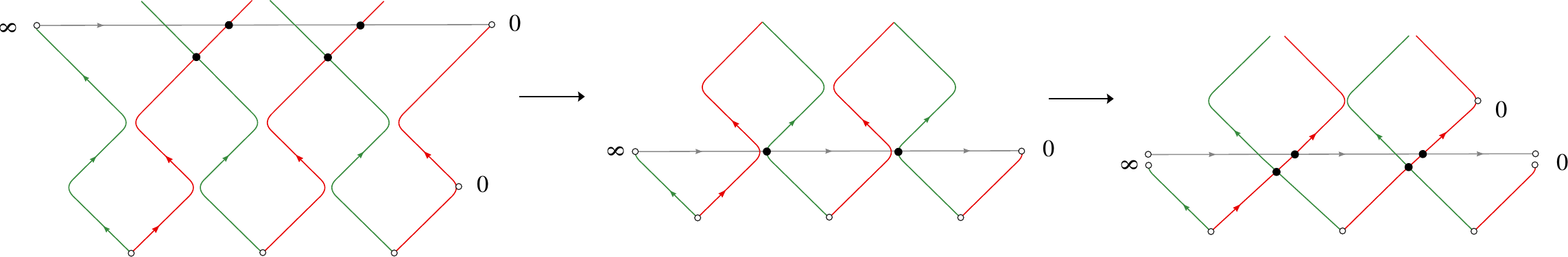}
\end{center}
\caption{The mirror excitation (grey line) is injected in the squared-lattice (green/red lines) by the layer operator of Fig.\ref{scatter_lay}. The excitation is pulled down through the layer, scattering with its modes and factoring out the first right (green) component and the last left (red) component. The lattice size is reduced by one unit.}
\label{eigen_scatterf}
\end{figure}

\begin{figure}[H]
\begin{center}
\includegraphics[scale=0.8]{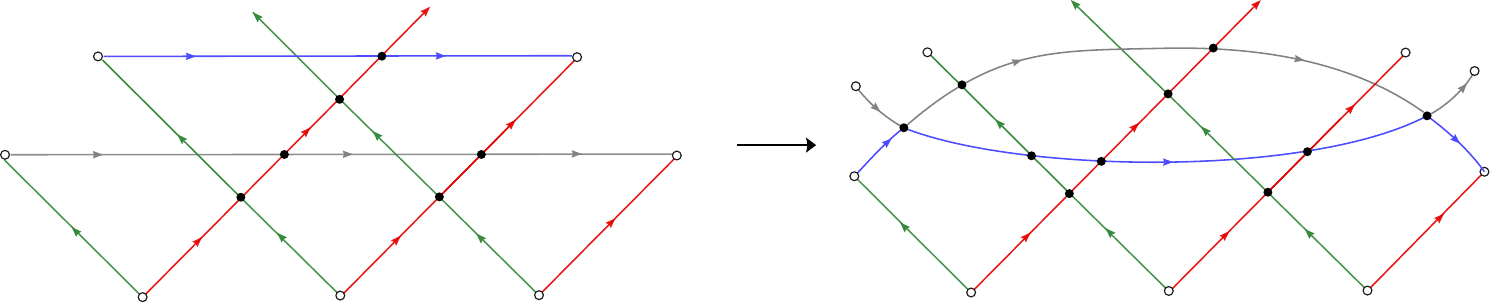}
\end{center}
\caption{\textbf{Left:} Scattering picture of the convolution of two layer operators $\mathbf{ \Lambda}_{3}(Y)\mathbf{ \Lambda}_2(Y')$. The grey and blue lines are the mirror excitations propagating across the lattice of left/right movers (red/green lines). \textbf{Right:} the two excitations are exchanged polling them across each other. This moves implies the appearance of a scattering matrix - black blob - each time two lines get to cross or its disappearance when they uncross.}
\label{exch_scatter}
\end{figure}

In fact, the latter describes the passage of a mirror excitation across a row of square lattice, at a price of energy $q_{a,k}(u,Y)$. We repeat the steps of the computation of the spectrum - for a spinless auxiliary space $\ell_a=\dot{\ell}_a=0$ - in Fig.\ref{eigen_scatterf}: the mirror excitation (grey line) is pulled down across the central part of the square lattice moving across the physical particles and scattering with them.
 The first right and the last left spinors in physical space are just pulled down together with the mirror excitation. The central left and right movers (red/green lines) are moved across each other, producing a correspondent scattering matrix. In the last step, we re-form the setup of the layer operator, but with one unit of length less, namely pushing out of the lattice the first right spinor and the last left spinor. The price of the moves/scattering in the lattice amounts to $q_{a,3}(u,Y)$ as it can be read out of the star-triangle scattering.
In a similar fashion one can prove \eqref{exchange_gen} for the product of two layer operators $\mathbf{\Lambda}_{k+1}(Y')\mathbf{ \Lambda}_k(Y)$ via the scattering picture, according to Fig.\ref{exch_scatter}.
The same technique can be applied to the overlap of layers $\mathbf{\bar \Lambda}_k(Y')\mathbf{ \Lambda}_k(Y)$, in order to derive the algebra \eqref{exch_orto}.
 In this case, as one of the operators is adjoint, the two excitations move in opposite directions, which leads to transposition in one of the two spaces of the scattering matrix $\mathbf{R}(Y,Y')^{t'}$. Pulling the two excitations across each other, we eventually recover two layers of length diminished by one, where one right and one left mode are singled out of the lattice, with the additional scattering matrices of the r.h.s. of \eqref{exch_orto} resulting from the moves.
\begin{figure}[H]
\begin{center}
\includegraphics[scale=0.87]{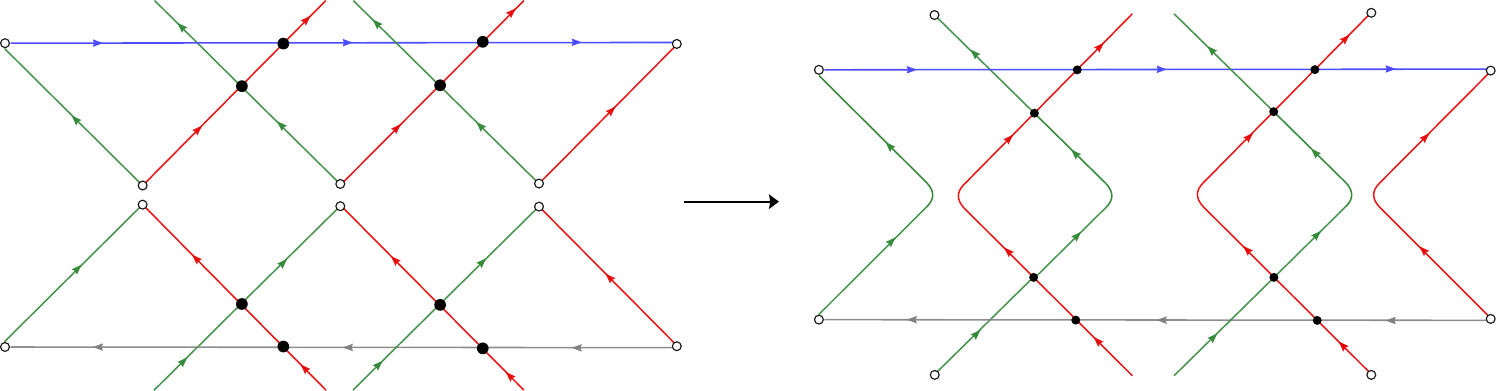}
\vspace{5mm}
\vspace{3mm}
\includegraphics[scale=0.9]{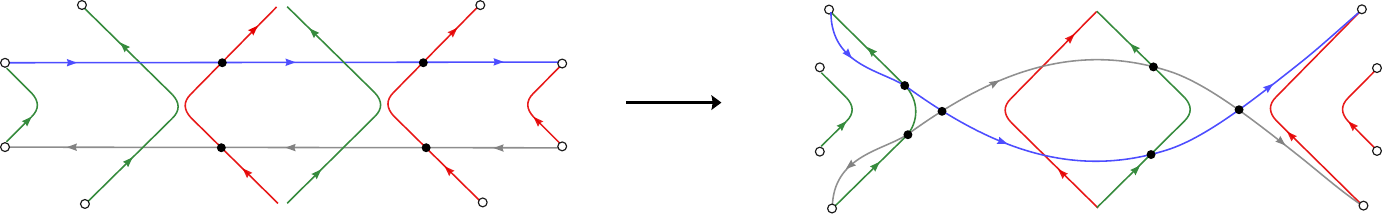}
\vspace{5mm}
\vspace{3mm}
\includegraphics[scale=0.9]{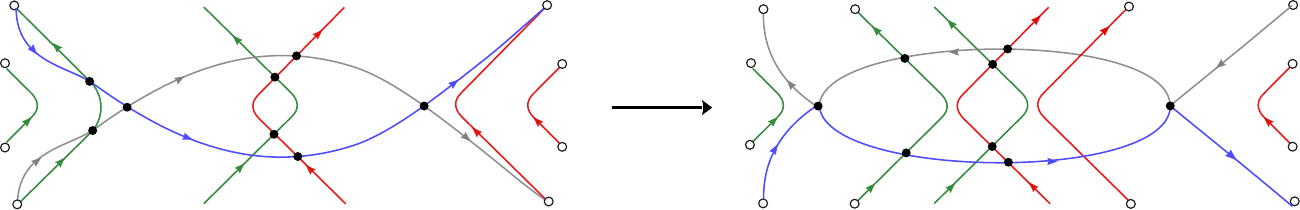}
\end{center}
\caption{Scattering of two conjugated mirror excitation across each other in the lattice, as it occurs in the overlap of layers $\mathbf{\bar \Lambda}_k(Y')\mathbf{ \Lambda}_k(Y)$, corresponding to the first panel. Blue and grey excitation are pulled down- and up-stairs respectively, giving rise to scattering (black blobs) among each other and with the left/right components of the chain's sites (red/green lines). The last panel is the scattering picture of the r.h.s. of \eqref{exch_orto}.}
\end{figure}
In conclusion of the section, we shall mention that the \emph{mirror-mirror} $S$-matrix for the scattering of the two mirror excitations $Y$ and $Y'$ can be read out of the r.h.s. of \eqref{exchange_gen}: the undotted indices are mixed by $\mathbf{R}(Y',Y)$ from the left, while the dotted indices are mixed from the right by $\mathbf{R}(Y,Y')=\mathbf{R}(Y',Y)^{\dagger}$. Finally, accounting for the dynamical factors $a_{n,n'}(i\nu-i\nu')$ and $a_{n,n'}(i\nu'-i\nu)$ associated to the two $\mathbf{R}$-matrices, the scattering process amounts to
\begin{align}
\begin{aligned}
\label{S_matrix}
\mathbf{S}_{n,n'}(i\nu-i\nu')=\mathcal S(Y,Y') \times \mathbf{R}(Y,Y') \otimes \dot{\mathbf{R}}(Y,Y')
\end{aligned}
\end{align}
with
\begin{align}
\begin{aligned}
\mathcal S(Y,Y')  &=a_{n,n'}(i\nu-i\nu')/a_{n',n}(i\nu'-i\nu) = \\&=\frac{\Gamma\left(1+i\nu-i\nu'+\frac{n-n'}{2}\right)\Gamma\left(i\nu-i\nu'+\frac{n-n'}{2}\right)}{\Gamma\left(1+i\nu'-i\nu+\frac{n-n'}{2}\right)\Gamma\left(i\nu'-i\nu+\frac{n-n'}{2}\right)}\frac{\left(1+i\nu-i\nu'+\frac{n+n'}{2}\right)}{\left(1+i\nu'-i\nu+\frac{n+n'}{2}\right)}\,,
\end{aligned}
\end{align}
and inherits unitarity, YBE and crossing property \eqref{R_cross} by the components $\mathbf{R}$. The $S$-matrix defined in \eqref{S_matrix} reproduces the part depending on the exchange momentum $\nu-\nu'$ of (2.32) in \cite{Basso_2019}. The missing phase of type $\phi_n(\nu)/\phi_{n'}(\nu')$ can be fixed by crossing symmetry. The matrix part of \eqref{S_matrix} is actually the spinor representation of the Yangian $R$-matrix of $SO(4)$ fused in the symmetric representations of irreducible tensors of rank $n$ and $n'$ as made explicit out by equations (4.19) and (4.22) of \cite{Derkachov:2020zvv}.
\section{Conclusions}
In this paper we put the base for a direct derivation of the \emph{hexagonalization} techniques of the $n$-point functions of in the conformal Fishnet theories. In fact, we have exhaustively discussed the construction of the eigenfunctions of the mirror channel of the spinning and inhomogeneous Fishnet lattice. The reductions of this general model describe Feynman integrals on the disk in the various, particular, realizations of CFTs of the fishnet type. Our derivation successfully extends the iterative logic of \cite{Derkachov:2014gya,Derkachov:2001yn}, and the properties of the wave-functions of the lattice excitations are mapped to those of their building blocks  - the \emph{layer operators} defined in various degrees of generality in section \ref{sect:eigenf}. The latter play the role of creation/annihilation operators of the fishnet's mirror excitations, showing the emergence of a scattering of mirror-excitations from the Feynman integrals, as noted by \cite{BassoFerrando} for the bi-scalar theory in any space-time dimension \cite{Kazakov:2018qez}. With these observations, we formulated a scattering interpretation for the integral star-triangle identities standing at the basis of our calculations, propedeutic to the next chapter of the series \cite{Olivucci_hex_II} . All the results are presented in detail and constitute the handbook for computing more overlaps of functions of many excitations, necessary to derive the \emph{hexagon} form factors.

The fishnet integrals on the disk have already attracted the focus of research about the integrability of higher-dimension field theory, in two main directions we shall mention. The first, is the Yangian symmetry of such diagrams \cite{Chicherin:2017frs, Chicherin:2017cns}, which is also applied to bootstrap techniques for their actual computation \cite{Loebbert_2020,Corcoran_2021}. The second reason is that reduction of specific $n$-point function to $4$-point functions - named {Basso-Dixon} integrals - reveal the remarkable property of the fishnets as a theory of \emph{ladder integrals} (and determinant thereof) \cite{Basso:2017jwq,Basso:2021omx}, furnishing at the same time valuable data-points for the bootstrap of $4$-point functions in the supersymmetric $\mathcal{N}=4$ SYM \cite{Coronado_2019,Coronado_2020,Kostov_2019}. 

In our work, the spectral equation of the transfer matrix of the fishnet lattice has been solved by quantum separation of variables (SoV) \cite{Sklyanin:1991ss, Sklyanin:1995bm}. In fact, we extended to the general setup of inhomogeneous and spinning fishnet the results obtained by the author and S.~Derkachov in \cite{Derkachov2020}. The approach to quantum integrability by separation of variables (SoV) for higher rank and non-compact spin chains - as the $SO(1,5)$ magnet of this paper - is an compelling topic that lead to recent progresses in multiple directions, for instance \cite{Ryan:2020rfk,Gromov:2020fwh,Cavaglia:2019pow}. In this respect it remains an open problem the SoV of the fishnet's direct channel, i.e. of the conformal chain with periodic boundaries that describes the dilation spectrum of local operators in the field theory. Moreover, recent results are aiming towards an application of the SoV to the computation of higher-point correlators \cite{Cavaglia:2021mft,Cavaglia:2021bnz}; the present series of papers belongs with full rights to this approach.

In conclusion, beside the main target of this series, we express the hope that our derivations may lead to:
\begin{enumerate}
\item a derivation of the SoV for the model with periodic boundaries, borrowing the logic of the $SL(n,\mathbb{C})$, $n=2,3$ results \cite{Derkachov:2001yn,Derkachov:2018ewi} to the conformal group in $4d$ ;
\item the explicit computation of Feynman integrals with higher-loops and number of external legs, via methods of conformal spin chain, and its application to the bootstrap of higher-point supersymmetric correlators (see for instance \cite{Vasco}), even in the $3d$ ABJM theory \cite{Aharony_2008};
\item the origin of quantum integrability in the undeformed $\mathcal{N}=4$ SYM theory, by the observation that analogue properties in the fishnets \emph{emerge} from the simple content of Feynman diagrams. Further progress in this direction would amount to the extension of the spin-chain techniques to the generalized fishnet described in \cite{Kazakov_2019} by Feynman integrals with a \emph{dynamical} lattice topology, hinting to ``a regular dynamical lattice structures for the planar Feynman diagrams of $\mathcal{N}=4$ SYM theory "\cite{Volodya_strings}.
\end{enumerate}
\section*{Acknowledgements}
The author thanks S.~Derkachov and G.~Ferrando for several discussion on the topic and fruitful collaboration on related projects. The author is grateful to P.~Vieira, V.~Kazakov for providing insights, questions and motivations.  
Research at the Perimeter Institute is supported in part by the Government of Canada through NSERC and by the Province of Ontario through MRI. This work was additionally supported by a grant from the Simons Foundation (Simons Collaboration on the Nonperturbative Bootstrap).
\appendix
\section{Left/right $SU(2)$ spinors}
\label{app:spinors}
The left/right spinors are two-component complex vectors $u_{a}, v_{\dot a} \in \mathbb{C}^2$, for $a,\dot a =1,2.$ The spinors $u_a$ and $v_{\dot a}$ are rotated by the standard $SU(2)$ matrices
\begin{equation}
u_a \mapsto M_{a}^{b} u_b\,,\,\,\, v_{\dot a} \mapsto M_{\dot a}^{\dot b} v_{\dot b}\,,\,\,\,\,\, M^{\dagger}M=MM^{\dagger} =\mathbbm{1}\,.
\end{equation}
The action $SO(4)$ on left/right spinors is realized via the identification $SO(4)\sim SU(2)\times SU(2)$. Given a matrix $\Lambda_{\mu}^{\nu} \in SO(4)$, we recall that it is realized in terms of Lie algebra generators $J_{\mu\nu}$ as
\begin{align}
\notag
&\Lambda(\boldsymbol{\omega}) = \exp\left(i \omega_{\alpha\beta}J^{\alpha \beta}\right)\,,\,\,\, \omega_{\alpha\beta} = -\omega_{\beta\alpha}\,,\\\label{so4_algebra}
&
[J^{\mu\nu},J^{\alpha \beta}] = \delta^{\nu\alpha} J^{\mu\beta}-\delta^{\mu\alpha} J^{\nu\beta}-\delta^{\mu\beta} J^{\nu\alpha}+\delta^{\nu\beta} J^{\mu\alpha}\,.
\end{align}
The action of $SO(4)$ on the left/right spinors is defined by the transformation property of the euclidean $\sigma$-matrices, defined in terms of the standard $\sigma_k$ as
\begin{equation}
\boldsymbol{\sigma}_0 =\mathbbm{1}\,,\,\boldsymbol{\sigma}_{k} = i\sigma_k\,,\,\overline{\boldsymbol{\sigma}}_{\mu}={\boldsymbol{\sigma}}_{\mu}^{\dagger}\,,
\end{equation}
and reads
\begin{align}
\begin{aligned}
\label{spin_rot_good}
\Lambda_{\nu}^{\,\mu} (\boldsymbol{\sigma}_{\mu})_{a}^{\dot a} \,=\,U_{a}^{b} (\boldsymbol{\sigma}_{\mu})_{b}^{\dot b} (V^{\dagger})_{\dot b}^{\dot a}\,,\,\,\,\Lambda_{\nu}^{\,\mu} (\boldsymbol{\bar \sigma}_{\mu})_{\dot a}^{ a} \,=\,V_{\dot a}^{\dot b} (\boldsymbol{\bar \sigma}_{\mu})_{\dot b}^{ b} (U^{\dagger})_{b}^{ a}\,.
\end{aligned}
\end{align}
The matrices $\boldsymbol{\sigma},\boldsymbol{\bar \sigma}$ satisfy the Clifford algebra
\begin{equation}
\label{cliff}
\boldsymbol{\sigma}_{\mu}\boldsymbol{\bar \sigma}_{\nu}+\boldsymbol{\sigma}_{\nu}\boldsymbol{\bar \sigma}_{\mu} = 2 \delta_{\mu\nu}\,.
\end{equation}
Given the $SO(4)$ group element defined by the parameters $\boldsymbol{\omega}=\{\omega_{\mu\nu}\}$, the matrices $U,V$ appearing in the formula \eqref{spin_rot_good} are defined as
\begin{equation}
 U(\boldsymbol{\omega}) = \exp(i\omega_{\mu\nu} S^{\mu\nu})\,,\,\,\,  V(\boldsymbol{\omega}) = \exp(i\omega_{\mu\nu} \bar{S}^{\mu\nu})\,,
\end{equation}
for
\begin{equation}
 S^{\mu\nu} = \frac{i}{4} \left(\boldsymbol{\sigma}^{\mu}\boldsymbol{\bar \sigma}^{\nu}-\boldsymbol{\sigma}^{\nu}\boldsymbol{\bar \sigma}^{\mu}\right)\,,\,\, \bar{S}_{\mu\nu}= \frac{i}{4} \left(\boldsymbol{\bar\sigma}^{\mu}\boldsymbol{ \sigma}^{\nu}-\boldsymbol{\bar \sigma}^{\nu}\boldsymbol{\sigma}^{\mu}\right)\,.
 \end{equation}
Both matrices ${S}_{\mu \nu}$ and ${\bar S}_{\mu \nu}$ satisfy the $so(4)$ algebra \eqref{so4_algebra} and in addition are hermitian, so that $U$ and $V$ are unitary $2\times 2$ matrices. In agreement with \eqref{spin_rot_good} and with the transposition rule ${\sigma}_2 \boldsymbol{\sigma}_{\mu} {\sigma}_2=\boldsymbol{\bar \sigma}^{t}_{\mu}$ one can check
\begin{equation}
\label{propert_2}
U^{\dagger} = U^{-1} = (\boldsymbol{\bar \sigma}_2 U \boldsymbol{ \sigma}_2 )^t\,,\,\,  (\boldsymbol{\bar\sigma}_2 S_{\mu\nu}\boldsymbol{ \sigma}_2 )^t = -S_{\mu\nu}\,,
\end{equation}
and the same properties with exchanged $\boldsymbol{\sigma}_2, \boldsymbol{\bar \sigma}_2$ hold for $V$ and $\bar{S}_{\mu\nu}$ respectively. 
The actions on left spinors and right spinors induced by $SO(4)$ are defined by the $M=U(\boldsymbol{\omega})$  and $M=V(\boldsymbol{\omega})$ respectively. The multiplication of left, right spinors by the matrix $i{ \sigma}_2=\boldsymbol{\sigma}_2$ defines the complex conjugate representations
 \begin{align}
 \begin{aligned}
 \label{charge_C}
 & u_{ a} \mapsto U(\boldsymbol{\omega})_{a}^{b} u_{b}\,\Longrightarrow \, (i{\sigma}_2 u)_{a} \mapsto U^*(\boldsymbol{\omega})_{a}^{b} (i{\sigma}_2 u)_{b} \,,\\  &v_{\dot a} \mapsto V(\boldsymbol{\omega})_{\dot a}^{\dot b} v_{\dot b}\,\Longrightarrow \, (i{\sigma}_2 v)_{\dot a} \mapsto V^*(\boldsymbol{\omega})_{\dot a}^{\dot b} (i{\sigma}_2 v)_{\dot b} \,.
 \end{aligned}
 \end{align}
The scalar product in the space of spinors $\mathbb{C}^2$ is the standard one on a complex linear space
\begin{equation}
\langle w|u\rangle = (w^{a})^* u_a = \sum_{a=1,2} (w_{a})^* u_a\,,\,\,\, \langle z|v\rangle = (z^{\dot a})^* v_{\dot a} = \sum_{a=1,2} (z_{\dot a})^* v_{\dot a}\,,
\end{equation}
and it is invariant under the $SO(4)$ left/right actions and respect to conjugation \eqref{charge_C} .

The generalization to symmetric spinors follows. We consider the generic element of the symmetric subspace $\text{Sym}_{\ell}[\mathbb{C}^2] \subset (\mathbb{C}^2)^{\otimes \ell}$
\begin{equation}
u_{\mathbf{a}} = u_{(a_1,\dots, a_{\ell})} =u_{(a_{\pi(1)},\dots, a_{\pi(\ell)})} \,,
\end{equation}
where $\pi$ is a permutation of $\ell$ indices. The spinor $u_{\mathbf{a}}$ is rotated by $SU(2)$ in the representation of spin ${\ell}/2$, i.e. the $\ell$-fold symmetric representation $\mathbf{(\boldsymbol{\ell}+1)}$. Given a matrix $U\in SU(2)$, the action reads
\begin{equation}
[U]_{\mathbf{a}}^{\mathbf{b}} \,u_{\mathbf{b}} \equiv \frac{1}{\ell!} \sum_{\pi} U_{\mathbf{a}_{\pi(1)}}^{\mathbf{b_1}}\cdots U_{\mathbf{a}_{\pi(\ell)}}^{\mathbf{b_{\ell}}} \,u_{(b_1\dots b_\ell)}  \,.
\end{equation}
The analogue definition apply to the $\dot{\ell}$-fold symmetric spinor $v_{\dot{\mathbf{a}}}$. The rule for charge conjugation follows from the symmetrization of \eqref{charge_C}\ \begin{equation}
 u_{\mathbf\mathbf{a}}  \mapsto [i{ \sigma}_2]_{\mathbf{ a}}^{\mathbf b} \,u_{\mathbf b}\,,\,\, v_{\dot{\mathbf a}}  \mapsto [i{\sigma}_2]_{\mathbf{\dot a}}^{\dot{\mathbf{b}}} \,v_{\dot{\mathbf{b}}}\,.
 \end{equation}

\section{Amputated Star-triangle and Interchange relation}
\label{app:str_amp}
\begin{figure}[H]
\begin{center}
\includegraphics[scale=1.2]{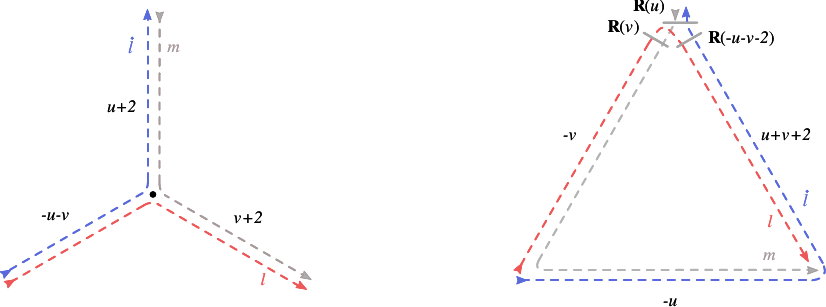}
\end{center}
\caption{Star-triangle identity \eqref{STR_ker} in the equivalent form obtained by the matrix $\mathbf{R}(u+v)$ in the star of Fig.\ref{STR_1} to the triangle by the identity $\mathbf{R}(u+v)^{t}\mathbf{R}^{t}(-u-v-2) \propto \mathbbm{1}$ explained in \eqref{id_transp}. This form of the identity is applied, for instance, in the computation of eigenfunction overlaps of sections \ref{exch_spin_inhom}, \ref{orto_spin_sect}.}
\label{STR_app}
\end{figure}
\begin{figure}[H]
\begin{center}
\includegraphics[scale=1.2]{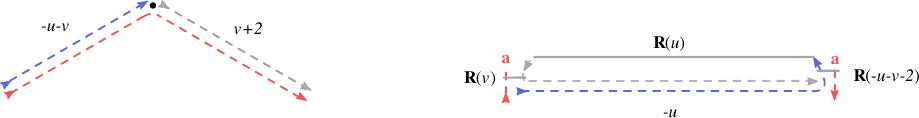}
\end{center}
\caption{Amputated star-triangle identity - also dubbed \emph{chain-rule} identity in \cite{Derkachov:2021rrf}. It is obtained from the star-triangle in the form of Fig.\eqref{STR_app} by a amputation of one leg of the star integral, achieved by sending the top vertex of the star to $\infty$ by a conformal inversion of coordinates around it. On the r.h.s. the indices $a$ are summed over, and stands for the product of matrices $\mathbf{R}_{\ell\dot{\ell}}(v)\mathbf{R}_{\ell m}(-u-v-2)$ in the space of $\ell$-fold symmetric spinors (red). This relation is applied multiple times in sect.\ref{sect:eigenf} for the computation of eigenvalues and especially in sections \ref{exch_spin_inhom} and \ref{orto_spin_sect}.}
\label{STR_amp_app}
\end{figure}
\begin{figure}[H]
\begin{center}
\includegraphics[scale=1.0]{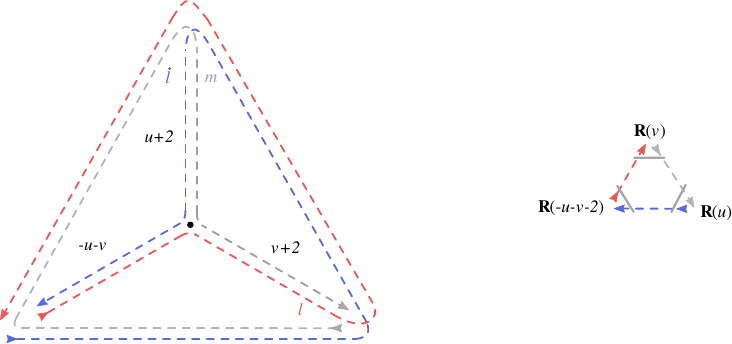}
\end{center}
\caption{Infinite dimensional representation of the finite dimensional product of $\mathbf{R}$-matrices $\mathbf{R}_{n\ell}^{t_n}(u)\mathbf{R}_{m\ell}(-u-v-2)\mathbf{R}^{t_n}_{nm}(v)$, obtained adding lines in the l.h.s. and r.h.s of the relation of Fig.\eqref{STR_app}, so that the triangle simplifies and cancels. The latter product is equivalent to one side of the Yang-Baxter equation for $SU(2)$ symmetric representations, rewriting the transposed matrices via crossing \eqref{R_cross}. Hence, this picture provides a Feynman diagram representation for the $SU(2)$ finite-dimensional YBE.}
\label{Finite_dim_R}
\end{figure}
\begin{figure}[H]
\begin{center}
\includegraphics[scale=1.0]{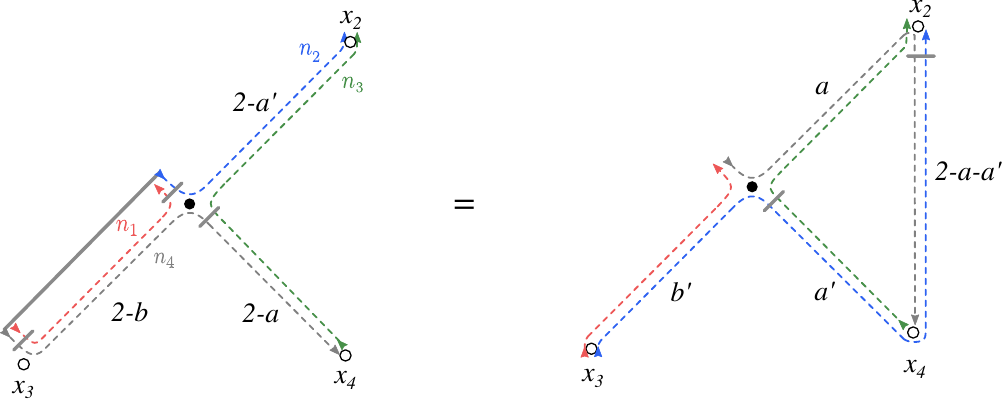}
\end{center}
\caption{Amputated interchange relation, obtained from (I) in Fig.\ref{interchanges} replacing the left (grey/blu/red) triangle in both sides of the equality by its amputated form as in Fig.\ref{STR_amp_app}. The amputation is achieved by sending the top left point to $\infty$ via conformal inversion around it. This relation is applied in the proof of exchange symmetry of section \ref{exch_spin_inhom}.}
\end{figure}
\section{Factorization of $\rmat$}
\label{app:R_fact}
We introduce here a factorization of the operator $\hat\rmat(u)$ into the product of two ``halves" $ \rmat_{12}(u) = {\rmat}_{12}^{-}(u){\rmat}^{+}_{12}(u)$,
that stands at the basis of the factorization of mirror transfer matrices of Fig.\ref{Q+Q-}:
\begin{equation}
\label{R+}
{\rmat}_{12}^{+}(u)=\frac{[\mathbf{\overline{x}}_{12}]^{\dot{\ell}_1}
\mathbf{R}_{\dot{\ell}_1\ell_1}(2-\Delta_1)
[\mathbf{x}_{12}]^{\ell_1}}{x_{12}^{2(\Delta_1-2)}}\,
\frac{[\mathbf{\overline{p}}_{1}]^{\dot{\ell}_2}
\mathbf{R}_{\dot{\ell}_2\dot{\ell}_1}(u-\Delta_{-})
[\mathbf{p}_{1}]^{\dot{\ell}_1}}{\hat{p}_{1}^{2(\Delta_{-}-u)}}\,
\frac{[\mathbf{\overline{x}}_{12}]^{\ell_1}
\mathbf{R}_{\ell_1\dot{\ell}_2}(u+\Delta_{+})
[\mathbf{x}_{12}]^{\dot{\ell}_2}}{x_{12}^{-2(u+\Delta_{+})}}\,,
\end{equation}
and
\begin{equation}
\label{R-}
{\rmat}_{12}^{-}(u)=
\frac{\mathbf{R}_{\dot{\ell}_1\ell_2}(u-\Delta_{+})
[\mathbf{x}_{12}]^{\ell_2}}{x_{12}^{2(\Delta_{+}-u)}}\,
\frac{[\mathbf{\overline{p}}_{2}]^{\ell_2}
\mathbf{R}_{\ell_2\ell_1}(u+\Delta_{-})
[\mathbf{p}_{2}]^{\ell_1}}{\hat{p}_{2}^{-2(u+\Delta_{-})}}\,
\frac{[\mathbf{\overline{x}}_{12}]^{\ell_1}
\mathbf{R}_{\ell_1\dot{\ell}_1}(\Delta_{1}-2)
[\mathbf{x}_{12}]^{\dot{\ell}_1}}{x_{12}^{2(2-\Delta_{1})}}\,.
\end{equation}
\begin{figure}[H]
\begin{center}
\includegraphics[scale=1.0]{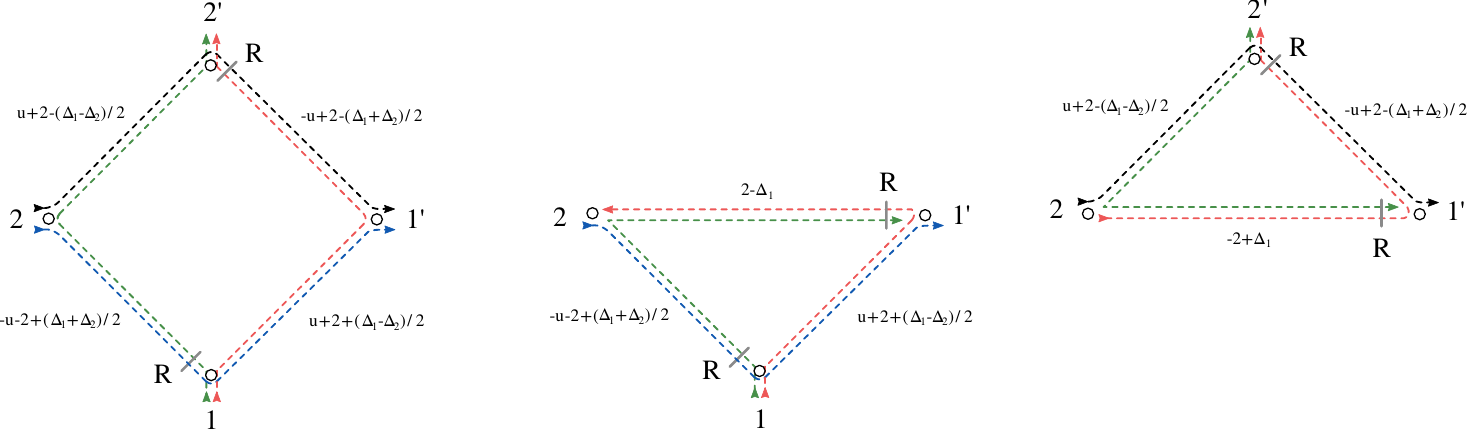}
\caption{\textbf{Left:} the kernel of the general $\rmat$-operator in diagram formalism. It can be factorized into the product of two integral kernels $\rmat_{\pm}$. \textbf{Center:} integral kernel of the operator $ \rmat_-(u)$ \textbf{Right:} integral kernel of the operator $ \rmat_+(u)$. When multiplied, the horizontal red/green lines in the two triangular kernels simplify to the identity. Indeed, their power is opposite and the $SU(2)$ matrices in the numerators are the inverse of each other, and their product reads $[\mathbf{x_{21'}}]^{\dot \ell_1}\mathbf{R}_{\ell_1\dot{\ell}_1}(\Delta_1-2)[\mathbf{\overline{x_{21'}}}]^{\ell_1}[\mathbf{x_{21'}}]^{\ell_1}\mathbf{R}_{\ell_1\dot{\ell}_1}(2-\Delta_1)[\mathbf{x_{21'}}]^{\dot \ell_1} =\mathbbm{1}$ - see \eqref{R_unit}. Hence, the kernel of $\rmat (u)$ is reconstructed.}
\end{center}
\label{R_fac}
\end{figure}
\section{Identity operator}
The identity operator on the space $\mathbb{V}_a$ of the representation $(
\Delta,\ell,\dot\ell)$ can be expressed in terms of Feynman integrals as a convolution of two conformal propagators of type $[\mathbf{x}]^{\ell} [\mathbf{\overline x}]^{\dot \ell} (x^2)^{-u}$, if their scaling dimensions $u_1$ and $u_2$ sum up to the dimension of spacetime $d=4$ . Let's start from the trivial equation
\begin{equation}
\label{identity_p}
\left(\hat p^{2u}[\mathbf{p}]^{\ell}\mathbf{R}_{\ell\dot\ell}(u)[\mathbf{\overline p}]^{\dot \ell}\right)\left(\hat p^{2u}[\mathbf{p}]^{\ell}\mathbf{R}_{\ell\dot\ell}(u)[\mathbf{\overline p}]^{\dot \ell}\right)^{-1} =\left(\hat p^{2u}[\mathbf{p}]^{\ell}\mathbf{R}_{\ell\dot\ell}(u)[\mathbf{\overline p}]^{\dot \ell}\right)\left( \hat p^{2(-u)}[\mathbf{\overline p}]^{\dot \ell}\mathbf{R}_{\dot \ell \ell}(-u)[\mathbf{p}]^{ \ell}\right)  = \mathbbm{1}_{\ell}\otimes \mathbbm{1}_{\dot \ell}\,,
\end{equation}
and going to the Fourier space we extend the integral representation given in \eqref{p_def} to the spinning case \cite{Derkachov:2021rrf}
\begin{align}
\label{Fspinor2}
&\int d^4 x e^{ipx}\,\frac{[\mathbf{\overline{x}}]^{\ell}
[\mathbf{x}]^{\dot{\ell}}}{x^{2(u+2)}} = a_{\ell\dot{\ell}}(u)\,
\hat p^{2u}\,[\mathbf{\overline{p}}]^{\ell}\mathbf{R}_{\ell\dot{\ell}}(u)
[\mathbf{p}]^{\dot{\ell}}\,.
\end{align}
Thus, \eqref{identity_p} becomes an identity of integral operators
\begin{equation}
\label{delta_integral}
\frac{1}{a_{\ell\dot\ell}(u)a_{\dot \ell \ell}(-u)}\int d^4 z \int d^4 w  \frac{[\mathbf{(\overline{x-z})(z-w)}]_{\mathbf{a}}^{\mathbf{b}}[\mathbf{(x-z)(\overline{z-w})}]_{\mathbf{\dot c}}^{\mathbf{\dot d}}}{(x-z)^{2(u+2)}(z-w)^{2(-u+2)}} \Phi_{\mathbf{b \dot d} }(w)\,=\, \Phi_{\mathbf{a \dot c} }(x)\,,
\end{equation}
for any function $\Phi_{\mathbf{a\dot c} }(x)$ belonging to the space $\mathbb{V}_a$. Last formula defines a representation of the Dirac's $\delta$ as convolution of conformal propagators
\begin{equation}
\label{delta_Dirac}
\delta^{(4)}(x-y)\delta_{\mathbf{a}}^{\mathbf{b}} \delta_{\mathbf{\dot c}}^{\mathbf{\dot d}}  =\frac{1}{a_{\ell\dot\ell}(u-2)a_{\dot \ell \ell}(2-u)}\int d^4 z  \frac{[\mathbf{(\overline{x-z})(z-y)}]_{\mathbf{a}}^{\mathbf{b}}[\mathbf{(x-z)(\overline{z-y})}]_{\mathbf{\dot c}}^{\mathbf{\dot d}}}{(x-z)^{2(u)}(z-w)^{2(4-u)}}\,, \end{equation}
which one can regard to the well-known scalar formula for spinning particles $\ell,\dot \ell\neq 0$.
\section{Proof of \eqref{twist_comm_1}}
\label{app:comm_tw}
The commutation relation $\left[\mathbf{Q}_{a}(u),\mathbf{Q}_{b}(v)\right]=0$ holds as a consequence of the two following properties of the kernel $\hat\rmat_{12}(x_1,x_2|y_1,y_2)$:
\begin{equation}
\label{twist_11}
\int d^4x_1  d^4x_2 \hat\rmat_{12}(x_1,x_2|y_1,y_2)(u) \,=\,C(u)\,\mathbf{R}_{{\ell_1} \ell_2}\left(u+\frac{\Delta_2-\Delta_1}{2}\right)\otimes \mathbf{R}_{\dot{\ell_1}\dot \ell_2}\left(u+\frac{\Delta_1-\Delta_2}{2}\right)\,,
\end{equation}
and 
\begin{equation}
\label{twist_12}
\hat\rmat(x_1,x_2|x_0,x_0) \,=\, C(u)\,\mathbf{R}_{{\ell_1} \ell_2}\left(u+\frac{\Delta_2-\Delta_1}{2}\right)\mathbf{R}_{\dot{\ell_1}\dot \ell_2}\left(u+\frac{\Delta_1-\Delta_2}{2}\right)\, \delta^{(4)}(x_0-x_1)\delta^{(4)}(x_0-x_2) \,,
\end{equation}
where
\begin{equation}
\label{coeff_C}
 C(u) = a_{\ell_1\ell_2}\left(u+\frac{\Delta_1-\Delta_2}{2}\right)a_{\dot\ell_1\dot \ell_2}\left(u-\frac{\Delta_1-\Delta_2}{2}\right)\frac{a_{\dot \ell_1\ell_2}\left(-u+2-\frac{\Delta_1+\Delta_2}{2}\right)}{a_{\dot \ell_1\ell_2}\left(u+2-\frac{\Delta_1+\Delta_2}{2}\right)}\,.
\end{equation}
The proof of \eqref{twist_11} is obtained by straightforward integration of the points $x_1$ and $x_2$ by the amputated star-triangle identity (in detail in Fig.\ref{STR_amp_app}). Starting from the kernel $\hat\rmat(x_1,x_2|y_1,y_2)$ depicted on the left of the next figure, the chain integration in $x_1$ delivers a triangle, depicted in the center. Furthermore, the chain integration in $x_2$ reduces the triangle to the edge $y_{12}$. The power of $y_{12}^2$ is in fact zero (picture on the right), and in addition the numerators simplify via the identity $[\mathbf{x\overline x}]=[\mathbf{\overline x x}]=\mathbbm{1}$. The only matrix structure left behind are the matrices $\mathbf{R}_{\ell_1\ell_2}$ and $\mathbf{R}_{\dot\ell_1\dot\ell_2}$.
\begin{figure}[H]
\begin{center}
\includegraphics[scale=0.9]{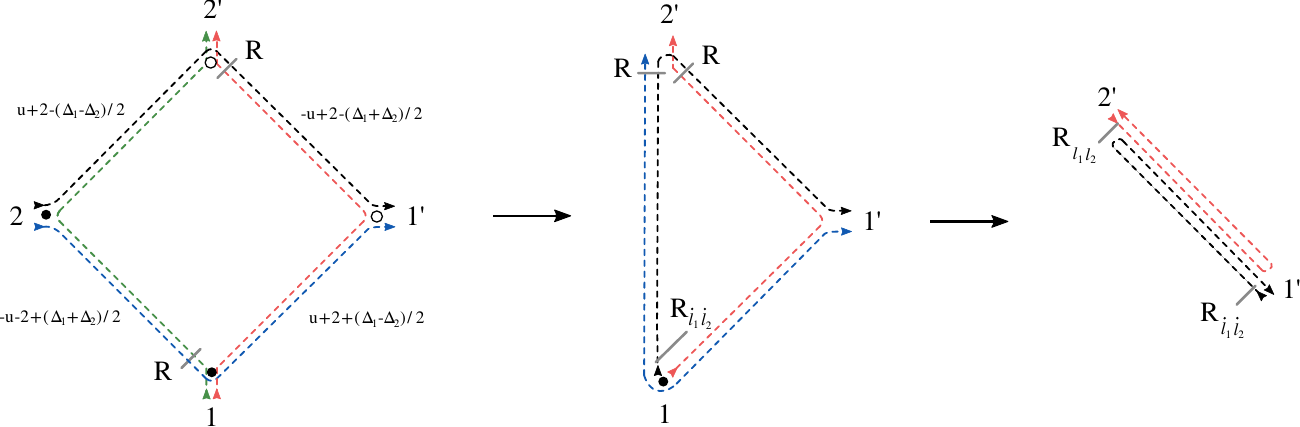}
\end{center}
\end{figure}
Finally, collecting the coefficients $a_{nm}(u)$ produced by the two star-triangle integrations, according to \eqref{STR_ker}, we recover the expression \eqref{coeff_C} $\square$.\\
The proof of \eqref{twist_12} is slighly more involved: we start from the kernel, depicted on the left and we rewrite it according to the decomposition $\hat \rmat_{\pm}$ in the product of two triangles (central picture), which are transformed into cubic vertices by the star-triangle duality (right picture), and this point, we set $y_1=y_2=x_0$
\begin{figure}[H]
\begin{center}
\includegraphics[scale=0.9]{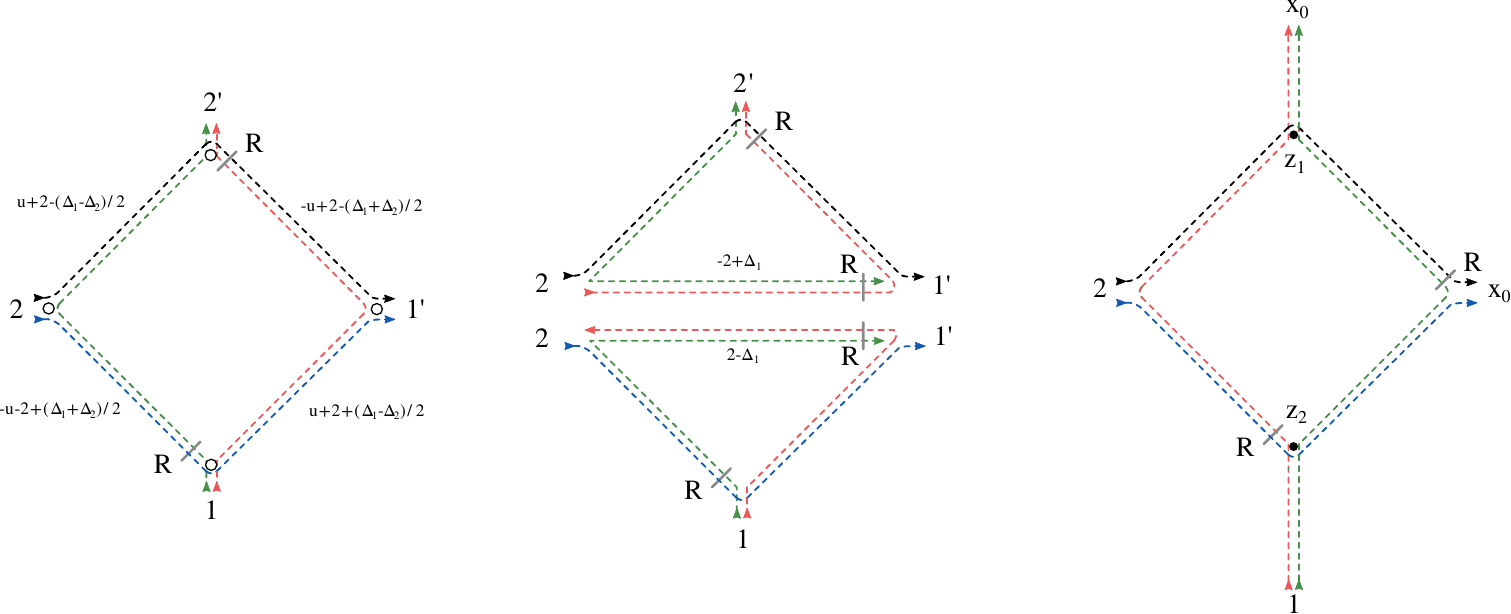}
\end{center}
\end{figure}
With such identification of points, the kernel reduces as depicted on the left of next figure. As the powers of the propagators convoluted in $z_1$ sum up to $d=4$, according to formula \eqref{delta_integral}, the result of the integration is the distribution $\delta^{(4)}(x_2-x_0)$, leaving with the diagram depicted in the centre. The latter is reduced to the one on the right by the simplification among the lines $z_2-x_0$, depicted on the right. The final step consist in the integration in $z_2$, that via \eqref{delta_integral} produces $\delta^{(4)}(x_1-x_0)$. \begin{figure}[H]
\begin{center}
\includegraphics[scale=0.9]{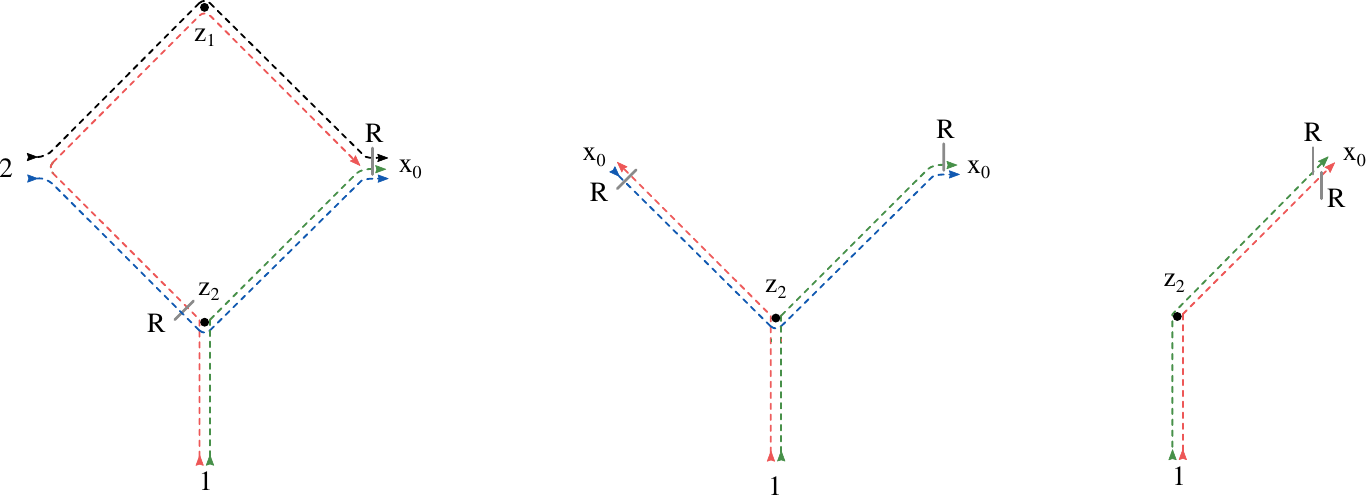}
\end{center}
\end{figure} 
The final result is a product of two $\delta$-functions, appearing in the r.h.s. of \eqref{twist_12}, and the coefficients resulting from the application of star-triangle dualities and of \eqref{delta_integral} is equal to $C(u)$ in \eqref{coeff_C} $\square$.
\section{Proof of \eqref{hatR_properties}}
\label{app:Rhat}
The proof of the first relation
\begin{align}
\begin{aligned}
\label{hatR_unit}
&\hat{\mathbf{R}}_k(Y,Y')\hat{\mathbf{R}}_k(Y',Y)= \mathbbm{1}\,,
\end{aligned}
\end{align}
is a consequence of the definition 
\begin{equation}
\label{hatR_def}
\hat{\mathbf{R}}_k(Y,Y')\equiv \frac{\left(1-i\nu' +\frac{n'- \ell_k}{2} \right)}{\left(1-i\nu'+\frac{n'+ \ell_k}{2} \right)} \frac{ \left(i\nu-1 +\frac{n- \ell_k}{2} \right)}{ \left(i\nu-1 +\frac{n+ \ell_k}{2} \right)} \left[ \mathbf{R}(0,Y')^{t_k} \mathbf{R}(Y',Y) \mathbf{R}(Y,-2i)^{t_k} \right]^{t_k}\,,
\end{equation}
and of the properties of the $SU(2)$ fused $\mathbf R$-matrix. First, one can verify the general matrix identity
\begin{equation}
\label{mat_transp}
\left[A^{t_2}_{12} B_{13} C_{23}^{t_2}\right]^{t_2} \left[ {A'}_{12}^{t_2} {B'}_{13} {C'}_{23}^{t_2}\right]^{t_2}=\left[A^{t_2}_{12} B_{13} C_{23}^{t_2} {A'}_{12}^{t_2} {B'}_{13} {C'}_{23}^{t_2}\right]^{t_2}\,,
\end{equation}
valid for matrices $A_{ij}$ acting on the tensor product of two generic linear spaces $V_i \otimes V_j$, and we apply it to the left hand side of \eqref{hatR_unit}
\begin{align}
\begin{aligned}
&\left[ \mathbf{R}(0,Y')^{t_k} \mathbf{R}(Y',Y) \mathbf{R}(Y,-2i)^{t_k} \right]^{t_k}\left[ \mathbf{R}(0,Y)^{t_k} \mathbf{R}(Y,Y') \mathbf{R}(Y',-2i)^{t_k} \right]^{t_k} =\\&=
\left[ \mathbf{R}(0,Y')^{t_k} \mathbf{R}(Y',Y) \mathbf{R}(Y,-2i)^{t_k} \mathbf{R}(0,Y)^{t_k} \mathbf{R}(Y,Y') \mathbf{R}(Y',-2i)^{t_k} \right]^{t_k}\,.
\end{aligned}
\end{align}
Then, we use the crossing \eqref{R_cross} and unitarity of $\mathbf{R}(u)$ and derive the following identity for transposed $\mathbf{R}$-matrices
\begin{align}
\begin{aligned}
\label{id_transp}
\mathbf{R}_{mn}^{t_n}(u)\mathbf{R}_{mn}^{t_n}(-u-2)& = r_{mn}(u)r_{mn}(-u-2) (\boldsymbol{\sigma}_2)^{\otimes n} \mathbf{R}_{mn}^{t}(-u-1)\mathbf{R}_{mn}^{t}(u+1)  (\boldsymbol{\sigma}_2)^{\otimes n} =\\&= r_{mn}(u)r_{mn}(-u-2)\,  \mathbbm{1} = \frac{\left(-u-1+\frac{m+n}{2} \right)\left(u+1+\frac{m+n}{2} \right)}{\left(-u-1+\frac{m-n}{2} \right)\left(-u-1+\frac{m-n}{2} \right)}\, \mathbbm{1}\,,
\end{aligned}
\end{align}
hence we apply it to the previous expression that simplifies drastically
\begin{align}
\begin{aligned}
&\left[ \mathbf{R}(0,Y')^{t_k} \mathbf{R}(Y',Y)  \mathbf{R}(Y,Y') \mathbf{R}(Y',-2i)^{t_k} \right]^{t_k}  \frac{\left(i\nu-1+\frac{{\ell}_k+n}{2} \right)\left(1-i\nu+\frac{{\ell}_k+n}{2} \right)}{\left(i\nu-1+\frac{{\ell}_k-n}{2} \right)\left(1-i\nu+\frac{{\ell}_k-n}{2} \right)} =\\&=\left[ \mathbf{R}(0,Y')^{t_k}\mathbf{R}(Y',-2i)^{t_k} \right]^{t_k}   \frac{\left(i\nu-1+\frac{{\ell}_k+n}{2} \right)\left(1-i\nu+\frac{{\ell}_k+n}{2} \right)}{\left(i\nu-1+\frac{{\ell}_k-n}{2} \right)\left(1-i\nu+\frac{{\ell}_k-n}{2} \right)} =\\&= \frac{\left(i\nu-1+\frac{{\ell}_k+n}{2} \right)\left(1-i\nu+\frac{{\ell}_k+n}{2} \right)}{\left(i\nu-1+\frac{{\ell}_k-n}{2} \right)\left(1-i\nu+\frac{{\ell}_k-n}{2} \right)} \frac{\left(1-i\nu'+\frac{{\ell}_k+n'}{2} \right)\left(i\nu'-1+\frac{{\ell}_k+n'}{2} \right)}{\left(1-i\nu'+\frac{{\ell}_k-n'}{2} \right)\left(i\nu'-1+\frac{{\ell}_k-n'}{2} \right)} \mathbbm{1} \,,
\end{aligned}
\end{align}
and taking into account the normalization factor in \eqref{hatR_def}, we recover the r.h.s. of \eqref{hatR_unit} $\square$.\\
\noindent\\
The second identity in \eqref{hatR_properties}
\begin{align}
\begin{aligned}
\label{hatR_2id}
&\hat{\mathbf{R}}_k (Y,Y')\hat{\mathbf{R}}_{k+1}(Y,Y'')\hat{\mathbf{R}}_k (Y',Y'')=\hat{\mathbf{R}}_{k+1} (Y',Y'')\hat{\mathbf{R}}_{k}(Y,Y'')\hat{\mathbf{R}}_{k+1} (Y,Y')\,,
\end{aligned}
\end{align}
is a natural consequence of the YBE for $\mathbf{R}(u)$. In order to prove it, we rewrite the l.h.s. of \eqref{hatR_2id} - stripped of the normalization prefactors of \eqref{hatR_def} and under the sign of transposition $t_k$ and $t_k+1$ - using the property \eqref{mat_transp}
\begin{align}
\begin{aligned}
\mathbf{R}(0,Y')^{t_k} \mathbf{R}(Y',Y) \mathbf{R}(Y,-2i)^{t_k} \mathbf{R}(0,Y'')^{t_{k+1}} \mathbf{R}(Y'',Y) \mathbf{R}(Y,-2i)^{t_{k+1}} \mathbf{R}(0,Y'')^{t_k} \mathbf{R}(Y'',Y') \mathbf{R}(Y',-2i)^{t_k}.
\end{aligned}
\end{align}
Secondly, we can move $ \mathbf{R}(0,Y'')^{t_{k+1}}$ and $\mathbf{R}(Y,-2i)^{t_{k+1}}$ to the left and right of the expression respectively, we use the Yang-Baxter equation to reshuffle the three central $\mathbf{R}$-matrices and finally we move $\mathbf{R}(0,Y'')^{t_k}$ and $\mathbf{R}(Y,-2i)^{t_k}$ to the left and right of the expression
\begin{align}
\begin{aligned}
\thinmuskip=-0.5mu
\medmuskip=-0.5mu
\thickmuskip=-0.5mu
& \mathbf{R}(0,Y'')^{t_{k+1}} \mathbf{R}(0,Y')^{t_k} \mathbf{R}(Y',Y) \mathbf{R}(Y,-2i)^{t_k} \mathbf{R}(Y'',Y) \mathbf{R}(0,Y'')^{t_k} \mathbf{R}(Y'',Y')\mathbf{R}(Y',-2i)^{t_k} \mathbf{R}(Y,-2i)^{t_{k+1}}\\&= \mathbf{R}(0,Y'')^{t_{k+1}} \mathbf{R}(0,Y')^{t_k} \mathbf{R}(Y',Y)\mathbf{R}(0,Y'')^{t_k} \mathbf{R}(Y'',Y) \mathbf{R}(Y,-2i)^{t_k}  \mathbf{R}(Y'',Y')\mathbf{R}(Y',-2i)^{t_k} \mathbf{R}(Y,-2i)^{t_{k+1}}\\&= \mathbf{R}(0,Y'')^{t_{k+1}} \mathbf{R}(0,Y')^{t_k} \mathbf{R}(0,Y'')^{t_k}\mathbf{R}(Y',Y) \mathbf{R}(Y'',Y) \mathbf{R}(Y'',Y') \mathbf{R}(Y,-2i)^{t_k} \mathbf{R}(Y',-2i)^{t_k} \mathbf{R}(Y,-2i)^{t_{k+1}}\,.
\end{aligned}
\end{align}
As third step, we recognize that the three central $\mathbf{R}$-matrices carrying arguments $Y,Y',Y''$ satisfy the Yang-Baxter equation, so we can re-write the last expression as
\begin{align}
\begin{aligned}
\mathbf{R}(0,Y'')^{t_{k+1}} \mathbf{R}(0,Y')^{t_k} \mathbf{R}(0,Y'')^{t_k}\mathbf{R}(Y'',Y')  \mathbf{R}(Y'',Y) \mathbf{R}(Y',Y)\mathbf{R}(Y,-2i)^{t_k} \mathbf{R}(Y',-2i)^{t_k} \mathbf{R}(Y,-2i)^{t_{k+1}}\,.
\end{aligned}
\end{align}
Let us select the five central $\mathbf{R}$-matrices in the last expression and proceed with a few steps. Namely, we insert the identity in the form of \eqref{id_transp} between $\mathbf{R}(Y'',Y')$ and $\mathbf{R}(Y'',Y)$ (the symbol $\propto$ stands for the coefficient in \eqref{id_transp} that in the next passages is neglected), and we use twice the Yang-Baxter relation in the form \eqref{YBE_trans}
\begin{align}
\begin{aligned}
\label{5_central}
&\mathbf{R}(0,Y'')^{t_k}\mathbf{R}(Y'',Y')\mathbf{R}(Y'',Y) \mathbf{R}(Y',Y)\mathbf{R}(Y,-2i)^{t_k} \propto\\&\propto\mathbf{R}(0,Y'')^{t_k}\mathbf{R}(Y'',Y') \mathbf{R}(Y',-2i)^{t_k}\mathbf{R}(0,Y')^{t_k} \mathbf{R}(Y'',Y) \mathbf{R}(Y',Y)\mathbf{R}(Y,-2i)^{t_k} =\\&= \mathbf{R}(Y',-2i)^{t_k}\mathbf{R}(Y'',Y') \mathbf{R}(0,Y'')^{t_k} \mathbf{R}(Y'',Y)\mathbf{R}(0,Y')^{t_k} \mathbf{R}(Y',Y)\mathbf{R}(Y,-2i)^{t_k} =\\&= \mathbf{R}(Y',-2i)^{t_k}\mathbf{R}(Y'',Y') \mathbf{R}(0,Y'')^{t_k} \mathbf{R}(Y'',Y)\mathbf{R}(Y,-2i)^{t_k}\mathbf{R}(Y',Y)\mathbf{R}(0,Y')^{t_k}  \,.
\end{aligned}
\end{align}
The matrices $\mathbf{R}(Y',-2i)^{t_k}$ and $\mathbf{R}(0,Y')^{t_k} $ get canceled once we insert back the last expression into the full one, and the cancelation produces twice the opposite of the neglected factor between first and second row of \eqref{5_central}, leading to
\begin{align}
\begin{aligned}
\propto \mathbf{R}(0,Y'')^{t_{k+1}}\mathbf{R}(Y'',Y') \mathbf{R}(0,Y'')^{t_k} \mathbf{R}(Y'',Y)\mathbf{R}(Y,-2i)^{t_k}\mathbf{R}(Y',Y) \mathbf{R}(Y,-2i)^{t_{k+1}}\,.
\end{aligned}
\end{align}
At this point it is enough to insert again the identity in the form of \eqref{id_transp} between $\mathbf{R}(Y'',Y')$ and $\mathbf{R}(Y'',Y)$, in the form $\mathbf{R}(Y',-2i)^{t_{k+1}}\mathbf{R}(0,Y')^{t_{k+1}}$, producing the same neglected factor that gets - all in all - canceled
\begin{equation}
\begin{aligned}
\thinmuskip=0mu
\medmuskip=0mu
\thickmuskip=0mu
&\mathbf{R}(0,Y'')^{t_{k+1}}\mathbf{R}(Y'',Y') \mathbf{R}(0,Y'')^{t_k} \mathbf{R}(Y'',Y)\mathbf{R}(Y,-2i)^{t_k}\mathbf{R}(Y',Y) \mathbf{R}(Y,-2i)^{t_{k+1}}\propto\\&\propto\mathbf{R}(0,Y'')^{t_{k+1}}\mathbf{R}(Y'',Y')\mathbf{R}(Y',-2i)^{t_{k+1}}\mathbf{R}(0,Y')^{t_{k+1}} \mathbf{R}(0,Y'')^{t_k} \mathbf{R}(Y'',Y)\mathbf{R}(Y,-2i)^{t_k}\times \\ &\qquad \qquad \qquad\qquad \qquad\qquad \qquad \qquad\qquad \qquad \qquad\qquad \qquad \qquad \qquad \times\mathbf{R}(Y',Y) \mathbf{R}(Y,-2i)^{t_{k+1}}=\\&=\mathbf{R}(0,Y'')^{t_{k+1}}\mathbf{R}(Y'',Y')\mathbf{R}(Y',-2i)^{t_{k+1}}\mathbf{R}(0,Y'')^{t_k} \mathbf{R}(Y'',Y)\mathbf{R}(Y,-2i)^{t_k}\times \\ &\qquad\qquad \qquad \qquad\qquad \qquad \qquad\qquad \qquad \qquad\qquad \qquad \qquad\times\mathbf{R}(0,Y')^{t_{k+1}} \mathbf{R}(Y',Y) \mathbf{R}(Y,-2i)^{t_{k+1}}\,.
\end{aligned}
\end{equation}
The last expression and the one we started from are - modulo the same normalization and the transpositions $t_k$, $t_{k+1}$ - the l.h.s and the r.h.s. of \eqref{hatR_2id} $\square$.
\bibliographystyle{nb}
\bibliography{biblio_v2}

\end{document}